\documentclass[a4paper,11pt]{article}
\usepackage{geometry,setspace}
\usepackage{hyperref}  
\usepackage{xcolor}
\hypersetup{
    colorlinks,
    linkcolor={blue!50!black},
    citecolor={blue!50!black},
    urlcolor={blue!80!black}
}
\usepackage{amsmath,amsfonts,amssymb,mathtools,mathrsfs}
\usepackage{theorem}
\usepackage{graphicx}
\usepackage{array}
\usepackage{enumerate}
\usepackage{latexsym}
\usepackage{fancyhdr}
\usepackage{sectsty}
\usepackage{url}
\usepackage{epstopdf}
\usepackage{booktabs}
\usepackage{tabularx,multirow,rotating}
\usepackage{natbib}
\usepackage[labelfont=sc]{caption}

\graphicspath{{C:/Temp/peter/}}
\newtheorem{theorem}{Theorem}

{
\theorembodyfont{\upshape}
\theoremheaderfont{\scshape}

}

{\theorembodyfont{\upshape}
\newtheorem{remark}{Remark}[section]
}

\newtheorem{lemma}{{\sc Lemma}}

\newenvironment{proof}[1][Proof]{\bigskip \noindent \textbf{#1:} }{\  \rule{0.5em}{0.5em}}

\theoremheaderfont{\bfseries}
\theorembodyfont{\upshape}
\sectionfont{\centering \sc}
\subsectionfont{\centering \sc}
\subsubsectionfont{\centering \sc}
\allsectionsfont{\centering \sc}
\allowdisplaybreaks
\bibpunct[:~]{(}{)}{;}{a}{,}{;}
\renewcommand{\cite}{\citet}
\setcitestyle{notesep={, }}

\begin{document}
\begin{center}
\vspace{-0.3cm}{\Large BOOTSTRAP INFERENCE FOR\ HAWKES\ AND\ GENERAL\ POINT
PROCESSES}\bigskip

\vspace{-0.3cm}
\renewcommand{\thefootnote}{}
\footnote{
\hspace{-7.2mm}
$^{a}%
$Department of Economics, Exeter Business School, UK, and Department of Economics, University of Bologna, Italy\\
$^{b}$School of Economics, University of Sydney, Australia\\
$^{c}$Department of Economics, University of Copenhagen, Denmark\\
$^{d}$Center for Bubble Studies, University of Copenhagen, Denmark\\
Correspondence to: Giuseppe Cavaliere, Department of Economics, University of Bologna,
Piazza Scaravilli 2, I-40126 Bologna, Italy. Email: giuseppe.cavaliere@unibo.it.}
\addtocounter{footnote}{-1}
\renewcommand{\thefootnote}{\arabic{footnote}}
\vspace{0.1cm}

Giuseppe Cavaliere$^{a}$, Ye Lu$^{b}$, Anders Rahbek$^{c}$ and Jacob
St\ae rk-\O stergaard$^{d}$
\bigskip

First Draft: March 2021; This version: September 2021
\end{center}
\bigskip

\par\begingroup\leftskip=0.8cm\rightskip=0.8cm
\small

\begin{center}
\textsc{Abstract}
\end{center}
Inference and testing in general point process models such as the Hawkes model
is predominantly based on asymptotic approximations for likelihood-based
estimators and tests. As an alternative, and to improve finite sample
performance, this paper considers bootstrap-based inference for interval
estimation and testing. Specifically, for a wide class of point process models
we consider a novel bootstrap scheme labeled `fixed intensity
bootstrap' (FIB), where the conditional intensity is kept fixed across
bootstrap repetitions. The FIB, which is very simple to implement and fast in
practice, extends previous ideas from the bootstrap literature on time series
in discrete time, where the so-called `fixed design' and `fixed
volatility' bootstrap schemes have shown to be particularly useful and
effective. We compare the FIB with the classic recursive bootstrap, which is
here labeled `recursive intensity bootstrap' (RIB). In RIB algorithms, the
intensity is stochastic in the bootstrap world and implementation of the
bootstrap is more involved, due to its sequential structure. For both
bootstrap schemes, we provide new bootstrap (asymptotic) theory which
allows to assess bootstrap validity, and propose a 
`non-parametric' approach based on resampling
time-changed transformations of the original waiting times. We also establish 
the link between the proposed bootstraps for point process models and
the related autoregressive conditional duration (ACD) models.
Lastly, we show
effectiveness of the different bootstrap schemes in finite samples through a
set of detailed Monte Carlo experiments, and
provide applications to both financial data and social media data
to illustrate the proposed methodology.
\bigskip

\bigskip

\bigskip

\numberwithin{equation}{section}
\par\endgroup\normalsize
\renewcommand{\thefootnote}{\arabic{footnote}}

\section{Introduction}

Point processes are well-known to be useful tools to characterize dynamics of
event occurrence times. This includes the homogeneous Poisson process where
the intensity process is constant over time, the inhomogeneous Poisson
process, where the intensity is a deterministic (or strictly exogenous)
time-varying function, as well as the class of `self-exciting' point
processes, such as the well-known and much applied Hawkes process. In
particular, for the Hawkes process, the conditional intensity process depends
on all past history of the events and thereby allows for (exponential or
fractional) memory features, similar to autoregressive or fractional
time-series processes in discrete time series econometrics. The self-exciting
class of models, which are the focus of this paper, were originally proposed
for modelling earthquake sequences; see \citet{O1988} and the references therein. Later, they have been put to
use in a wide range of applications such as financial transactions
\citep{B2007, BH2009}, financial contagion \citep{ASCDL2015}, monetary policy
\citep{DMD2002}, criminal fights and relations \citep{MSBST2011}, forecasting
electricity price spikes \citep{CHH2015} and the rich literature on social
network information diffusion \citep{RLMX2017}, among others. Hawkes processes
are also closely related to the class of autoregressive conditional duration [ACD]
models of \citet{ER1998}, which are well known and much used in
financial economics; see Sections \ref{sec: MLE} and
\ref{Sec on bootstrapping ACD} below for the relation between the two classes
of processes.

Inference for self-exciting point process models is generally performed
through classic, likelihood-based asymptotic inference and testing\footnote{As
an alternative, the general methods of moments (GMM) has also been used, see
e.g. \citet{ASCDL2015}.}, as originally discussed in \citet{O1978}. However,
see e.g. \citet{R2018} and \citet{WSJ2010}, the finite sample performance of
asymptotic inference is not always satisfactory. This is in general the case
because the finite sample distributions of the estimators are often very
skewed and far from the Gaussian asymptotic distribution.

In this framework, a key motivation for the results presented in the paper is
to provide a simple to implement, and theoretically well-grounded, bootstrap
approach to inference in self-exciting point process models. We do this by
providing six main contributions.

The first contribution is to propose a novel (non-)parametric bootstrap scheme
for such point process models, which we label as `fixed intensity bootstrap'
(FIB). The FIB is simple and fast to implement in practice -- particularly so
when compared to existing (recursive) applications of the bootstrap. The key
difference between the new and the classic bootstrap schemes is how to
generate the sequence of waiting times in the bootstrap world. Specifically,
while for standard, recursive bootstrap schemes, the bootstrap event times are
generated recursively through the past bootstrap events, for the novel
bootstrap scheme the bootstrap event times are generated using a `fixed'
conditional intensity function, which entirely depends on the event times in
the original world. Therefore, the FIB contrasts with existing implementations
of the bootstrap, see e.g. \citet{ELL2011} and \citet{SNCWWSB2011}, which
utilize a (possibly highly complex and time consuming) sequential update of
the bootstrap conditional intensities.

The second contribution is to provide bootstrap (first-order) asymptotic
theory, including establishing bootstrap validity for inference and testing in
point process models for both the novel (FIB) bootstrap and for the classic
recursive bootstrap (for which no theory exists in the literature). We show
that the bootstrap based on the FIB is valid under regularity conditions which
are milder than those required for validity of recursive bootstrap schemes
(hereafter, RIB).

The third contribution is to introduce novel `non-parametric' implementations
of the FIB and RIB schemes, which are based on resampling time-changed
transformations of the original waiting times, rather than generating the
(transformed) waiting times through a parametric model (usually the
exponential distribution), as is in the literature. These implementations are
likely to be robust to model misspecifications which generate non-exponential
transformed waiting times. We show how to scale the original time-changed
waiting times properly and to resample them; we also show that, in the
homogeneous case, validity of the implied bootstrap follows from a time-change
functional central limit theorem derived by \citet{B1968} which, as far as we
are aware, has never been applied to the bootstrap of self-exciting point
process models.

The fourth contribution of the paper is a detailed Monte Carlo simulation
study on the performance of the bootstrap for self-exciting Hawkes processes.
Possibly due to the high computational costs involved in the implementation of
a simulation study for the bootstrap in this framework, to the best of our
knowledge studies like ours have not been attempted in the literature. We show
that for Hawkes processes with exponential kernels, the coverage probabilities
of confidence intervals based on the Gaussian asymptotic approximation may be
well below the nominal level. In contrast, the bootstrap is able to correct
this and, in particular, FIB implementations are particularly well performing
in terms of coverage probabilities.

The fifth contribution is to provide two real data examples where we
illustrate the key differences between asymptotic and the various bootstrap
inference methods in applications. The first refers to the problem of modeling
and predicting extreme financial results, see \citet{ELL2011}. We use this
example, including the data sample considered in \citet{ELL2011}, to compare
the outcome of the four different bootstrap schemes discussed in the paper.
The second example is based on social media data and considers the flows of
tweets and re-tweets proceeding and following a political announcement.
Specifically, using recent tweets related to the COVID-19 pandemic in Denmark,
we show how bootstrap-based inference is able to detect structural breaks (in
the mean intensity as well as in the decay rate of intensity) induced by the
announcement, which may not be detected based on asymptotic inference.

Sixth, we discuss the link between the proposed bootstraps based on the point
process representation and bootstrap inference for autoregressive conditional 
duration (ACD) models. 
We establish the relation between our proposed bootstrap schemes and bootstrap
algorithms based on the ACD representation. Specifically, we show that our
recursive bootstrap corresponds to a residual-based bootstrap in the ACD world
\citep[as discussed e.g. in][]{PS2021}, with the crucial
difference that the number of events generated through our scheme is random,
rather than being fixed. This is a key improvement, as our bootstrap ensures
that the sum of the bootstrap waiting times always cover the original time
interval. We also show that residual-based implementation of the bootstrap in
the ACD world corresponds to our proposed non-parametric bootstrap with
re-sampling based on the (estimated) transformed waiting times. Finally, we
discuss the relation between our proposed fixed intensity bootstrap and a
bootstrap in the ACD framework, which is novel in the literature, where the
conditional duration in the bootstrap world is fixed to the estimated
conditional duration in the original sample.

\subsection*{Structure of the paper}

The paper is organized as follows. In Section~\ref{sec: MLE} likelihood-based
analysis for point processes inference is presented, and in
Section~\ref{sec: bootstrap} the novel fixed intensity, as well as the
recursive intensity, bootstraps are discussed, with theory and validity
results in Section~\ref{sec BS validity}. Non-parametric bootstrap is discussed in Section~\ref{sec NPAR}, and the relation between our bootstraps and the bootstraps for ACD models is discussed in Section~\ref{Sec on bootstrapping ACD}. Section~\ref{Sec MC sim} provides a Monte Carlo
study of the different schemes. Section~\ref{sec: empirical} contains two
empirical illustrations, and Section~\ref{sec: concl} concludes. All proofs
are contained in the Appendix.

\subsection*{Notation}

We use the counting process $N(t)$ to characterize the total number of events
occurring before and including time $t$, with $N(s,t]$ and $N[s,t)$ the
numbers of events in the interval $(s,t]$ and $[s,t)$, respectively, for
$s<t$. For a right-continuous natural filtration $(\mathcal{F}_{t}%
)_{t\in\mathbb{R}}$ of a continuous time stochastic process, we denote by
$\mathcal{F}_{t-}$ the left limit of $\mathcal{F}_{t}$, which contains all the
information before but not including time $t$. We use $\mathbb{I}(\cdot)$ to
denote the indicator function, and define $\mathbb{R}^{+}:=(0,\infty)$ and
$\mathbb{R}_{+}:=[0,\infty)$. For $x\in\mathbb{R}$, $\lfloor x\rfloor
:=\max_{z\in\mathbb{Z}}\{z\leq x\}$. For the bootstrap, as is standard, we
denote by $P^{\ast}$ the probability measure induced by the bootstrap;
expectation and variance computed under $P^{\ast}$ are denoted by $E^{\ast}$
and $V^{\ast}$, respectively. For a sequence $X_{T}^{\ast}$ computed on the
bootstrap data, $X_{T}^{\ast}\overset{p^{\ast}}{\rightarrow}_{p}0$ or
$X_{T}^{\ast}=o_{p}^{\ast}(1)$, in probability, denote that $P^{\ast}%
(|X_{T}^{\ast}|>\epsilon)\rightarrow0$ in probability for any $\epsilon>0$;
$X_{T}^{\ast}=O_{p}^{\ast}(1)$, in probability, denotes that there exists a
$c>0$ such that $P^{\ast}(|X_{T}^{\ast}|>c)\rightarrow0$ in probability; with
$X_{T}^{\ast}\overset{d^{\ast}}{\rightarrow}_{p}X$ (weak convergence in
probability) we mean that $E^{\ast}(g(X_{T}^{\ast}))\overset{p}{\rightarrow
}E(g(X))$ for all continuous bounded functions $g$, in each case as
$T\rightarrow\infty$. Finally, $\mathcal{N}$ denotes a Gaussian random
variable and, for $\mu>0$, $\mathcal{E}(\mu)$ denotes an exponential random
variable with mean $1/\mu$.

\section{Likelihood-based analysis of the point process}\label{sec: MLE} 

We discuss here likelihood-based estimation for a general
class of point process models. For later use when establishing asymptotic
validity of the bootstrap, we state explicit sufficient conditions for classic
likelihood-based asymptotic theory. Precisely, and as in \citet{O1978}, we
establish consistency and limiting distributions of likelihood-based
estimators, as well as the related (likelihood ratio) test statistics.

\subsection{The model}

By assumption, the observed event times are realizations from a univariate
point process, i.e. a collection $\{t_{i}\}_{i=1}^{\infty}$, $t_{i}>0$, of
stochastic event times with associated waiting times (or durations),
$w_{i}:=t_{i}-t_{i-1}$, for $i=1,2,...$ with $t_{0}:=0$; see e.g.
\citet{DVJ2003} for an introduction to point processes. The point process can
be equivalently characterized by the continuous-time counting process
\begin{equation}
N(t):=\sum_{i\geq1}\mathbb{I}(t_{i}\leq t), \label{eq: N(t)}%
\end{equation}
for $t\geq0$, with associated filtration $(\mathcal{F}_{t}),t\geq0$ where
$\mathcal{F}_{t}$ is the $\sigma$-field generated by $\{N(s),s\leq t\}$.

In addition, and as used here predominantly, a regular point process is
uniquely defined by its conditional intensity process, $\lambda(t)$, $t\geq0$,
which captures the instantaneous conditional probability of event
occurrences\footnote{Note that the limit on the right-hand side of
\eqref{eq: intens1} is assumed to exist, such that the conditional
distribution of the waiting times is continuous.} and is defined as
\begin{equation}
\lambda(t):=\lim_{\delta\rightarrow0^{+}}\frac{1}{\delta}P(N[t,t+\delta
)>0|\mathcal{F}_{t-}). \label{eq: intens1}%
\end{equation}
Observe that, as the point process is assumed to be regular and orderly,
$\lambda(t)$ essentially captures the instantaneous conditional probability of
observing a single event at each time $t$.

A key example used throughout is the `self-exciting' Hawkes point process,
where the conditional intensity is given by
\begin{equation}
\lambda(t)
=\mu+\int_{-\infty}^{t}\gamma(t-s)dN(s)
=\mu+\sum_{t_{i}<t}\gamma(t-t_{i}),
\label{eq: Hawkes example}
\end{equation}
where the $\mu>0$ is the baseline intensity and $\gamma(t)$ is the so-called
kernel function, which typically is either exponential,
\begin{equation}
\gamma(x)=\alpha\exp(-\beta x), \label{eq: exp kernel}%
\end{equation}
or following a power law,
\begin{equation}
\gamma(x)=\alpha(x+\beta)^{-\delta}, \label{eq: power law}%
\end{equation}
where $\alpha,\beta,\delta\geq0$.

Note as the sum in (\ref{eq: Hawkes example}) is over all events $t_{i}$ prior
to $t$, the Hawkes process has infinite (or long) memory. In contrast, if
$\gamma(t)=0$, $\lambda(t)=\mu>0$, then the point process reduces to a
homogeneous Poisson process which has i.i.d. exponentially distributed waiting
times $w_{i}$ with rate $\mu$, that is, the $w_{i}$'s are i.i.d.
$\mathcal{E}(\mu)$ distributed. Likewise, an example of counting process with
finite memory (or, $q$ `lags'), sometimes referred to as a `Wold process', is
given by
\begin{equation}
\lambda(t)=\mu+\gamma(t-t_{N(t-)},...,t-t_{N(t-)-q+1}), \label{eq: Wold}%
\end{equation}
where $\gamma(\cdot)$ is a mapping from $\mathbb{R}_{+}^{q}$ to $\mathbb{R}%
_{+}$. A specific example is given by
\[
\gamma(t-t_{N(t-)},...,t-t_{N(t-)-q+1};\theta)=\sum_{i=1}^{q}\gamma
_{i}(t-t_{N(t-)-i+1};\theta),
\]
with $\gamma_{i}(\cdot)$ being exponential or power law kernel functions as in
\eqref{eq: exp kernel} and \eqref{eq: power law} for $i=1,2,...,q$. Notice
that for $q=1$, this is an example of a renewal process, with associated
i.i.d. waiting times $w_{i}$ which are not exponentially distributed.

The class of self-exciting point process models is also linked to the ACD
model of \citet{ER1998}, which is based on the following dynamic
equation for the waiting times $w_i:=t_i-t_{i-1}$ between events:
\begin{equation}
w_{i}=\psi_{i}\varepsilon_{i}, \quad
\psi_{i}=E(w_{i}|w_{i-1},...,w_{0})=\psi(w_{i-1},...,w_{0}),
\label{eq acd model}
\end{equation}
where the $\varepsilon_{i}$'s are strictly positive i.i.d. random variables
with mean one. The ACD model can be given a point process representation;
specifically, the conditional intensity associated to the model, see \citet{ER1998}, takes the form
\begin{equation}
\lambda(t)=\lambda_{\varepsilon}\left(\frac{t-t_{N(t-)}}
{\psi_{N(t-)+1}}\right)\frac{1}{\psi_{N(t-)+1}},
\label{eq acd intensity}
\end{equation}
where $\lambda_{\varepsilon}(\cdot)=p_{\varepsilon}(\cdot)/S_{\varepsilon
}(\cdot)$, with $p_{\varepsilon}$ and $S_{\varepsilon}$ denoting the pdf and
the survival function of $\varepsilon_{i}$, respectively. A simple example is
the ACD(1) with exponential errors,
\[
w_{i}=\psi_{i}\varepsilon_{i},\quad
\psi_{i}=\omega+\alpha w_{i-1},\quad
\varepsilon_{i}\sim\mathcal{E}(1),
\]
with intensity given by the piecewise constant function
\[
\lambda(t)=\frac{1}{\psi_{N(t-)+1}}
=\frac{1}{\omega+\alpha w_{N(t-)}}
=\frac{1}{\omega+\alpha(t_{N(t-)}-t_{N(t-)-1})},
\]
which is a special case of a Wold process with two `lags'. Should
$\varepsilon_{i}$ be a continuous, non-exponentially distributed random
variable, it follows from (\ref{eq acd intensity}) that the intensity of the
ACD(1) takes the form
\[
\lambda(t)
=\lambda_{\varepsilon}\left(\frac{t-t_{N(t-)}}{\omega+\alpha w_{N(t-)}}\right)  \frac{1}{\omega+\alpha w_{N(t-)}}.
\]
Finally, with $\alpha=0$ the ACD reduces to a renewal process with intensity
$\lambda(t)=\lambda_{\varepsilon}(\omega^{-1}(t-t_{N(t-)}))\omega^{-1}$.

\subsection{Likelihood-based estimation}

For the statistical analysis we assume that the conditional intensity
$\lambda(t)$ in (\ref{eq: intens1}) is parameterized by a finite-dimensional
vector of unknown parameters $\theta\in\Theta\subseteq\mathbb{R}^{d}$,
$d:=\dim\theta$. To emphasize the dependence of the intensity on $\theta$, we
write $\lambda(t;\theta)$ and, for the associated counting process,
$N(t;\theta)$. For notational convenience, when evaluated at the true value,
we write $\lambda(t;\theta_{0})=:\lambda(t)$ and $N(t;\theta_{0})=:N(t)$.

Consider a sample of event times $t_{1},t_{2},\ldots,t_{n_{T}}$ observed in a
time interval $[0,T]$, with $n_{T}=N(T)$ the total number of events in the
interval. Standard arguments as in \citet{DVJ2003} imply that the joint
log-likelihood function $\ell_{T}(\theta)$ can be written as
\begin{align}
\ell_{T}(\theta)
&=\int_{0}^{T}\log\lambda(t;\theta)dN(t)-\Lambda(T;\theta)
=\sum_{i=1}^{n_T}\left(\log\lambda(t_i;\theta)-\int_{t_{i-1}}^{t_{i}}\lambda(t)dt\right) \label{eq log lik compact}
\end{align}
where $\Lambda(\cdot;\theta)$ is the so-called integrated intensity, given by
\begin{equation}
\Lambda(t;\theta):=\int_{0}^{t}\lambda(s;\theta)ds,
\label{eq integrated intensity function}
\end{equation}
and we assume $t_{n_{T}}=T$ (such that $T$ coincides with the last event time) 
in deducing the second equality in \eqref{eq log lik compact}.
The maximum likelihood estimator (MLE) $\hat{\theta}_{T}$ is defined by
\begin{equation}
\hat{\theta}_{T}:=\arg\max_{\theta\in\Theta}\ell_{T}(\theta).
\label{eq MLE def}%
\end{equation}
For the Hawkes model, $\lambda(t;\theta)$ is given by \eqref{eq: intens1} with
$\theta=(\mu,\alpha,\beta)^{\prime}$ for the exponential kernel in
\eqref{eq: exp kernel} and $\theta=(\mu,\alpha,\beta,\eta)^{\prime}$ for the
power law kernel in \eqref{eq: power law}. Then the log-likelihood
function $\ell_{T}(\theta)$ in \eqref{eq log lik compact} becomes
\begin{equation}
\ell_{T}(\theta)=\sum_{i=1}^{n_{T}}\left(\log\Big(\mu+\sum\limits_{j<i}
\gamma(t_{i}-t_{j};\theta)\Big)-\int_{t_{i-1}}^{t_{i}}\Big(\mu
+\sum\limits_{j<i}\gamma(t-t_{j};\theta)\Big)dt\right).
\label{eq: Hawkes Lik}
\end{equation}

Note that for the special case of a homogeneous Poisson process where
$\lambda(t)=\mu$, then $\theta=\mu$, and the log-likelihood simplifies to
$\ell_{T}(\theta)=n_{T}\log\theta-T\theta$. Hence, in this special case the
MLE has the closed form $\hat{\theta}_{T}=n_{T}/T$.

\subsection{Asymptotic theory}

For the asymptotic theory of $\hat{\theta}_{T}$ we assume that the information
set $\mathcal{F}_{t}$ is defined as the $\sigma$-field generated by
$\{N(s,t],-\infty<s\leq t\}$. Under mild requirements
\citep[e.g.][Assumption C]{O1978}, the analysis presented below extends to the
case where $\mathcal{F}_{t}=\{N(s),0\leq s\leq t\}$. Likewise, we assume for
simplicity that $t_{n_{T}}=T$.

A key role in the asymptotic analysis here -- as well as for the novel
bootstrap asymptotics below -- is played by the Doob-Meyer decomposition of
$N(t)$ in \eqref{eq: N(t)}, which is given by
\[
N(t)=M(t)+A(t).
\]
Here $M$ is a square integrable continuous-time $\mathcal{F}_{t}$-local
martingale and $A(t)$ is the compensator of $N(t)$, which in this case is
given by the integrated intensity $\Lambda(t;\theta_{0})=\Lambda(t)$ in
\eqref{eq integrated intensity function}; that is, $A(t)=\int_{0}^{t}%
\lambda(s)ds=\Lambda(t)$. By definition,
\begin{equation}
M(t)=N(t)-\Lambda(t),\ t\geq0, \label{eq def of M}%
\end{equation}
is a continuous-time martingale, and we may write $E(dM(t)|\mathcal{F}%
_{t-})=E(dN(t)-\lambda(t)dt|\mathcal{F}_{t-})=0$ for any $t>0$. Since
$\lambda(t)$ is $\mathcal{F}_{t-}$-measurable, it follows that
$E(dN(t)|\mathcal{F}_{t-})=\lambda(t)dt$ \citep[see also][p. 250]{O1978}
which will be used repeatedly throughout for both the standard and
the bootstrap asymptotic analyses. Furthermore, we make the following
technical assumptions. \bigskip

\noindent\textsc{Assumption 1}

\medskip\noindent(a) The parameter space $\Theta\subseteq\mathbb{R}^{d}$ is
compact with $\theta_{0}\in\Theta_{0}\subset\Theta$;

\medskip\noindent(b) For $\theta\in\Theta_{0}$, with $N(\cdot;\theta)$
denoting the counting process $N(\cdot)$ indexed by $\theta$, $N(\cdot
;\theta)$ is an orderly point process with stationary and ergodic increments.
Moreover, $E\big(\sup_{m\geq1}m(N(t+1/m;\theta)-N(t;\theta
))^{2}\big)<\infty$;

\medskip\noindent(c) the intensity process $\lambda(t;\theta)$ satisfies the
following conditions almost surely: (i) it is predictable (left-continuous)
for all $\theta$, continuous in $\theta$ and strictly positive; (ii) for all
$\theta$, $|\lambda(t;\theta)|\leq\xi_{1}(\theta)$ with $E(\xi_{1}(\theta
)^{2})<\infty$; (iii) $\lambda(t;\theta_{1})=\lambda(t;\theta_{2})$ if and
only if $\theta_{1}=\theta_{2}$. \bigskip

Notice that Assumption 1(c) implies in particular that $\log\lambda(t;\theta)$
has a finite second order moment.

Consistency of the MLE is given in the next theorem from \citet{O1978}.

\begin{theorem}
[Ogata, 1978]\label{Th consistency} Under Assumption 1, $\hat{\theta}%
_{T}\overset{p}{\rightarrow}\theta_{0}$.
\end{theorem}

For the analysis of the score and the information, and for establishing the
asymptotic normality of the MLE, we make use of the Assumption 2 below, where
we use the following notation: for any function $f(t;\theta)$ of $\theta$ (and
$t$), $f(t):=f(t;\theta_{0})$, $\partial_{\theta}f(t;\theta)=\partial
f(t;\theta)/\partial\theta$ and $\partial_{\theta_{0}}f(t)=\left.  \partial
f(t;\theta)/\partial\theta\right\vert _{\theta=\theta_{0}}$ (and similarly for
higher order and partial derivatives). \bigskip

\noindent\textsc{Assumption 2}

\medskip\noindent(a) The intensity process $\lambda(t;\theta)$ satisfies the
following conditions almost surely: (i) $\lambda(t;\theta)$ is continuously
differentiable with respect to $\theta$ up to order three, for all $t\geq0$;
(ii) $E((\partial_{\theta_{i}}\lambda(t;\theta))^{2})<\infty$ and
$E((\partial_{\theta_{i},\theta_{j}}^{2}\lambda(t;\theta))^{2})<\infty$ for
all $\theta$;

\medskip\noindent(b) With $h(t;\theta):=\lambda(t;\theta)^{-1}(\partial
_{\theta}\lambda(t;\theta))(\partial_{\theta}\lambda(t;\theta))^{\prime}$ and
$I(\theta):=E(h(t;\theta))$, it holds that $I(\theta_{0})>0$ and each element
of $h(t):=h(t;\theta_{0})$ has finite variance;

\medskip\noindent(c) With $N_{\epsilon}(\vartheta)$ denoting a neighborhood of
$\vartheta$, for all $\vartheta\in\Theta$,
\[
\sup_{\theta\in N_{\epsilon}(\vartheta)}|\partial_{\theta_{i},\theta
_{j},\theta_{k}}^{3}\lambda(t;\theta)|\leq c_{ijk}(t),\quad\sup_{\theta\in
N_{\epsilon}(\vartheta)}|\partial_{\theta_{i},\theta_{j},\theta_{k}}^{3}%
\log\lambda(t;\theta)|\leq d_{ijk}(t)
\]
where $c_{ijk}(t),d_{ijk}(t)$ are stationary and ergodic processes with
$E(c_{ijk}(t))<\infty$ and $E(\lambda(t)^{2}d_{ijk}^{2}(t))<\infty$. \bigskip

Note that Assumption 2(c) differs from standard requirements as in
\citet{O1978} which address uniformity of the Hessian.

Next, let $S_{T}(\theta):=\partial_{\theta}\ell_{T}(\theta)$ and $H_{T}%
(\theta):=\partial_{\theta}^{2}\ell_{T}(\theta)$ denote the score and the
Hessian, respectively. Using \eqref{eq log lik compact} and
\eqref{eq def of M} it holds that
\begin{align}
S_{T}(\theta_{0})  &  =\int_{0}^{T}\partial_{\theta_{0}}\log\lambda
(t)dM(t),\label{eq score at theta0}\\
H_{T}(\theta_{0})  &  =\int_{0}^{T}\partial_{\theta_{0}}^{2}\log
\lambda(t)dM(t)-\int_{0}^{T}h(t)dt. \label{eq HEssian at theta0}%
\end{align}
Applying \citet[][Lemma 1]{JR2004} the following theorem holds, where we
here also consider the distribution of the likelihood ratio test statistic
$LR_T(\theta_{0})$ for a simple null hypothesis $H_{0}:\theta=\theta_{0}$.

\begin{theorem}
\label{Th asy normality}Under Assumption 1 and 2(a),(b),
\[
T^{-1/2}S_{T}(\theta_{0})\overset{d}{\rightarrow}\mathcal{N}(0,I(\theta
_{0}))\quad\text{and}\quad-T^{-1}H_{T}(\theta_{0})\overset{p}{\rightarrow
}I(\theta_{0}).
\]
Moreover, if also Assumption 2(c) holds, then
\[
T^{1/2}(\hat{\theta}_{T}-\theta_{0})\overset{d}{\rightarrow}\mathcal{N}%
(0,I(\theta_{0})^{-1})
\]
and
\begin{align}
LR_{T}(\theta_{0}):=2(\ell_{T}(\hat{\theta}_{T})-\ell_{T}(\theta_{0}))\overset{d}{\rightarrow}\chi_{d}^{2}.
\label{eq LR test}
\end{align}
\end{theorem}

\section{The bootstrap}

\label{sec: bootstrap} We discuss here two bootstrap schemes. The first
bootstrap, which is novel, is denoted as the `fixed intensity bootstrap'
(FIB). The FIB as proposed here builds on ideas from the `fixed design
bootstrap' in regression (and time series) models
\citep[see e.g.][]{W1986,GK2004}, as well as the so-called `fixed volatility
bootstrap' in conditional volatility modelling \citep[see][]{CPR2018}, in the
sense that the bootstrap intensity function is fixed across bootstrap
repetitions. The second scheme, which has been applied in e.g. \citet{ELL2011}
and \citet{SNCWWSB2011}, is here denoted the `recursive intensity bootstrap'
(RIB). As will be discussed later, in practice the FIB is simpler and faster
to implement than the RIB, in addition to being valid under milder regularity
conditions. Since no theory exists for either of the FIB and the RIB schemes,
in Section~\ref{sec BS validity} we establish validity of the bootstrap for both.

\subsection{Random time change}

\label{sec time change} A key property we employ in defining our bootstrap
algorithms is that using the integrated intensity to transform the original
event times $\{t_{i}\}$ to another sequence of event times $\{s_{i}\}$ gives a
homogeneous Poisson process with unit intensity. Equivalently, the original
and non i.i.d. waiting times $\{w_{i}\}$, $w_{i}=t_{i}-t_{i-1}$, can be
transformed into new waiting times $\{v_{i}\}$, $v_{i}=s_{i}-s_{i-1}$, which
are i.i.d. $\mathcal{E}\left(  1\right)  $, see also \citet{DVJ2003}.

The time change transformation $t_{i}\mapsto s_{i}$ is given by
\begin{equation}
s_{i}(\theta):=\Lambda(t_{i};\theta), \label{eq TFT(theta)}%
\end{equation}
where the integrated intensity $\Lambda(t;\theta)$ is defined in
(\ref{eq integrated intensity function}). Moreover, with
\begin{equation}
\Lambda(t_{i},t_{i-1};\theta):=\Lambda(t_{i};\theta)-\Lambda(t_{i-1}%
;\theta)=\int_{t_{i-1}}^{t_{i}}\lambda(t;\theta)dt, \label{eq: piece Lam}%
\end{equation}
the associated transformed waiting times $v_{i}(\theta)$ are given by
\begin{equation}
v_{i}(\theta):=s_{i}(\theta)-s_{i-1}(\theta)=\Lambda(t_{i},t_{i-1};\theta),
\label{eq TFTwait}%
\end{equation}
for $i=1,2,...$. By definition, at the true value $\theta_{0}$ the transformed
waiting times $v_{i}:=v_{i}(\theta_{0})=\Lambda(t_{i},t_{i-1};\theta_{0})$ are
i.i.d. $\mathcal{E}\left(  1\right)  $, such that the transformed event times,
$s_{i}:=s_{i}(\theta_{0})$, form a homogeneous Poisson process with unit
intensity. For the Hawkes process in
\eqref{eq: Hawkes example}, $\lambda(t)=\mu+\sum_{t_{j}<t}\gamma(t-t_{j})$,
and hence
\begin{align}
v_{i}=\Lambda(t_{i},t_{i-1};\theta_{0})  &  =\int_{t_{i-1}}^{t_{i}}%
\Big(\mu+\sum\nolimits_{t_{j}<t}\gamma(t-t_{j})\Big)dt
\label{eq: Hawks change}\\
&  =\mu w_{i}+\int_{0}^{w_{i}}\sum\nolimits_{j<i}\gamma(w+t_{i-1}%
-t_{j})dw\nonumber
\end{align}
which, for the exponential kernel in (\ref{eq: exp kernel}), reduces to
\begin{align*}
v_{i}=\Lambda(t_{i},t_{i-1};\theta_{0})  &  =\mu w_{i}+\frac{\alpha}{\beta
}\sum\nolimits_{j<i}\big(e^{-\beta(t_{i-1}-t_{j})}-e^{-\beta(t_{i}-t_{j}%
)}\big)\\
&  =\mu w_{i}+\frac{\alpha}{\beta}(1-e^{-\beta w_{i}})\sum\nolimits_{j<i}%
e^{-\beta(t_{i-1}-t_{j})}.
\end{align*}
For the implementation of the bootstrap, the reverse time transformation
$s_{i}\mapsto t_{i}$ is of key interest. Specifically, consider initially a
sequence $\{v_{i}\}$ of waiting times in the transformed time scale, generated
as i.i.d. and $\mathcal{E}\left(  1\right)  $-distributed. Then, under the
true model, we can (numerically) invert the mapping \eqref{eq: Hawks change}
and generate the $i$th waiting time $w_{i}$ (or, equivalently, the $i$th event
time) recursively in terms of the $i$th waiting time in transformed time scale
$v_{i}$ and the past event times $\{t_{j},j=1,...,i-1\}$. The recursion is
initiated by generating the first waiting time $w_{1}$ as $w_{1}=\Lambda
^{-1}(v_{1};\theta_{0})$, where $v_{1}$ is the first ($\mathcal{E}\left(
1\right)  $-distributed) waiting time in transformed time scale.


As detailed below, for both the FIB and RIB schemes, we
generate i.i.d. random event times in the transformed time scale, which are
next transformed to the original time scale using the intensity dynamics
estimated from the data. The key difference between the two algorithms is
whether the transformation from transformed to original event times is
\emph{fixed} or \emph{sequential} (and hence random) across bootstrap samples.

\subsection{Fixed intensity bootstrap} \label{sec: FIB} 

Given a sample of event times $\{t_{i}\}_{i=1}^{n_{T}}$ in $[0,T]$, fix the
bootstrap true value parameter, $\theta_{T}^{\ast}$. As is standard, one may
for example set $\theta_{T}^{\ast}=\hat{\theta}_{T}$, the unrestricted MLE
based on $\{t_{i}\}_{i=1}^{n_{T}}$; for hypothesis testing, one may also set
$\theta_{T}^{\ast}=\tilde{\theta}_{T}$, the MLE restricted by the null hypothesis.

For the FIB, where the intensity is kept fixed across replications, denote the intensity process implied by the bootstrap true value as $\hat{\lambda}(t):=\lambda(t;\theta_{T}^{*})$, and the corresponding integrated intensity process as 
\begin{align}
\hat{\Lambda }(t):=\int_{0}^{t}\lambda(u;\theta_T^*)du.
\label{eq:lambda-FIB}
\end{align}
By definition, $\hat{\lambda}(t)$ and
$\hat{\Lambda}(t)$ depend on the original data through the observed event
times $\{t_{i}\}_{i=1}^{n_{T}}$ and bootstrap true value $\theta_{T}^{*}$.
Therefore, by construction, $\hat{\lambda}(t)$ and $\hat{\Lambda}(t)$ are
known and fixed conditionally on the data. \bigskip

\noindent\textsc{Algorithm 1 (FIB)}

\medskip\noindent(i) Generate a (conditionally on the original data) i.i.d.
sample $\{v_{i}^{\ast}\}$ of bootstrap transformed waiting times from the
$\mathcal{E}\left(  1\right)  $ distribution; the bootstrap transformed event
times are then given by $\{s_{i}^{\ast}\}$ where $s_{i}^{\ast}=\sum_{j=1}%
^{i}v_{j}^{\ast}$.

\medskip\noindent(ii) Construct the bootstrap event times in the original time
scale as
\[
t_{i}^{\ast}=\hat{\Lambda}^{-1}(s_{i}^{\ast}),
\]
for $i=1,\ldots,n_{T}^{\ast}$, where $\hat\Lambda$ is defined in \eqref{eq:lambda-FIB} and the number of bootstrap events
$n_{T}^{\ast}$ is
\[
n_{T}^{\ast}:=\max\{k:s_{k}^{\ast}\leq\hat{\Lambda}(T)\}=\max\left\{
k:t_{k}^{\ast}\leq T\right\}  \text{;}%
\]
the associated bootstrap counting process is $N^{\ast}(t):=\sum_{i\geq
1}\mathbb{I}(t_{i}^{\ast}\leq t)$, for $t\in\lbrack0,T]$.

\medskip\noindent(iii) Define the bootstrap MLE as $\hat{\theta}_{T}^{*}%
:=\arg\max_{\theta\in\Theta}\ell_{T}^{*}(\theta)$ with bootstrap
log-likelihood
\begin{align}
\ell_{T}^{*}(\theta)&:=\int_{0}^{T}\log\lambda(t;\theta)dN^{*}%
(t)-\Lambda(T;\theta) \notag\\
&=\sum_{i=1}^{n_{T}^{*}}\log\lambda(t_{i}^{*};\theta)-\int_{0}^{T}%
\lambda(t;\theta)dt. \label{eq FIB loglik}%
\end{align}
\smallskip

\noindent Some remarks are in order.

\begin{remark}
$\overset{}{\text{ }}$

\medskip\noindent(i) As is standard, the distribution of $T^{1/2}(\hat{\theta
}_{T}-\theta_{0})$ is approximated by the empirical distribution
(conditionally on the original data) of $T^{1/2}(\hat{\theta}_{T}^{*}%
-\theta_{T}^{*})$ where $\theta_{T}^{*}=\tilde{\theta}_{T}$ for the restricted
bootstrap and $\theta_{T}^{*}=\hat{\theta}_{T}$ for the unrestricted
bootstrap. Moreover, the bootstrap analog of the LR statistic in
\eqref{eq LR test} is given by $\operatorname*{LR}_{T}^{*}(\theta_{T}%
^{*}):=2(\ell_{T}(\hat{\theta}_{T}^{*})-\ell_{T}(\theta_{T}^{*}))$.

\medskip\noindent(ii) Notice that in the FIB log-likelihood
\eqref{eq FIB loglik} the last term $\int_{0}^{T}\lambda(t;\theta)dt$ only
depends on the original data and hence is non-random upon conditioning on the
original data.

\medskip\noindent(iii) A key feature of the FIB is that, since the bootstrap
waiting times in the transformed time are i.i.d. $\mathcal{E}\left(  1\right)
$-distributed, conditionally on the original data the bootstrap counting
process $N^{\ast}(t)$ is an inhomogeneous Poisson process with time-varying
intensity $\hat{\lambda}(t)$, $t\in\lbrack0,T]$. Bootstrap algorithms
specifically designed for inhomogeneous Poisson processes have been proposed
in \cite{CHP1996}. In contrast, despite (conditionally on the original data)
the bootstrap sample follows an inhomogeneous Poisson bootstrap process, our
FIB allows inference in a more general class of point processes.

\medskip\noindent(iv) One of the main features of the FIB is that its
implementation is straightforward and fast. Specifically, draws of the
bootstrap sample are obtained easily, since it is only required to invert the observed (strictly increasing) function $\hat{\Lambda}$. Similarly, computation of the
bootstrap likelihood and estimator is straightforward as $\lambda(t;\theta)$
is a function of the original data only. \hfill$\square$
\end{remark}

\subsection{Recursive intensity bootstrap} \label{sec: RIB} 

The RIB resembles the recursive bootstrap in time series
models, see e.g. \citet{CR2021} for a review. Thus, and in contrast to the
FIB, the RIB conditional intensity, denoted here by $\lambda^{*}(t;\theta)$,
is constructed using the functional form of the original intensity
$\lambda(t;\theta)$, but in terms of recursively obtained bootstrap event
times $t_{i}^{*}$. This entails that, for any $\theta\in\Theta$, $\lambda
^{*}(t;\theta)$ is a random process, even conditionally on the original data,
and hence differs from the FIB intensity, which is fixed across bootstrap
repetitions. Note also that the recursively obtained bootstrap intensity
process $\lambda^{*}(t;\theta)$ inherits the same properties, in terms of e.g.
differentiability with respect to $\theta$, of the original intensity process
$\lambda(t;\theta)$.

We define $\lambda^{*}(t):=\lambda^{*}(t;\theta_{T}^{*})$ and
\begin{equation}
\Lambda^{*}(t):=\int_{0}^{t}\lambda^{*}(u;\theta_{T}^{*})du.
\label{eq: RIB intens}%
\end{equation}
The RIB is then defined as follows.\bigskip

\noindent\textsc{Algorithm 2 (RIB)}

\medskip\noindent(i) As in Algorithm 1.

\medskip\noindent(ii) For $i=1,...,n_{T}^{\ast}$, construct the bootstrap
event times $t_{i}^{\ast}$ in the original time scale recursively (see
Remark~\ref{Rem on Lambda* in the RB} below) as
\[
t_{i}^{\ast}=\Lambda^{\ast-1}(s_{i}^{\ast}),
\]
for $i=1,\ldots,n_{T}^{\ast}$, where the number of bootstrap events
$n_{T}^{\ast}$ is
\[
n_{T}^{\ast}:=\max\{k:s_{k}^{\ast}\leq\Lambda^{\ast}(T)\}=\max\left\{
k:t_{k}^{\ast}\leq T\right\}
\]
and $\Lambda^{\ast}(t)$ is defined in \eqref{eq: RIB intens}; the associated
bootstrap counting process is $N^{\ast}(t):=\sum_{i\geq1}\mathbb{I}%
(t_{i}^{\ast}\leq t)$, for $t\in\lbrack0,T]$.

\medskip\noindent(iii) Define the bootstrap MLE as $\hat{\theta}_{T}^{*}%
:=\arg\max_{\theta\in\Theta}\ell_{T}^{*}(\theta)$ with bootstrap
log-likelihood
\begin{align}
\ell_{T}^{\ast}(\theta)  &  :=\int_{0}^{T}\log\lambda^{\ast}(t;\theta
)dN^{*}(t)-\Lambda^{\ast}(T;\theta)\nonumber\\
&  =\sum_{i=1}^{n_{T}^{*}}\log\lambda^{*}(t_{i};\theta)-\int_{0}^{T}%
\lambda^{*}(t;\theta)dt. \label{eq RB loglik}%
\end{align}

\begin{remark} \label{Rem on Lambda* in the RB}
$\overset{}{\text{ }}$

\medskip\noindent(i) Note that, unlike the FIB, $\lambda^*(\cdot;\theta)$ in \eqref{eq RB loglik} is random, even conditional on the data.

\medskip\noindent(ii) In the second step of Algorithm 2, the $t_{i}^{\ast}$'s are generated recursively by using the bootstrap event times in transformed time $s_{1}^{\ast},\ldots,s_{n_{T}^{\ast}}^{\ast}$ obtained in the first step. Specifically, the first bootstrap event time $t_{1}^{\ast}$ is obtained as the solution of $s_{1}^{\ast}=\Lambda^{\ast}(t_{1}^{\ast})=\int_{0}^{t_{1}^{\ast}}%
\lambda^{\ast}(u)du$. Next, given $t_{1}^{\ast}$, we obtain $t_{2}^{\ast}$ as
the solution to $s_{2}^{\ast}=s_{1}^{\ast}+\int_{t_{1}^{\ast}}^{t_{2}^{\ast}%
}\lambda^{\ast}(u)du$. Likewise, for $i>2$, $t_{i}^{\ast}$ is the solution to
$s_{i}^{\ast}=s_{i-1}^{\ast}+\int_{t_{i-1}^{\ast}}^{t_{i}^{\ast}}\lambda
^{\ast}(u)du$ given $t_{1}^{\ast},\dots,t_{i-1}^{\ast}$.\hfill$\square$
\end{remark}

\section{Validity of bootstrap inference}\label{sec BS validity}

In this section, we establish bootstrap asymptotic validity for the FIB and
RIB bootstrap schemes outlined above. As emphasized the bootstrap true
parameter is assumed to be consistent, $\theta_{T}^{*}\rightarrow_{p}%
\theta_{0}$, which holds, e.g., for the particular choices where $\theta_{T}^{*}
=\hat{\theta}_{T}$ (unrestricted bootstrap) or $\theta_{T}^{*}=\tilde{\theta
}_{T}$ (restricted bootstrap) under the null.

Throughout, we let $\mathcal{F}_{t}^{\ast}$ denote the $\sigma$-field
generated by $\{N^{\ast}(s),$ $0\leq s\leq t\}$ and $\mathcal{F}_{t-}^{\ast}$
be its left limit. Notice that, since the distribution of $N^{\ast}$ depends
on $T$, formally we have an array $\mathcal{F}_{T,t}^{\ast}:=\{N_{T}^{\ast
}(s),$ $0\leq s\leq t\leq T$, $T\geq0\}$; for simplicity, in the following we
suppress the dependence on $T$ and write $N_{T}^{\ast}(t)$ and $\mathcal{F}%
_{T,t}^{\ast}$ simply as $N^{\ast}(t)$ and $\mathcal{F}_{t}^{\ast}$.

\subsection{Preliminaries}

As for the non-bootstrap asymptotic analysis, define the bootstrap martingale%
\[
M^{\ast}(t)=N^{\ast}(t)-\Lambda_{N^{\ast}}(t).
\]
Here $\Lambda_{N^{\ast}}(t)$ is the integrated conditional intensity of either
the FIB or the RIB bootstrap process $N^{\ast}(t)$ (see also Remark
\ref{para bootstrap preliminary}(i)) and hence it corresponds to the bootstrap
compensator of $N^{\ast}(t)$ conditionally on the data. Consequently,
$M^{\ast}(t)$ is a continuous-time $\mathcal{F}_{t}^{\ast}$ local martingale
conditionally on the data. Moreover, for any process $\xi^{\ast}(t)$ which
(conditionally on the original data) is predictable with respect to
$\mathcal{F}_{t}^{\ast}$, the (Stieltjes) stochastic integral process
\begin{equation}
Y^{\ast}(t):=\int_{0}^{t}\xi^{\ast}(u)dM^{\ast}(u)=\int_{0}^{t}\xi^{\ast
}(u)[dN^{\ast}(u)-d\Lambda_{N^{\ast}}(u)] \label{eq def of why t}%
\end{equation}
is also (conditionally on the original data) a continuous-time martingale.

\begin{remark}
\label{para bootstrap preliminary} $\overset{}{\text{ }}$

\medskip\noindent(i) For both bootstrap algorithms, the bootstrap waiting
times $\{v_{i}^{\ast}\}$ in the transformed time scale are i.i.d.
$\mathcal{E}\left(  1\right)  $, and the transformation to the original time
scale is continuous. Therefore, the conditional distributions of the bootstrap
waiting times are absolutely continuous, and hence the bootstrap process
$N^{\ast}(t)$ has well-defined integrated intensity function which is given
by
\[
\Lambda_{N^{\ast}}(t)=\int_{0}^{t}\hat{\lambda}(u)du=\hat{\Lambda}(t)
\]
for the FIB, and
\[
\Lambda_{N^{\ast}}(t)=\int_{0}^{t}\lambda^{\ast}(u)du=\Lambda^{\ast}(t)
\]
for the RIB. As the conditional intensity process, the integrated intensity
$\Lambda_{N^{\ast}}(t)$ depends on the original sample; it is non-random in
the bootstrap world for the FIB, and depends on the past bootstrap event times
$t_{1}^{\ast},\dots,t_{N^{\ast}(t-)}^{\ast}$ for the RIB; see also
Remark~\ref{Rem on Lambda* in the RB}.

\medskip\noindent(ii) In some cases, the theoretical arguments are simplified
by working in transformed time rather than in the original time. Specifically,
consider the bootstrap counting process in the transformed time, given by
$Q^{\ast}(s):=\sum_{i\geq1}\mathbb{I}(s_{i}^{\ast}\leq s)$. From Algorithm
1(i), which defines $s_{i}^{\ast}=s_{i-1}^{\ast}+v_{i}^{\ast}$ where
$\{v_{i}^{\ast}\}$ are i.i.d. $\mathcal{E}\left(  1\right)  $ random
variables, the cdf of each event time $s_{i}^{\ast},i=1,2,\dots,$
conditionally on the past event times is given by
\begin{align}
\mathcal{F}_{s_{i}^{\ast}}(s|\mathcal{F}_{s_{i-1}^{\ast}})  &  :=P(s_{i}%
^{\ast}\leq s|\mathcal{F}_{s_{i-1}^{\ast}})\nonumber\\
&  =P(v_{i}^{\ast}\leq s-s_{i-1}^{\ast}|\mathcal{F}_{s_{i-1}^{\ast}%
})=1-e^{-(s-s_{i-1}^{\ast})}, \label{F-si*-pbs}%
\end{align}
which is a continuous function for $s>s_{i-1}^{\ast}$.

\medskip\noindent(iii) For both the fixed intensity and recursive intensity
bootstraps, $Q^{\ast}$ is a homogeneous Poisson process with unit intensity,
and the probability measure induced by $Q^{\ast}$ is independent of the
original data. For the FIB, the process $Q^{\ast}$ is related to the bootstrap
counting process $N^{\ast}$ through the relation
\[
N^{\ast}(t)=\sum\nolimits_{i\geq1}\mathbb{I}(t_{i}^{\ast}\leq t)=\sum
\nolimits_{i\geq1}\mathbb{I}(s_{i}^{\ast}\leq\hat{\Lambda}(t))=:Q^{\ast}%
(\hat{\Lambda}(t))
\]
and, equivalently, $Q^{\ast}(s)=N^{\ast}(\hat{\Lambda}^{-1}(s))$. Using
$Q^{\ast}$, we can write the integral in \eqref{eq def of why t} as
\[
Y^{\ast}(t)=\int_{0}^{\hat{\Lambda}(t)}\xi(\hat{\Lambda}^{-1}(s))dM_{Q}^{\ast
}(s),
\]
where $M_{Q}^{\ast}(s):=Q^{\ast}(s)-s$ is a continuous-time martingale
independent of the original data. For the RIB the formulas above are similar,
with $\hat{\Lambda}(\cdot)$ replaced by $\Lambda^{\ast}(\cdot)$.\hfill
$\square$
\end{remark}

\subsection{Validity of the FIB}

We first consider the FIB. From the bootstrap log-likelihood defined in
\eqref{eq FIB loglik}, we derive the corresponding bootstrap score and
Hessian,
\begin{align*}
S_{T}^{\ast}(\theta)  &  =\int_{0}^{T}\xi(t;\theta)(dN^{\ast}(t)-\lambda
(t;\theta)dt)\quad\text{and}\\
H_{T}^{\ast}(\theta)  &  =\int_{0}^{T}\zeta(t;\theta)(dN^{\ast}(t)-\lambda
(t;\theta)dt)-\int_{0}^{T}h(t;\theta)dt,
\end{align*}
where $\xi(t;\theta):=\partial_{\theta}\log\lambda(t;\theta)$, $\zeta
(t;\theta):=\partial_{\theta}^{2}\log\lambda(t;\theta)$ and $h(t;\theta)$ is
defined in Assumption 2.

Notice that $S_{T}^{\ast}(\theta)$ and $H_{T}^{\ast}(\theta)$ depend on the
bootstrap data only through $N^{\ast}(t)$ which, conditionally on the original
data, is an inhomogeneous Poisson point process with fixed conditional
intensity given by $\hat{\lambda}(t)=\lambda(t;\theta_{T}^{\ast})$. With
$M^{\ast}(t):=N^{\ast}(t)-\hat{\Lambda}(t)=N^{\ast}(t)-\Lambda(t;\theta
_{T}^{\ast})$, the score and Hessian evaluated at the bootstrap true value
$\theta_{T}^{\ast}$ can be rewritten as
\begin{align}
S_{T}^{\ast}(\theta_{T}^{\ast})  &  =\int_{0}^{T}\hat{\xi}(t)dM^{\ast
}(t)\text{ \ \ \ and}\label{eq score at the true value FIB}\\
H_{T}^{\ast}(\theta_{T}^{\ast})  &  =\int_{0}^{T}\hat{\zeta}(t)dM^{\ast
}(t)-\int_{0}^{T}\hat{h}(t)dt, \label{eq Hessian at the true value FIB}%
\end{align}
where $\hat{\xi}(t)=\xi(t;\theta_{T}^{\ast})$, $\hat{h}(t)=h(t;\theta
_{T}^{\ast})$ and $\hat{\zeta}(t)=\zeta(t;\theta_{T}^{\ast})$.

Using the fact that $M^{\ast}$ is a martingale, we prove in the appendix the
following lemma, which requires only a mild strengthening of the assumptions
in Theorem~\ref{Th asy normality}.

\begin{lemma}
\label{Lemma BS score and Hessian} Under the assumptions of Theorem
\ref{Th asy normality}, provided that, additionally, (i) $\theta_{T}^{\ast
}\overset{p}{\rightarrow}\theta_{0}$; for all $\theta$, (ii)\ $E((\partial
_{\theta_{i}}\lambda(t;\theta))^{3})<\infty$ and (iii) either $\lambda
_{T}\left(  t;\theta\right)  \geq\lambda_{L}>0$, a.s., or $\partial
_{\theta_{i}}\log\lambda(t;\theta)\leq c<\infty$, a.s., it holds that%
\begin{align}
T^{-1/2}S_{T}^{\ast}(\theta_{T}^{\ast})  &  \overset{d^{\ast}}{\rightarrow
}_{p}\mathcal{N(}0,I(\theta_{0})),\label{eq asymptotics for FIB score}\\
H_{T}^{\ast}(\theta_{T}^{\ast})  &  =-\int_{0}^{T}\hat{h}(t)dt+o_{p}^{\ast
}(1)\overset{p^{\ast}}{\rightarrow}_{p}-I(\theta_{0})
\label{eq asymptotics for FIB Hessian}%
\end{align}
where $I(\theta_{0})$ is defined in Assumption 2.
\end{lemma}

The following theorem shows the first-order validity of the FIB and of the
associated likelihood ratio test.

\begin{theorem}
\label{Th asy validity of the FIB} Under the conditions of Lemma
\ref{Lemma BS score and Hessian}, as $T\rightarrow\infty$, it holds that
\begin{equation}
\sup_{x\in\mathbb{R}}\left\vert P^{\ast}(T^{1/2}(\hat{\theta}_{T}^{\ast
}-\theta_{T}^{\ast})\leq x)-P(T^{1/2}(\hat{\theta}_{T}-\theta_{0})\leq
x)\right\vert \rightarrow_{p}0. \label{eq FIB BS consistency}%
\end{equation}
Moreover, for the bootstrap likelihood-ratio statistic it holds that
\begin{equation}
LR_{T}^{\ast}(\theta_{T}^{\ast}):=2(\ell_{T}^{\ast}(\hat{\theta}_{T}^{\ast
})-\ell_{T}(\theta_{T}^{\ast}))\overset{d^{\ast}}{\rightarrow}_{p}\chi_{d}%
^{2}. \label{eq FIB BS LR test}
\end{equation}
\end{theorem}

\subsection{Validity of the RIB}

For the RIB, the bootstrap score and Hessian at the bootstrap true value
$\theta_{T}^{\ast}$ mimic their counterparts on the original data, see
\eqref{eq score at theta0}--\eqref{eq HEssian at theta0}. Specifically, with
$\lambda^{\ast}(t):=\lambda^{\ast}(t;\theta_{T}^{\ast})$ and $M^{\ast
}(t)=N^{\ast}(t)-\int_{0}^{t}\lambda^{\ast}(t)dt$,
\begin{align}
S_{T}^{\ast}(\theta_{T}^{\ast})  &  =\int_{0}^{T}\xi^{\ast}(t)dM^{\ast
}(t),\quad\xi^{\ast}(t):=\partial_{\theta_{T}^{\ast}}\log\lambda^{\ast
}(t),\label{eq RIB score at true value}\\
H_{T}^{\ast}(\theta)  &  =\int_{0}^{T}\zeta^{\ast}(t)dM^{\ast}(t)-\int_{0}%
^{T}h^{\ast}(t)dt,\quad\zeta^{\ast}(t):=\partial_{\theta_{T}^{\ast}}^{2}%
\log\lambda^{\ast}(t), \label{eq RIB hessian at true value}%
\end{align}
where $h^{\ast}(t):=h^{\ast}(t;\theta_{T}^{\ast})$. The next lemma shows that
the RIB score and Hessian mimic the large sample properties of the original
score and Hessian. It requires an additional assumption, see
\eqref{eq extra condition RIB validity} below, which is not required for the
FIB. In order to introduce it, we emphasize that the quantity $h(t;\theta)$ in
Assumption 2 depends on the data generating process, and hence on the true
parameter $\theta_{0}$. That is, $h(t;\theta)=h_{\theta_{0}}(t;\theta)$.

The proof is based on the fact that for any fixed $T$ and conditionally on the
data, the bootstrap sample can be made stationary.

\begin{lemma}
\label{Lemma RIB BS score and Hessian} Under the assumptions of Theorem
\ref{Th asy normality}, provided $\theta_{T}^{\ast}\overset{p}{\rightarrow
}\theta_{0}$ and, for $i,j=1,...,d$,
\begin{equation}
\sup_{\vartheta,\theta\in\Theta_{0}}|h_{\vartheta;i,j}(t;\theta)|\leq
e_{i,j}(t), \label{eq extra condition RIB validity}%
\end{equation}
where $h_{\vartheta;i,j}(t;\theta)=\partial^{2}h_{\vartheta}(t;\theta
)/\partial\theta_{i}\partial\theta_{j}<\infty$ and $E(e_{i,j}(t))<\infty$, it
holds that
\[
T^{-1/2}S_{T}^{\ast}(\theta_{T}^{\ast})\overset{d^{\ast}}{\rightarrow}%
_{p}\mathcal{N}(0,I(\theta_{0})),-H_{T}^{\ast}(\theta_{T}^{\ast})=\int_{0}%
^{T}h^{\ast}(t)dt+o_{p}^{\ast}(1)\overset{p^{\ast}}{\rightarrow}_{p}%
I(\theta_{0})
\]
with $I(\theta_{0})$ defined is Assumption 2.
\end{lemma}

For bootstrap consistency, we modify Assumption 2(c) as follows.\bigskip

\noindent\textsc{Assumption} 2(c$^{\ast}$)

\noindent Assumption 2(c) holds with $c_{ijk}(t)=c_{ijk}(t;\theta_{0})$ and
$d_{ijk}(t)=d_{ijk}(t;\theta_{0})$ replaced by $\sup_{\theta\in\Theta_{0}%
}(c_{ijk}(t;\theta))$ and $\sup_{\theta\in\Theta_{0}}(d_{ijk}(t;\theta))$,
respectively. \bigskip

The modification is necessary in order to bound the third order derivatives of
the RIB likelihood.

\begin{theorem}
\label{Th asy validity of the RIB} Under the conditions of Lemma
\ref{Lemma RIB BS score and Hessian}, with Assumption 2(c) replaced by
2(c$^{\ast}$), \eqref{eq FIB BS consistency} and \eqref{eq FIB BS LR test} hold.
\end{theorem}

\begin{remark}
Condition \eqref{eq extra condition RIB validity} is required to show
convergence of the bootstrap score in a neighborhood of the true value
$\theta_{0}$. This is specific of the bootstrap and not necessary to show
convergence of the original score at the true value. For proving convergence
of the bootstrap Hessian in a neighborhood of $\theta$, no extra conditions
are needed, as under the bounds on the terms entering the third derivative of
the likelihood function, see Assumption 2(c), such convergence is already
implied, as shown in \citet[proof of Theorem~3]{O1978}.\hfill$\square$
\end{remark}

\section{Non-parametric FIB and RIB}

\label{sec NPAR} In the presented parametric bootstrap, bootstrap event times
are obtained in transformed time scale by cumulating randomly-generated i.i.d.
$\mathcal{E}\left(  1\right)  $ waiting times. This was motivated by the fact
that waiting times $v_{i}=v_{i}(\theta_{0})$ in \eqref{eq TFTwait} are i.i.d.
$\mathcal{E}\left(  1\right)  $-distributed for $i=1,...,n_{T}$ and, moreover,
with $\theta_{T}^{\ast}=\theta_{0}+o_{p}(1)$,
\begin{equation}
\hat{v}_{i}:=\Lambda(t_{i};\theta_{T}^{\ast})-\Lambda(t_{i-1};\theta_{T}%
^{\ast})=v_{i}+o_{p}(1). \label{vi hat def}%
\end{equation}
However, in the case of a misspecified model, it may be the case that the
transformed waiting times $\hat{v}_{i}$ are not exponentially distributed
(asymptotically). Therefore, we consider here the point process bootstrap
equivalent of the well-known residual-based i.i.d. bootstrap in discrete time
series models. Specifically, after the point process model is fit to data, the
residuals to resample from can be taken as the waiting times in transformed
time scale, i.e. $\hat{v}_{i}$, $i=1,...,n_{T}$. Then, the bootstrap waiting
times in transformed time can be generated as an i.i.d sample from the sample
$\{\hat{v}_{i}\}_{i=1}^{n_{T}}$. This algorithm is denoted here as the
`non-parametric bootstrap', and can be implemented for both FIB and RIB
bootstraps, see below.

For the bootstrap in conditional mean and variance time series models, the
residuals are typically centered and/or scaled prior to the implementation of
the bootstrap. Similarly, here the waiting times $\hat{v}_{i}$ need to be
properly standardized, such that the bootstrap transformed waiting times
$v_{i}^{\ast}$ match (as a minimum) the mean of the $\mathcal{E}\left(
1\right)  $ distribution, i.e. $E^{\ast}(v_{i}^{\ast})=1$. This is achieved by
sampling from $\hat{v}_{i}^{c}$ given by
\begin{equation}
\hat{v}_{i}^{c}:=\frac{\hat{v}_{i}}{\bar{v}_{T}},\quad i=1,\dots,n_{T},
\label{recenter}%
\end{equation}
where $\bar{v}_{T}:=n_{T}^{-1}\sum_{j=1}^{n_{T}}\hat{v}_{j}$. Note that
$\hat{v}_{i}^{c}>0$ for all $i$, and therefore a random draw from $\{\hat
{v}_{i}^{c}\}_{i=1}^{n_{T}}$ has, conditionally on the original data, unit
expected value, i.e. $E^{\ast}(v_{i}^{\ast})=n_{T}^{-1}\sum_{i=1}^{n_{T}}%
\hat{v}_{i}^{c}=1$.

With the transformed waiting times $\{\hat{v}_{i}^{c}\}_{i=1}^{n_{T}}$ defined
in \eqref{recenter}, the proposed non-parametric bootstrap algorithm is as
follows.\bigskip

\noindent\textsc{Algorithm 3 (Non-Parametric Bootstrap)}

\medskip\noindent(i) Generate a sample $\{v_{i}^{*}\}$ of bootstrap
transformed waiting times by resampling with replacement from $\{\hat{v}%
_{i}^{c}\}_{i=1}^{n_{T}}$, such that
\begin{equation}
v_{i}^{\ast}=\hat{v}_{u_{i}^{\ast}}^{c},\quad\text{for }i=1,2,\dots
\label{dist-vi*2}%
\end{equation}
where $u_{i}^{\ast}$ is an i.i.d. discrete uniformly distributed sequence on
$\{1,\dots,n_{T}\}$. The bootstrap transformed event times are then given by
$s_{i}^{\ast}=\sum_{j=1}^{i}v_{j}^{\ast}$.

\medskip\noindent(ii)-(iii) as in Algorithm 1 or Algorithm 2 depending on
whether it is a fixed intensity or recursive intensity bootstrap.

\begin{remark}
\label{Rem on NonPar BS alg} $\overset{}{\text{ }}$

\medskip\noindent(i) As mentioned, a crucial step of the non-parametric
bootstrap is the rescaling of the waiting times in transformed time scale. By
doing as above, it holds that $v_{i}^{\ast}>0$, a.s., $E^{\ast}(v_{i}^{\ast
})=n_{T}^{-1}\sum_{i=1}^{n_{T}}\hat{v}_{i}^{c}=1$, and, moreover, $V^{\ast
}(v_{i}^{\ast})\rightarrow_{p}1$. Apart from matching the mean and,
asymptotically, the variance of the $\mathcal{E}\left(  1\right)  $
distribution, scaling is a key ingredient to center the bootstrap score around
$0$. Additionally, the convergence of the variance of the bootstrap waiting
times to unity guarantees that, in large sample, the variance of the bootstrap
score matches the inverse of the bootstrap information.

\medskip\noindent(ii) Without rescaling it holds that $E^{\ast}(v_{i}^{\ast
})\rightarrow_{p}1$ and $V^{\ast}(v_{i}^{\ast})\rightarrow_{p}1$. However,
this is not enough for the bootstrap score to be centered around $0$, because
unless $E^{\ast}(v_{i}^{\ast}-1)=o_{p}(T^{-1/2})$ the bootstrap score will
have a non-zero (and random) mean driven by the term $T^{1/2}(E^{\ast}%
(v_{i}^{\ast})-1)$. This is well-known for the bootstrap in time series
models, where if the residuals are not centered, their $O_{p}(T^{-1/2})$
sample mean will induce randomness in the limit distribution of the bootstrap
statistics \citep{CNR2015,CG2020}.\hfill$\square$
\end{remark}

To provide an intuition about validity of this bootstrap and about the
importance of rescaling, consider a simple Poisson process model with
intensity $\lambda(t)=\theta$, where interest is in inference on $\theta$
using the (unrestricted) bootstrap. Recall that the log-likelihood for the
original sample is $\ell_{T}(\theta)=\int\log\theta dN(t)-\int\theta
dt=n_{T}\log\theta-T\theta$, with associated bootstrap score $\theta^{-1}\int
dN(t)-T=\theta^{-1}n_{T}-T$, which leads to the unique MLE, $\hat{\theta}%
_{T}=n_{T}/T$. To implement the non-parametric bootstrap, consider the
transformed waiting times, see Section~\ref{sec time change}, which in this
case are given by $\hat{v}_{i}=\hat{\theta}_{T}w_{i}$, with $w_{i}%
=t_{i}-t_{i-1}$ the original observed waiting times. The non-parametric
bootstrap generates the $v_{i}^{\ast}$'s by initially resampling from the
rescaled $\hat{v}_{i}^{c}$ defined in \eqref{recenter}; next, the $v_{i}%
^{\ast}$'s are transformed back in the original time scale using the inverse
mapping $w_{i}^{\ast}=v_{i}^{\ast}/\hat{\theta}_{T}$. This leads to the
bootstrap event times $t_{i}^{\ast}:=\sum_{j=1}^{i}w_{j}^{\ast}$ with
associated bootstrap counting process $N^{\ast}(t):=\sum_{i\geq1}%
\mathbb{I}(t_{i}^{\ast}\leq t)$. The bootstrap likelihood and score are then
given by $\ell_{T}^{\ast}(\theta)=\int\log\theta dN^{\ast}(t)-\int\theta
dt=n_{T}^{\ast}\log\theta-T\theta$ and $S_{T}^{\ast}(\theta)=\theta^{-1}\int
dN^{\ast}(t)-T=\theta^{-1}n_{T}^{\ast}-T$, respectively, where as earlier
$n_{T}^{\ast}$ denotes the total number of events, $n_{T}^{\ast}=\max
\{k:\sum_{1}^{k}w_{i}^{\ast}\leq T\}$.

Consider next the bootstrap score at the true value $\theta_{T}^{\ast}%
=\hat{\theta}_{T}$,
\begin{equation}
S_{T}^{\ast}(\hat{\theta}_{T})=\hat{\theta}_{T}^{-1}n_{T}^{\ast}%
-T=-\hat{\theta}_{T}^{-1}\sum_{i=1}^{n_{T}^{\ast}}(\hat{\theta}_{T}w_{i}%
^{\ast}-1)=-\hat{\theta}_{T}^{-1}\sum_{i=1}^{n_{T}^{\ast}}(v_{i}^{\ast}-1).
\label{eq score for NPB Poisson case}%
\end{equation}
Because of the rescaling in \eqref{recenter}, $E^{\ast}(v_{i}^{\ast}-1)=0$.
This is a key feature for the bootstrap score to mimic the large-sample
behavior of the original score. In contrast, without rescaling, the bootstrap
mean of $v_{i}^{\ast}-1$ would be of order $O_{p}(n_{T}^{-1/2})=O_{p}%
(T^{-1/2})$ (the order being sharp), thereby introducing an asymptotically
non-negligible (random) bias term in the distribution of the bootstrap score.

In order to analyze the large sample properties of the non-parametric
bootstrap score, it is important to observe that a standard (bootstrap version
of the) CLT cannot be applied to \eqref{eq score for NPB Poisson case} because
the number of terms in the sum is itself random. That is, $S_{T}^{\ast}%
(\hat{\theta}_{T})$ is a randomly selected partial sum. Its behavior can
however be analyzed by considering the following FCLT for i.i.d. waiting
times, which for non-bootstrap sequences is due to \cite{B1968} (the extension
to bootstrap random variables is straightforward and is omitted for brevity).

\begin{theorem}
\label{Billingsley's FCLT for count processes} Let $u_{1}^{\ast},u_{2}^{\ast
},\ldots$ be bootstrap random variables which, conditionally on the original
data, are i.i.d. with mean $1$, variance $\hat{\kappa}_{T}$ (being a function
of the original data) and a.s. positive. For $T>0$, define with $s\in
\lbrack0,1]$ the c\`{a}dl\`{a}g process
\[
n_{T}^{\ast}(s):=\max\left\{  k\geq0:\sum\nolimits_{i=1}^{k}u_{i}^{\ast}%
\leq\left\lfloor Ts\right\rfloor \right\}  .
\]
Assume that, as $T\rightarrow\infty$, $\hat{\kappa}_{T}\rightarrow_{p}%
\kappa>0$ and that a bootstrap FCLT holds for $\{u_{i}^{\ast}\}$, i.e.
\[
\frac{1}{T^{1/2}}\sum_{i=1}^{\lfloor Ts\rfloor}\left(  \frac{u_{i}^{\ast}%
-1}{\hat{\kappa}_{T}^{1/2}}\right)  \overset{d^{\ast}}{\rightarrow}_{p}B(s),
\]
with $B(\cdot)$ a standard Brownian motion. It then holds that, as
$T\rightarrow\infty$,
\[
\frac{n_{T}^{\ast}(s)-\lfloor Ts\rfloor}{\sqrt{\hat{\kappa}_{T}T}}%
\overset{d^{\ast}}{\rightarrow}_{p}B(s).
\]
\end{theorem}
\medskip

By using the fact that $\hat{\theta}_{T}$ is consistent and that the sample
variance of the transformed waiting times converges to one, an immediate
application of Theorem~\ref{Billingsley's FCLT for count processes} yields
that
\[
T^{-1/2}S_{T}^{\ast}(\hat{\theta}_{T})\overset{d^{\ast}}{\rightarrow}%
_{p}\mathcal{N}(0,\theta_{0}^{-1}),
\]
which matches the asymptotic distribution of the original score. For the
Hessian,
\[
T^{-1}H_{T\ }^{\ast}(\hat{\theta}_{T})=T^{-1}n_{T}^{\ast}/\hat{\theta}_{T}%
^{2}\overset{p^{\ast}}{\rightarrow}_{p}1/\theta_{0}\text{,}%
\]
as $T^{-1}n_{T}^{\ast}=\hat{\theta}_{T}^{\ast}=\theta_{0}+o_{p^{\ast}}\left(
1\right)  $, in probability, applying again
Theorem~\ref{Billingsley's FCLT for count processes}. By standard arguments
this implies that
\[
T^{1/2}(\hat{\theta}_{T}^{\ast}-\theta_{T}^{\ast})\overset{d^{\ast}%
}{\rightarrow}_{p}\mathcal{N}(0,\theta_{0})\text{,}%
\]
and similarly, $LR_{T}^{\ast}(\theta_{T}^{\ast})\overset{d^{\ast}}%
{\rightarrow}_{p}\chi_{1}^{2}$. The general (non-Poisson) case is more
involved due to the fact that conditionally on the data the bootstrap waiting
times in transformed time scale have a discrete distribution. Although this
feature is not crucial in the Poisson case, the general case involves the
analysis of random terms of the form $\int\xi(t)dN^{\ast}(t)$ and an explicit
calculation of the compensator of $N^{\ast}(t)$.

We conclude by noticing that, as shown in the next section, the non-parametric
bootstrap performs as well as the parametric bootstrap.

\section{Relation with the bootstrap for ACD models}
\label{Sec on bootstrapping ACD}

In this section we discuss the relation between our proposed bootstrap
algorithms and theory and extant results on the bootstrap for ACD models; see in particular \citet{FG2005}, \citet{GKS2015}, \citet{PHS2016} and \citet{PS2017, PS2021} for the related class of multiplicative error models [MEM].

Consider, initially, the exponential ACD process [EACD] which, by
(\ref{eq acd model}) and (\ref{eq acd intensity}), has intensity%
\[
\lambda\left(  t\right)  =\frac{1}{\psi_{N\left(  t-\right)  +1}}%
=\lambda\left(  t-t_{N\left(  t-\right)  },t-t_{N\left(  t-\right)
-1},....\right)  ,
\]
and associated integrated intensity
\begin{equation}
\Lambda\left(  t\right)  
=\sum_{i=1}^{N(t-)}\int_{t_{i-1}}^{t_{i}}
\lambda(u)du+\int_{N(t-)}^{t}\lambda(u)du
=\sum_{i=1}^{N(t-)}\frac{w_{i}}{\psi_{i}}+\frac{t-t_{N(t-)}}
{\psi_{N(t-)+1}}.\label{eq lambda for EACD}%
\end{equation}
Note that, to simplify notation, we omit here the dependence on $\theta$
parametrizing the intensity function and hence $\psi_{i}$.

It follows that our proposed RIB algorithms are related to recursive
bootstraps in the ACD framework. To see this, recall that for the parametric
RIB, we first generate the sequence of transformed waiting times $\left\{
v_{i}^{\ast}\right\}  $ as i.i.d. $\mathcal{E}\left(  1\right)  $, while for
the non-parametric RIB we resample from the original (standardized)
transformed waiting times $v_{i},i=1,...,n_{T}$ which, using
(\ref{eq lambda for EACD}), are given by $v_{i}=w_{i}/\psi_{i}$ in the case
where $\theta_{T}^{\ast}=\theta_{0}$ without loss of generality. Next, the
bootstrap waiting times $w_{i}^{\ast}$ are generating recursively as
\begin{align}
w_{i}^{\ast}  &  =\Lambda^{\ast-1}(t_{i}^{\ast},t_{i-1}^{\ast})=\psi_{i}^{\ast}v_{i}^{\ast},\quad i=1,...,n_{T}^{\ast},
\label{eq bootstrap recursion for EACD}
\end{align}
with 
\begin{align*}
n_{T}^{\ast}&=\max\{k:\sum\nolimits_{i=1}^{k}w_{i}^{\ast}\leq T\},
\end{align*}
which is equivalent to a recursive bootstrap for the EACD model (either
parametric or non-parametric). Therefore, the theory we develop in this paper
can also be used to establish bootstrap validity for EACD models. A crucial
difference between the recursive bootstrap for MEM is that in
(\ref{eq bootstrap recursion for EACD}) the number of event times $n_{T}%
^{\ast}$ is random, $\sum\nolimits_{i=1}^{n_{T}^{\ast}}w_{i}^{\ast}\leq T$,
such that the event times fall within the interval $[0,T]$. In contrast, in
the recursive MEM case $n_{T}^{\ast}=n_{T}$, which implies that $\sum
\nolimits_{i=1}^{n_{T}}w_{i}^{\ast}$ can be much smaller or even larger than
$T$.

For the case of ACD with non-exponentially distributed errors $\varepsilon
_{i}$, the intensity is (\ref{eq acd intensity}) with corresponding integrated
intensity%
\begin{equation}
\Lambda\left(  t_{i},t_{i-1}\right)  =\int_{t_{i-1}}^{t_{i}}\lambda\left(
u\right)  du=-\log S_{\varepsilon}\left(  \frac{w_{i}}{\psi_{i}}\right)
\text{,} \label{eq lambda ti timinus one acd}%
\end{equation}
where $S_{\varepsilon}$ is one minus the cdf of $\varepsilon_{i}$. In the
parametric case, with $v_{i}^{\ast}$ drawn as $\mathcal{E}(1)$, using
(\ref{eq lambda ti timinus one acd}) in the bootstrap world, we recursively
obtain
\begin{equation}
w_{i}^{\ast}=\psi_{i}^{\ast}S_{\varepsilon}^{-1}\left(  \exp\left(
-v_{i}^{\ast}\right)  \right)  =\psi_{i}^{\ast}\varepsilon_{i}^{\ast}\text{,}
\label{eq wistar ACD}%
\end{equation}
where
\begin{equation}
\varepsilon_{i}^{\ast}=S_{\varepsilon}^{-1}\left(  \exp\left(  -v_{i}^{\ast
}\right)  \right)  =S_{\varepsilon}^{-1}\left(  U_{i}^{\ast}\right)  \text{, }
\label{eq eps star acd}%
\end{equation}
with the $U_{i}^{\ast}$'s being i.i.d. uniform in $[0,1]$; hence, apart for the
random stopping time, our RIB covers the recursive bootstrap for ACD.

For the non-parametric case, existing bootstrap algorithms for ACD generate
the errors $\varepsilon_{i}^{\ast}$ by resampling the residuals $\varepsilon
_{i}=w_{i}/\psi_{i}$, while the RIB first generates the bootstrap waiting
times in transformed scale $v_{i}^{\ast}$ by resampling the estimated $v_{i}$;
these are later used to generated the bootstrap errors $\varepsilon_{i}^{\ast
}$ and then $w_{i}^{\ast}$, see (\ref{eq eps star acd}) and
(\ref{eq wistar ACD}) above.

Finally, consider our (either parametric or non-parametric) FIB applied to
the ACD. It would be tempting to think that our FIB would correspond to a
`fixed conditional expected duration' bootstrap in the ACD world, where the
bootstrap waiting times are generated as
\begin{equation}
w_{i}^{\ast}=\hat{\psi}_{i}\varepsilon_{i}^{\ast}, \quad i=1,...,n_{T},
\label{eq NEW bootstrap for ACD}
\end{equation}
where $\hat{\psi}_{i}$ is the $i$-th estimated conditional expected duration
on the original data. Although this algorithm, which resembles the fixed
volatility bootstrap for ARCH processes proposed in \citet{CPR2018}
and has not been investigated previously in the literature,
seems to be an interesting development, it does not correspond to our
FIB algorithm. In particular, the FIB uses the inverse of the estimated
(integrated) intensity, say $\hat{\Lambda}^{-1}$, to transform the bootstrap
$v_{i}^{\ast}$ into the bootstrap waiting times $w_{i}^{\ast}$ and generates a
number of event times which is random in the bootstrap world; in contrast, a
bootstrap based on (\ref{eq NEW bootstrap for ACD}) generates a number of
events, given by $n_{T}$, which is fixed in the bootstrap world.

\begin{remark}
In terms of validity of our FIB and RIB when applied to the ACD models, the
regularity conditions in Assumptions 1 and 2 are straightforward to verify. In
terms of the condition (iii) in Lemma \ref{Lemma BS score and Hessian}, while
$\lambda_{T}(t)$ cannot be bounded from below, it trivially
holds that the log-derivative of $\lambda_{T}\left(  t\right)  $ is bounded
for classic ACD($p,q$) models. \hfill$\square$
\end{remark}

\section{Monte Carlo Simulations}\label{Sec MC sim}

In this section we consider the finite sample properties of
asymptotic and bootstrap-based confidence intervals and hypothesis tests for
the well-known and much used case of a Hawkes process. By considering a
detailed simulation study based on the exponential kernel, we analyze how the
bootstrap compares to asymptotic inference for different values of key
quantities such as the `branching ratio' (defined below) and the decaying rate
of the memory of past events. We consider both the RIB and the proposed FIB
schemes, parametric as well as non-parametric.

\subsection{Model and implementation}\label{sec MC model} 

In the simulations, we consider the Hawkes process with
exponential kernel function, $\gamma(x;\alpha,\beta)=\alpha e^{-\beta x}$ and
conditional intensity
\[
\lambda(t;\theta)=\mu+\sum_{t_{i}<t}\gamma(t-t_{i};\alpha,\beta),
\]
with $\theta=(\mu,\alpha,\beta)^{\prime}$, $\mu,\alpha,\beta>0$, see also
\eqref{eq: exp kernel}. Here $\mu$ is the baseline intensity; $\alpha$ is the
jump size of the intensity when a new event occurs; $\beta$ is the exponential
decaying rate, which determines how fast the memory of past events declines to
zero. In terms of $\alpha$ and $\beta$, a key quantity is the branching ratio
\begin{equation}
a:=\alpha/\beta=\int_{0}^{\infty}\gamma(x;\alpha,\beta)dx=\int_{0}^{\infty
}\alpha e^{-\beta x}dx, \label{eq: branch rat}%
\end{equation}
which describes how quickly the number of events increases\footnote{More
precisely, in the Poisson cluster representation of the self-exciting point
process \citep{HO1974}, the branching ratio defines the expected number of
direct offsprings spawned by an `immigrant' event.}. Moreover, with
$\mu,\alpha,\beta>0$, stationarity of the Hawkes process requires the
branching ratio to satisfy $0<a<1$, in which case the mean intensity $m$ is
well defined and given by
\[
m:=E(\lambda(t))=\frac{\mu}{1-a}.
\]
Hence, the stationary region is given by $\{\theta=(\mu,\alpha,\beta)^{\prime
}\in\mathbb{R}\times\mathbb{R}\times\mathbb{R}:\mu>0,0<\alpha<\beta\}$. A few
remarks about the simulation scheme are as follows.
\medskip

\begin{remark}
\label{Rem on MC}\mbox{}

\medskip\noindent(i) We simulate the event times $\{t_{i}\}_{i=1}^{n_{T}}$ of
the Hawkes process using the `thinning algorithm' of \citet{LS1979} and
\citet{O1981}, which allows to simulate a general regular point processes
characterized by any conditional intensity. Other options, such as the
time-change method described in Section~\ref{sec time change}
\citep[see also][]{O1979}, the efficient sampling algorithm by exploring the
Markov property of the exponential kernel \citep{DZ2013}, and the `stochastic
reconstruction' method \citep{ZOVJ2004} are also available in the literature.

\medskip\noindent(ii) One important issue in simulating data in the time
interval $[0,T]$ (as well as in likelihood estimation) is how to treat the
events before and at time $t_{0}=0$ due to the `infinite memory' of the
simulated exponential intensity. In our simulations, we make use of a burn-in
period $[-M,0)$, with $M>0$ arbitrarily large (and no events prior to time
$-M$), and assume that data prior to $t_{0}=0$ are available for estimation.
Accordingly, in the bootstrap world, the bootstrap event times prior to time
$t_{0}$ are fixed to the original event times. We anticipate that the results
do not substantially change without burn-in period, provided the time span $T$
is large enough, see also \citet{O1979}, \cite{R2013} and \cite{RLMX2017}.

\medskip\noindent(iii) As is well-known, see e.g. \citet{ELL2011}, to avoid
numerical issues in estimations it is advisable to reparameterize the kernel
function as $\gamma(x;a,\beta)=a\beta e^{-\beta x}$ where $a=\alpha/\beta$ is
the branching ratio defined above, such that
\[
\lambda(t;\theta)=\mu+a\beta\sum_{t_{i}<t}e^{-\beta(t-t_{i})}%
\]
where $\theta=(\mu,a,\beta)^{\prime}$. The associated likelihood function of
$n_{T}$ event times observed in $[0,T]$ is given by
\[
\ell_{T}(\theta)=\sum_{i=1}^{n_{T}}\log\Big(\mu+a\beta\sum_{t_{j}<t_{i}%
}e^{-\beta(t_{i}-t_{j})}\Big)-\mu T-a\beta\int_{0}^{T}\sum_{t_{i}<t}%
e^{-\beta(t-t_{i})}dt.
\]
We employ this parameterization in our simulations.

\medskip\noindent(iv) The MLE $\hat{\theta}_{T}$ is obtained by maximizing the
likelihood function over the set $\mathbb{R}\times\mathbb{R}\times\mathbb{R}$,
i.e., without imposing the stationarity assumption in estimation. Therefore,
it can be the case that for certain samples $\hat{\theta}_{T}$ falls outside
the stationarity region (e.g., the estimated branching ratio $\hat{a}$ exceeds
unity). In such a case, recursive versions of the bootstrap based on
$\hat{\theta}_{T}$ would generate non-stationary bootstrap
samples\footnote{Interestingly, the issue is not crucial for the proposed
fixed intensity bootstrap.}. Therefore, as in \citet{CRT2012} and
\cite{S2006}, prior to the implementation of the bootstrap we check whether
$\hat{\theta}_{T}$ is within the stationarity region. Also it is checked
whether the Hessian evaluated at $\hat{\theta}_{T}$ is negative definite. We
refer to this step as `sanity check' [SC] and report statistics on this below.
In our Monte Carlo experiment, samples for which SC fails are discarded, and
the total number of Monte Carlo samples reported corresponds to the number of
valid samples.\hfill$\square$
\end{remark}

We simulate three stationary Hawkes processes (denoted by Models 1--3) with
true parameters $\theta_{0}$ set as follows. For all simulated processes, the
mean intensity is set to unity ($m_{0}=1$), while different levels of the
branching ratio $a_{0}=\alpha_{0}/\beta_{0}$ are considered; specifically, we
set $a_{0}\in\{0.2,0.5,0.8\}$. For each simulation, we consider three
parameterizations (see A--C below) to allow different jump sizes and decaying
behavior of the intensity. In all cases, we consider samples over $[0,T]$ for
$T\in\{50,100,200\}$ with initial burn-in period $[-M,0)$ for $M=500$. The
number of valid Monte Carlo replications (see Remark \ref{Rem on MC}(iv)) is
$10,000$, and the number of bootstrap repetitions is $B=199$.

The parameter configurations are summarized in Table~\ref{tab:SCheck} along
with the (Monte Carlo) probabilities that the SC fails. It can be noticed that
the probabilities of SC failure are severely high only for Model 1A when
$T=50$. This is because the number of events generated for $T=50$ is extremely
volatile and the likelihood of observing samples with a small number of events
(hence, not informative enough for estimating the model reasonably well) is
indeed high. Another reason is that, as is known, it is hard to precisely
estimate the parameters when the true parameters $\alpha_{0}$ and $\beta_{0}$
are close to the zero boundary and $T$ is small. The reparameterization by
branching ratio helps to resolve some numerical issues in estimation, as
discussed in Remark~\ref{Rem on MC}(iii) but the improvement is not sufficient
when the branching ratio itself is also low as in the case of Model 1A.
Nevertheless, despite the quite extreme parameter setting of Model 1A, we
decided to keep it in our Monte Carlo simulation for completion.

\begin{table}[t!]
\caption{\textsc{Monte Carlo parameter configuration with associated empirical probabilities that the SC fails.}}
\smallskip 

\label{tab:SCheck}
\centering
\begin{tabular}
[c]{ccccccccccc} 
\toprule
& & & & & & Branching ratio & & \multicolumn{3}{c}{Probability of
SC\ failure}\\
\multicolumn{2}{c}{Model} & & $\mu_{0}$ & $\alpha_{0}$ & $\beta_{0}$ &
$a_{0}=\alpha_{0}/\beta_{0}$ & & $T=50$ & $T=100$ & $T=1000$\\\hline
\multicolumn{1}{l}{$\overset{}{1}$} & \multicolumn{1}{l}{A} & &
\multicolumn{1}{r}{$0.8$} & \multicolumn{1}{r}{$0.2$} & \multicolumn{1}{r}{$1$%
} & $0.2$ & & \multicolumn{1}{r}{$0.268$} & \multicolumn{1}{r}{$0.134$} &
\multicolumn{1}{r}{$0.044$}\\
\multicolumn{1}{l}{} & \multicolumn{1}{l}{B} & & \multicolumn{1}{r}{$0.8$} &
\multicolumn{1}{r}{$1.0$} & \multicolumn{1}{r}{$5$} & $0.2$ & &
\multicolumn{1}{r}{$0.049$} & \multicolumn{1}{r}{$0.007$} &
\multicolumn{1}{r}{$0.001$}\\
\multicolumn{1}{l}{} & \multicolumn{1}{l}{C} & & \multicolumn{1}{r}{$0.8$} &
\multicolumn{1}{r}{$5.0$} & \multicolumn{1}{r}{$25$} & $0.2$ & &
\multicolumn{1}{r}{$0.008$} & \multicolumn{1}{r}{$0.002$} &
\multicolumn{1}{r}{$0.000$}\\
\multicolumn{1}{l}{$\overset{}{2}$} & \multicolumn{1}{l}{A} & &
\multicolumn{1}{r}{$0.5$} & \multicolumn{1}{r}{$0.5$} & \multicolumn{1}{r}{$1$%
} & $0.5$ & & \multicolumn{1}{r}{$0.042$} & \multicolumn{1}{r}{$0.006$} &
\multicolumn{1}{r}{$0.000$}\\
\multicolumn{1}{l}{} & \multicolumn{1}{l}{B} & & \multicolumn{1}{r}{$0.5$} &
\multicolumn{1}{r}{$2.5$} & \multicolumn{1}{r}{$5$} & $0.5$ & &
\multicolumn{1}{r}{$0.001$} & \multicolumn{1}{r}{$0.000$} &
\multicolumn{1}{r}{$0.000$}\\
\multicolumn{1}{l}{} & \multicolumn{1}{l}{C} & & \multicolumn{1}{r}{$0.5$} &
\multicolumn{1}{r}{$12.5$} & \multicolumn{1}{r}{$25$} & $0.5$ & &
\multicolumn{1}{r}{$0.000$} & \multicolumn{1}{r}{$0.000$} &
\multicolumn{1}{r}{$0.000$}\\
\multicolumn{1}{l}{$\overset{}{3}$} & \multicolumn{1}{l}{A} & &
\multicolumn{1}{r}{$0.2$} & \multicolumn{1}{r}{$0.8$} & \multicolumn{1}{r}{$1$%
} & $0.8$ & & \multicolumn{1}{r}{$0.025$} & \multicolumn{1}{r}{$0.000$} &
\multicolumn{1}{r}{$0.000$}\\
\multicolumn{1}{l}{} & \multicolumn{1}{l}{B} & & \multicolumn{1}{r}{$0.2$} &
\multicolumn{1}{r}{$4.0$} & \multicolumn{1}{r}{$5$} & $0.8$ & &
\multicolumn{1}{r}{$0.005$} & \multicolumn{1}{r}{$0.000$} &
\multicolumn{1}{r}{$0.000$}\\
\multicolumn{1}{l}{} & \multicolumn{1}{l}{C} & &
\multicolumn{1}{r}{$\underset{}{0.2}$} & \multicolumn{1}{r}{$20.0$} &
\multicolumn{1}{r}{$25$} & $0.8$ & & \multicolumn{1}{r}{$0.006$} &
\multicolumn{1}{r}{$0.000$} & \multicolumn{1}{r}{$0.000$}\\
\bottomrule
\end{tabular}
\end{table}

For each parameter configuration and sample size, we report the coverage
probabilities (estimated over the Monte Carlo replications) of confidence
intervals at the $95\%$ nominal level, using both asymptotic and bootstrap
methods. Asymptotic confidence intervals for the individual parameters as well
as the (joint) confidence ellipsoid are based on the sample Hessian. We also
report the coverage of (asymptotic and bootstrap) confidence intervals for the
branching ratio, $a=\alpha/\beta$. For bootstrap confidence intervals we
consider the naive percentile interval method.

Finally, we also report the (null) empirical rejection probabilities of
likelihood ratio tests for the hypothesis $H_{0}:\theta=\theta_{0}$. For the
bootstrap tests, we implement the unrestricted bootstrap (i.e., without the
null imposed on the bootstrap sample); results for the restricted bootstrap
(i.e., with the null imposed on the bootstrap sample) do not differ substantially.

\subsection{Results}\label{sec MC CIs}

The coverage probabilities of the asymptotic and bootstrap
confidence intervals [CI] for individual parameters are presented in
Table~\ref{tab:individual-interval}. We can see that, in general, the
asymptotic CIs suffer from the problem of undercoverage for almost all models
and sample spans, and this fact is particularly severe for some of the cases.
In contrast, the bootstrap methods, especially the FIB, powerfully correct
these distortions.

\begin{table}[t!]
\scriptsize
\caption{\textsc{Coverage of asymptotic and bootstrap confidence intervals for $\mu, \alpha, \beta$, and $a = \alpha/\beta$.}}
\smallskip

\label{tab:individual-interval}
\setlength{\tabcolsep}{0.65em}
\setlength{\extrarowheight}{0.1pt}
\begin{tabular}{cc|cccc|cccc|cccc} 
\toprule
& 
& \multicolumn{4}{c|}{Model 1A}
& \multicolumn{4}{c|}{Model 1B}
& \multicolumn{4}{c}{Model 1C}\\ 
&
& $\mu$ & $\alpha$ & $\beta$ & $a$ 
& $\mu$ & $\alpha$ & $\beta$ & $a$ 
& $\mu$ & $\alpha$ & $\beta$ & $a$ 
\\ \hline
$T=50$ 
& Asym 
& $94.9$ & $99.0$ & $91.9$ & $96.5$
& $92.6$ & $92.5$ & $88.4$ & $91.9$
& $92.2$ & $89.5$ & $89.4$ & $92.5$\\
& PRFB 
& $93.2$ & $93.0$ & $93.4$ & $95.8$
& $95.9$ & $96.0$ & $93.7$ & $96.4$
& $94.8$ & $95.9$ & $94.0$ & $96.7$\\
& NPFB 
& $92.5$ & $93.0$ & $93.4$ & $95.5$
& $95.5$ & $95.9$ & $93.5$ & $96.3$
& $94.2$ & $96.2$ & $94.1$ & $96.5$\\
& PRRB 
& $98.3$ & $95.5$ & $92.3$ & $99.1$
& $95.0$ & $97.2$ & $94.0$ & $97.5$
& $94.1$ & $96.9$ & $94.7$ & $96.9$\\
& NPRB 
& $97.2$ & $94.1$ & $90.7$ & $98.1$
& $93.5$ & $94.9$ & $89.5$ & $95.3$
& $90.3$ & $94.1$ & $89.8$ & $95.0$\\
\hline
$T=100$ 
& Asym 
& $95.0$ & $96.3$ & $90.1$ & $93.8$
& $93.8$ & $90.9$ & $88.9$ & $90.8$
& $93.8$ & $91.9$ & $91.8$ & $93.8$\\
& PRFB 
& $94.0$ & $94.9$ & $94.4$ & $95.8$
& $95.2$ & $96.1$ & $94.8$ & $96.0$
& $94.7$ & $96.1$ & $95.0$ & $96.2$\\
& NPFB 
& $93.7$ & $94.8$ & $94.3$ & $95.7$
& $94.9$ & $95.9$ & $94.5$ & $95.9$
& $94.1$ & $96.2$ & $95.2$ & $96.3$\\
& PRRB 
& $97.9$ & $96.3$ & $93.9$ & $98.1$
& $95.0$ & $96.8$ & $95.0$ & $96.5$
& $92.8$ & $96.7$ & $95.5$ & $96.3$\\
& NPRB 
& $96.7$ & $94.2$ & $92.0$ & $96.5$
& $93.4$ & $94.3$ & $92.1$ & $94.5$
& $91.9$ & $94.5$ & $92.7$ & $95.0$\\
\hline
$T=200$ 
& Asym 
& $94.0$ & $92.0$ & $87.6$ & $91.0$
& $93.7$ & $92.3$ & $90.7$ & $91.9$
& $94.2$ & $93.2$ & $93.1$ & $94.3$\\
& PRFB 
& $94.3$ & $95.2$ & $94.3$ & $95.7$
& $94.8$ & $95.5$ & $94.6$ & $95.4$
& $94.4$ & $95.1$ & $94.9$ & $95.2$\\
& NPFB 
& $94.3$ & $95.1$ & $94.3$ & $95.4$
& $94.5$ & $95.6$ & $94.7$ & $95.5$
& $94.1$ & $95.0$ & $95.0$ & $95.1$\\
& PRRB 
& $97.2$ & $95.9$ & $94.4$ & $97.1$
& $94.2$ & $96.0$ & $95.3$ & $95.7$
& $93.6$ & $95.3$ & $95.5$ & $95.2$\\
& NPRB  
& $96.2$ & $93.4$ & $93.0$ & $95.2$
& $93.1$ & $94.9$ & $92.9$ & $94.3$
& $93.1$ & $94.2$ & $92.7$ & $94.5$\\
\midrule
& 
& \multicolumn{4}{c|}{Model 2A}
& \multicolumn{4}{c|}{Model 2B}
& \multicolumn{4}{c}{Model 2C}\\ 
&
& $\mu$ & $\alpha$ & $\beta$ & $a$ 
& $\mu$ & $\alpha$ & $\beta$ & $a$ 
& $\mu$ & $\alpha$ & $\beta$ & $a$\\\hline
$T=50$ 
& Asym 
& $91.7$ & $92.2$ & $93.6$ & $88.7$
& $92.1$ & $92.2$ & $93.5$ & $92.0$
& $92.5$ & $90.7$ & $93.4$ & $92.0$\\
& PRFB 
& $96.1$ & $97.0$ & $95.0$ & $96.6$
& $95.5$ & $96.8$ & $95.2$ & $96.2$
& $95.5$ & $95.6$ & $94.8$ & $93.8$\\
& NPFB
& $95.6$ & $97.1$ & $95.0$ & $96.5$
& $95.2$ & $97.0$ & $95.3$ & $96.0$
& $95.0$ & $95.6$ & $95.2$ & $93.6$\\
& PRRB 
& $93.8$ & $98.5$ & $94.3$ & $88.3$
& $91.7$ & $98.2$ & $96.8$ & $91.0$
& $90.4$ & $96.6$ & $97.1$ & $89.2$\\
& NPRB 
& $90.8$ & $95.3$ & $92.6$ & $85.6$
& $90.0$ & $95.2$ & $93.6$ & $88.6$
& $89.8$ & $93.8$ & $94.1$ & $87.8$\\\hline
$T=100$ 
& Asym 
& $92.6$ & $91.8$ & $93.0$ & $91.0$
& $93.5$ & $92.5$ & $93.8$ & $93.2$
& $93.4$ & $93.2$ & $94.5$ & $93.8$\\
& PRFB 
& $96.0$ & $96.5$ & $95.2$ & $96.3$
& $95.3$ & $95.3$ & $94.7$ & $94.7$
& $95.1$ & $95.0$ & $94.7$ & $94.2$\\
& NPFB 
& $95.8$ & $96.6$ & $95.2$ & $96.3$
& $95.0$ & $95.6$ & $95.3$ & $94.5$
& $94.8$ & $95.1$ & $95.1$ & $94.3$\\
& PRRB 
& $94.1$ & $98.0$ & $95.6$ & $91.3$
& $92.9$ & $95.8$ & $95.9$ & $91.7$
& $92.1$ & $95.3$ & $96.0$ & $91.9$\\
& NPRB
& $92.0$ & $95.1$ & $94.2$ & $89.4$
& $91.2$ & $94.2$ & $93.6$ & $90.3$
& $91.4$ & $93.6$ & $93.1$ & $90.8$\\\hline
$T=200$ 
& Asym 
& $93.5$ & $93.4$ & $94.0$ & $92.3$
& $94.6$ & $94.4$ & $94.7$ & $94.8$
& $94.2$ & $94.6$ & $95.1$ & $93.9$\\
& PRFB 
& $95.3$ & $95.8$ & $94.9$ & $95.2$
& $95.0$ & $95.1$ & $94.9$ & $95.3$
& $94.8$ & $95.2$ & $94.9$ & $94.1$\\
& NPFB 
& $95.5$ & $95.8$ & $94.9$ & $95.3$
& $95.1$ & $95.3$ & $94.6$ & $94.8$
& $94.9$ & $95.1$ & $94.6$ & $93.8$\\
& PRRB 
& $94.5$ & $96.7$ & $95.4$ & $92.3$
& $94.0$ & $95.3$ & $95.3$ & $93.6$
& $93.5$ & $95.3$ & $95.3$ & $92.8$\\
& NPRB
& $92.3$ & $94.2$ & $94.0$ & $90.7$
& $92.5$ & $93.7$ & $92.7$ & $92.4$
& $92.5$ & $93.7$ & $92.6$ & $91.7$\\
\midrule
& 
& \multicolumn{4}{c|}{Model 3A}
& \multicolumn{4}{c|}{Model 3B}
& \multicolumn{4}{c}{Model 3C}\\ 
&
& $\mu$ & $\alpha$ & $\beta$ & $a$ 
& $\mu$ & $\alpha$ & $\beta$ & $a$ 
& $\mu$ & $\alpha$ & $\beta$ & $a$\\\hline
$T=50$ 
& Asym 
& $90.5$ & $91.1$ & $95.2$ & $86.5$
& $89.5$ & $90.0$ & $95.1$ & $87.1$
& $88.1$ & $89.7$ & $94.8$ & $87.6$\\
& PRFB 
& $88.9$ & $98.1$ & $95.2$ & $92.1$
& $89.6$ & $97.2$ & $95.4$ & $92.3$
& $90.0$ & $93.6$ & $95.9$ & $91.7$\\
& NPFB
& $87.7$ & $97.9$ & $95.2$ & $92.2$
& $88.9$ & $96.7$ & $95.0$ & $92.1$
& $91.4$ & $94.8$ & $94.0$ & $91.2$\\
& PRRB 
& $89.7$ & $98.9$ & $94.7$ & $72.5$
& $88.6$ & $97.2$ & $94.6$ & $80.1$
& $88.9$ & $95.7$ & $97.8$ & $80.0$\\
& NPRB 
& $86.6$ & $95.2$ & $96.0$ & $72.2$
& $87.9$ & $94.3$ & $96.3$ & $77.4$
& $90.3$ & $93.9$ & $96.2$ & $87.6$\\\hline
$T=100$ 
& Asym 
& $92.6$ & $92.5$ & $95.0$ & $90.9$
& $92.0$ & $92.6$ & $94.8$ & $91.2$
& $91.4$ & $92.1$ & $95.0$ & $90.7$\\
& PRFB 
& $95.4$ & $96.3$ & $94.7$ & $94.6$
& $95.3$ & $96.1$ & $94.4$ & $93.0$
& $95.0$ & $94.7$ & $94.2$ & $92.2$\\
& NPFB 
& $95.2$ & $96.5$ & $95.0$ & $94.1$
& $95.3$ & $96.0$ & $94.5$ & $92.4$
& $94.5$ & $94.6$ & $94.1$ & $91.6$\\
& PRRB 
& $92.2$ & $96.1$ & $95.4$ & $80.6$
& $90.3$ & $96.0$ & $96.9$ & $81.0$
& $89.1$ & $95.1$ & $97.3$ & $81.2$\\
& NPRB
& $88.6$ & $94.0$ & $95.1$ & $76.0$
& $89.7$ & $92.8$ & $95.3$ & $78.9$
& $92.6$ & $94.5$ & $95.5$ & $88.4$\\\hline
$T=200$ 
& Asym 
& $93.6$ & $93.6$ & $94.9$ & $92.9$
& $93.3$ & $93.7$ & $95.1$ & $93.4$
& $93.1$ & $93.1$ & $94.9$ & $92.6$\\
& PRFB 
& $95.1$ & $95.6$ & $95.1$ & $93.7$
& $95.2$ & $95.2$ & $94.9$ & $93.8$
& $94.9$ & $94.0$ & $94.0$ & $92.7$\\
& NPFB 
& $94.9$ & $95.7$ & $94.9$ & $93.5$
& $95.1$ & $95.0$ & $95.0$ & $93.4$
& $95.1$ & $94.6$ & $94.5$ & $92.9$\\
& PRRB 
& $92.7$ & $95.4$ & $96.1$ & $84.8$
& $92.8$ & $94.8$ & $96.9$ & $95.3$
& $91.6$ & $94.1$ & $96.2$ & $86.2$\\
& NPRB
& $90.6$ & $92.8$ & $94.1$ & $82.8$
& $91.3$ & $92.6$ & $94.3$ & $85.1$
& $95.1$ & $94.6$ & $94.5$ & $92.9$\\
\bottomrule
\end{tabular}
\smallskip

Note: Nominal coverage rate is $95\%$. PRFB, NPFB, PRRB, and NPRB refer to parametric fixed intensity, non-parametric fixed intensity, parametric recursive intensity and non-parametric recursive intensity bootstraps.
\end{table}

\begin{sidewaystable}[htp!]
\small
\caption{\textsc{Coverage of (asymptotic and bootstrap) confidence ellipsoids.}}
\smallskip 

\label{tab:ellipsoid}
\setlength{\tabcolsep}{0.6em}
\setlength{\extrarowheight}{2pt}
\begin{tabular}{cc|cc|cc|cc|cc|cc|cc|cc|cc|cc} 
\toprule
& Model
& \multicolumn{2}{c|}{1A}
& \multicolumn{2}{c|}{1B}
& \multicolumn{2}{c|}{1C}
& \multicolumn{2}{c|}{2A}
& \multicolumn{2}{c|}{2B}
& \multicolumn{2}{c|}{2C}
& \multicolumn{2}{c|}{3A}
& \multicolumn{2}{c|}{3B}
& \multicolumn{2}{c} {3C}\\ 
&
& $\theta$ & $\tilde\theta$ 
& $\theta$ & $\tilde\theta$ 
& $\theta$ & $\tilde\theta$ 
& $\theta$ & $\tilde\theta$ 
& $\theta$ & $\tilde\theta$ 
& $\theta$ & $\tilde\theta$ 
& $\theta$ & $\tilde\theta$ 
& $\theta$ & $\tilde\theta$ 
& $\theta$ & $\tilde\theta$\\\hline
$T=50$ 
& Asym 
& $81.7$ & $89.6$
& $85.0$ & $84.4$ 
& $85.3$ & $84.9$ 
& $86.4$ & $84.7$ 
& $88.8$ & $87.5$ 
& $88.8$ & $88.9$
& $86.5$ & $83.9$ 
& $86.4$ & $84.8$ 
& $84.7$ & $83.9$ \\
& PRFB 
& $99.6$ & $99.5$
& $98.1$ & $97.9$ 
& $96.0$ & $95.4$ 
& $99.5$ & $99.1$ 
& $97.5$ & $96.9$ 
& $95.2$ & $95.5$
& $99.3$ & $98.8$ 
& $98.3$ & $97.4$ 
& $97.2$ & $96.5$ \\
& NPFB 
& $99.5$ & $99.4$
& $97.7$ & $97.4$ 
& $95.9$ & $95.3$
& $99.4$ & $98.9$ 
& $97.2$ & $96.7$ 
& $95.3$ & $95.4$ 
& $99.2$ & $98.4$ 
& $97.7$ & $96.5$ 
& $97.8$ & $97.1$ \\
& PRRB 
& $99.6$ & $99.6$
& $97.2$ & $97.4$ 
& $95.6$ & $95.2$ 
& $99.4$ & $99.0$ 
& $97.5$ & $97.7$ 
& $96.6$ & $96.9$
& $98.0$ & $97.3$ 
& $98.7$ & $97.9$ 
& $98.2$ & $97.9$ \\
& NPRB 
& $99.5$ & $99.1$
& $96.3$ & $96.1$ 
& $93.0$ & $92.3$ 
& $99.4$ & $98.6$ 
& $95.9$ & $96.4$ 
& $94.5$ & $95.8$ 
& $98.1$ & $97.9$ 
& $97.3$ & $97.7$ 
& $97.1$ & $96.8$ \\
\hline
$T=100$ 
& Asym 
& $84.6$ & $86.7$
& $85.9$ & $84.7$ 
& $89.4$ & $89.4$ 
& $87.8$ & $87.9$ 
& $90.7$ & $90.3$ 
& $92.0$ & $92.2$ 
& $90.1$ & $88.7$ 
& $90.3$ & $89.6$ 
& $90.1$ & $90.0$ \\
& PRFB 
& $99.4$ & $98.9$
& $97.0$ & $96.6$ 
& $96.1$ & $95.9$ 
& $98.5$ & $97.9$ 
& $95.7$ & $95.5$ 
& $94.6$ & $94.5$
& $98.1$ & $97.5$ 
& $95.2$ & $94.3$ 
& $93.7$ & $93.0$ \\
& NPFB 
& $99.2$ & $98.7$
& $97.1$ & $96.7$ 
& $96.2$ & $96.1$ 
& $98.4$ & $97.8$ 
& $95.4$ & $95.2$ 
& $94.3$ & $94.2$ 
& $97.8$ & $97.1$ 
& $94.6$ & $93.4$ 
& $92.9$ & $92.3$ \\
& PRRB 
& $99.0$ & $99.0$
& $96.0$ & $96.1$ 
& $96.2$ & $96.2$ 
& $98.0$ & $98.3$ 
& $95.7$ & $96.3$ 
& $95.4$ & $95.7$ 
& $97.0$ & $96.8$
& $97.2$ & $97.2$ 
& $97.0$ & $97.1$ \\
& NPRB 
& $99.0$ & $98.6$
& $95.4$ & $95.5$ 
& $94.1$ & $95.0$ 
& $97.9$ & $98.2$ 
& $94.8$ & $95.7$ 
& $94.9$ & $95.9$ 
& $97.3$ & $97.6$ 
& $95.8$ & $96.4$ 
& $95.9$ & $96.1$\\
\hline
$T=200$ 
& Asym 
& $81.3$ & $83.4$ 
& $87.9$ & $87.2$ 
& $97.1$ & $92.1$ 
& $90.7$ & $91.2$ 
& $93.7$ & $93.6$ 
& $93.6$ & $93.6$ 
& $92.1$ & $91.9$ 
& $92.4$ & $92.5$ 
& $91.9$ & $92.3$ \\
& PRFB 
& $98.4$ & $98.0$ 
& $96.1$ & $95.8$ 
& $95.4$ & $95.5$ 
& $97.5$ & $96.9$ 
& $95.5$ & $94.9$ 
& $94.6$ & $94.5$ 
& $95.9$ & $95.4$ 
& $94.0$ & $93.8$ 
& $93.1$ & $93.0$ \\
& NPFB 
& $98.2$ & $97.8$ 
& $96.0$ & $95.9$ 
& $95.3$ & $95.5$ 
& $97.3$ & $96.5$ 
& $95.5$ & $95.0$ 
& $94.6$ & $94.4$ 
& $95.3$ & $94.9$ 
& $93.7$ & $93.3$ 
& $93.1$ & $92.9$ \\
& PRRB 
& $97.2$ & $98.1$ 
& $95.2$ & $95.6$ 
& $95.2$ & $95.7$ 
& $96.4$ & $96.7$ 
& $95.4$ & $95.5$ 
& $95.2$ & $95.1$ 
& $95.5$ & $96.0$ 
& $95.5$ & $95.7$ 
& $95.2$ & $95.3$ \\
& NPRB  
& $97.3$ & $97.9$ 
& $94.7$ & $95.4$ 
& $94.6$ & $95.6$ 
& $96.6$ & $96.6$ 
& $95.6$ & $96.1$ 
& $95.8$ & $95.9$ 
& $95.4$ & $96.0$ 
& $95.1$ & $95.8$ 
& $95.2$ & $95.4$ \\
\bottomrule
\end{tabular}
\smallskip 

Note: Nominal coverage rate is $95\%$; $\theta=(\mu,\alpha,\beta)'$ and $\tilde\theta=(\mu,a,\beta)'$.  PRFB, NPFB, PRRB, and NPRB refer to parametric fixed intensity, non-parametric fixed intensity, parametric recursive intensity and non-parametric recursive intensity bootstraps.
\end{sidewaystable}

\begin{table}[t!] 
\caption{\textsc{Empirical rejection probabilities (in percentage) of the $5\%$ asymptotic and unrestricted bootstrap likelihood-ratio tests.}}
\smallskip 

\label{tab:LR-test}
\setlength{\tabcolsep}{0.77em}
\begin{tabular}{cc|ccc|ccc|ccc} 
\toprule
& Model 
& 1A & 1B & 1C & 2A & 2B & 2C & 3A & 3B & 3C
\\ \hline
$T=50$ 
& Asym 
& $3.5$ & $4.1$ & $5.6$ & $4.3$ & $6.1$ & $6.1$ & $8.3$ & $7.9$ & $8.1$\\
& PRFB 
& $2.7$ & $3.1$ & $4.5$ & $3.0$ & $4.5$ & $4.8$ & $4.7$ & $4.9$ & $4.6$\\
& NPFB 
& $3.2$ & $3.5$ & $4.7$ & $3.6$ & $4.9$ & $5.2$ & $5.8$ & $5.7$ & $4.5$\\
& PRRB 
& $2.8$ & $3.4$ & $4.8$ & $3.1$ & $4.9$ & $5.3$ & $5.2$ & $5.2$ & $4.5$\\
& NPRB 
& $3.1$ & $4.6$ & $6.7$ & $3.5$ & $5.3$ & $5.6$ & $5.5$ & $5.3$ & $5.3$\\
\hline
$T=100$ 
& Asym 
& $3.3$ & $4.5$ & $5.4$ & $5.4$ & $6.0$ & $5.3$ & $7.0$ & $6.4$ & $6.2$\\
& PRFB 
& $2.8$ & $3.9$ & $4.7$ & $4.1$ & $5.2$ & $4.6$ & $4.6$ & $4.8$ & $4.7$\\
& NPFB 
& $3.0$ & $4.0$ & $5.1$ & $4.6$ & $5.2$ & $5.1$ & $4.9$ & $5.6$ & $5.4$\\
& PRRB 
& $2.8$ & $3.9$ & $5.2$ & $5.0$ & $4.7$ & $5.0$ & $5.1$ & $5.1$ & $4.6$\\
& NPRB 
& $2.9$ & $4.3$ & $5.5$ & $3.9$ & $4.6$ & $4.2$ & $4.9$ & $5.1$ & $5.0$\\
\hline
$T=200$ 
& Asym 
& $4.2$ & $5.7$ & $5.4$ & $5.2$ & $4.9$ & $5.3$ & $5.8$ & $5.5$ & $5.7$\\
& PRFB 
& $3.8$ & $5.0$ & $5.1$ & $4.7$ & $4.6$ & $4.8$ & $4.6$ & $4.7$ & $5.0$\\
& NPFB 
& $4.0$ & $4.9$ & $5.1$ & $4.7$ & $4.8$ & $5.0$ & $5.1$ & $4.9$ & $5.2$\\
& PRRB 
& $3.7$ & $5.1$ & $5.1$ & $5.0$ & $4.7$ & $5.0$ & $4.7$ & $4.6$ & $4.9$\\
& NPRB 
& $3.5$ & $4.9$ & $4.7$ & $4.0$ & $3.7$ & $3.8$ & $3.9$ & $3.8$ & $4.4$\\
\bottomrule
\end{tabular}
\medskip

{\small Note: The null hypothesis is $H_0:\theta=\theta_0$, where $\theta=(\mu,\alpha,\beta)'$. The bootstrap is based on unrestricted parameter estimation. PRFB, NPFB, PRRB, and NPRB refer to parametric fixed intensity, non-parametric fixed intensity, parametric recursive intensity and non-parametric recursive intensity bootstraps.}
\end{table}

Below we provide a summary of the problems related to the asymptotic CIs for
each individual parameter (branching ratio $a$, baseline intensity $\mu$,
intensity jump size $\alpha$ and decay rate $\beta$).

\smallskip\noindent(i) The undercoverage of the asymptotic CI for the
branching ratio is severe in finite sample for all Models 1--3. The coverage
deteriorates as the true value of branching ratio increases (moving from Model
1 to 3), and as the true values of $\alpha$ and $\beta$ decrease (moving from
Model C to A). Accordingly, the performance of the asymptotic CI for the
branching ratio is the worst for Model 3A, where the coverage probability is
$86.5\%$ for $T=50$. Larger $\alpha_{0}$ and $\beta_{0}$ seem to improve the
coverage rate of the branching ratio, and this improvement is the most
significant for Model 1 where the branching ratio is low.

\smallskip\noindent(ii) The asymptotic CI for the baseline intensity $\mu$
performs poorly in finite samples when $\mu_{0}$ is low. Note that for Model
3, where $\mu_{0}=0.2$, the empirical coverage probabilities are $90.5\%$,
$89.5\%$, and $88.1\%$ for Model 3A, 3B and 3C, respectively, when $T=50$. In
contrast, these probabilities are all above $90\%$ for Models 1 and 2.

\smallskip\noindent(iii) The problem of undercoverage deteriorates when
$\alpha_{0}$ is larger (moving from Model A to C). There are no significant
changes in the coverage of $\alpha$ over different values of the branching
ratio. Improvements in the coverage of $\alpha$ seem to come only from
increasing the sample span $T$. In general, the coverage is acceptable.

\smallskip\noindent(iv) The undercoverage of $\beta$ is severe for Model 1
with small branching ratio, but the coverage rate improves noticeably as
branching ratio increases, and as sample span $T$ increases. In particular,
the asymptotic CI coverage for $\beta$ is almost perfect for Models 3A--C even
when $T=50$. The performance is independent of the value of $\beta$.
\smallskip

In contrast to the coverage of asymptotic CIs, which show evident finite
sample distortions, the empirical coverage probabilities of the bootstrap
percentile intervals based on the fixed intensity scheme (for both the
parametric and non-parametric methods, labelled `PRFB' and `NPFB' in
Table~\ref{tab:individual-interval}) are very close to the nominal level, for
almost all simulation models and even when the sample span is very short
($T=50$). The only exceptions are for the coverage of branching ratio in Model
3, where the coverage probabilities of the parametric FIB and the
non-parametric FIB are slightly below $95\%$, the nominal level. Nevertheless,
the CIs of the two recursive intensity bootstraps, although performing
generally better than asymptotic CIs for the coverage of parameter $\alpha$
and $\beta$, share similar features of finite sample distortion as asymptotic
CIs. For instance, the coverage of $\mu$ deteriorates as $\mu_{0}$ decreases
(for both the parametric and non-parametric RIBs); the coverage of $\beta$ is
much below the nominal level for Model 1 where branching ratio is low, while
it converges to the nominal level as the branching ratio increases (for the
RIB); finally, we observe that the coverage of the branching ratio
deteriorates when the branching ratio increases.

Unreported simulations of average lengths of the $95\%$ asymptotic and
bootstrap confidence intervals for each parameter show that bootstrap
confidence intervals are not significantly wider than asymptotic confidence
intervals, except for Models 1A and 1B in which the parameter settings are
relatively more extreme, or when the sample span is short ($T=50$). The wider
bootstrap confidence intervals reveal the higher uncertainty associated to
parameter estimation, and is in line with existing literature on the
bootstrap.

Table~\ref{tab:ellipsoid} presents the joint coverage rate of the asymptotic
and bootstrap confidence ellipsoids [CE], for both parameterizations
$\theta=(\mu,\alpha,\beta)^{\prime}$ and $\tilde{\theta}=(\mu,a,\beta
)^{\prime}$. Noticeably, here the benefit of using bootstrap methods to
improve the finite sample joint coverage is way more than evident. The
performance of the asymptotic CEs is clearly unsatisfactory: For all nine
models, the empirical coverage probabilities of the asymptotic CEs are below
$89\%$ when $T=50$; despite a gradual improvement of the coverage rates as
sample span $T$ increases, the coverage rates when $T=200$ are still below the
nominal level for all models (the joint coverage probabilities of Model 1B are
even less than $88\%$ when $T=200$). On the contrary, all bootstrap methods
produce the joint CEs that cover the true parameters with probabilities very
close to the nominal level,\footnote{We do observe that there is some tendency
of over-coverage of the bootstrap joint CEs for Model 2A and 3A for relatively
short sample spans.} across different models and different sample spans.

Finally, in Table~\ref{tab:LR-test} we report the empirical rejection
probabilities of the asymptotic and unrestricted bootstrap likelihood-ratio
tests for the null hypothesis $H_{0}:\theta=\theta_{0}$. In general, both the
asymptotic and bootstrap tests perform satisfactorily well in terms of size,
especially when $T=100$ and $200$. Nevertheless, we do notice that the
asymptotic test tends to be oversized for larger values of the branching
ratio. This can be seen by inspecting the rejection probabilities of the
asymptotic test on $H_{0}$ for Model 3 (which has the largest branching ratio,
$a=0.8$) for $T=50,100$. In particular, the asymptotic test is severely
oversized for all three sub-models of Model 3, particularly so when $T=50$. In
contrast, we do not see much variability of the bootstrap empirical rejection
probabilities across different models or sample spans -- they are all very
close to the nominal level (slightly conservative in some cases).

\section{Empirical illustrations\label{sec: empirical}}

To illustrate how the proposed bootstrap schemes work in applications, we
consider two empirical examples. The first consists of `extreme occurrences'
in US stock market data, as measured by empirical quantiles of the Dow Jones
Index, see \citet{ELL2011}. We use this application to compare the four
different bootstrap schemes discussed in the paper. Next, we analyze recent
Danish COVID-19 tweets using the non-parametric FIB. We illustrate how
bootstrap confidence intervals reveal the presence of a structural break in
the parameters, whereas confidence intervals based on the asymptotic Gaussian
approximation do not.

\subsection{Dow Jones Index}

\begin{figure}[t!]
\centering
\includegraphics[width=.9\textwidth]{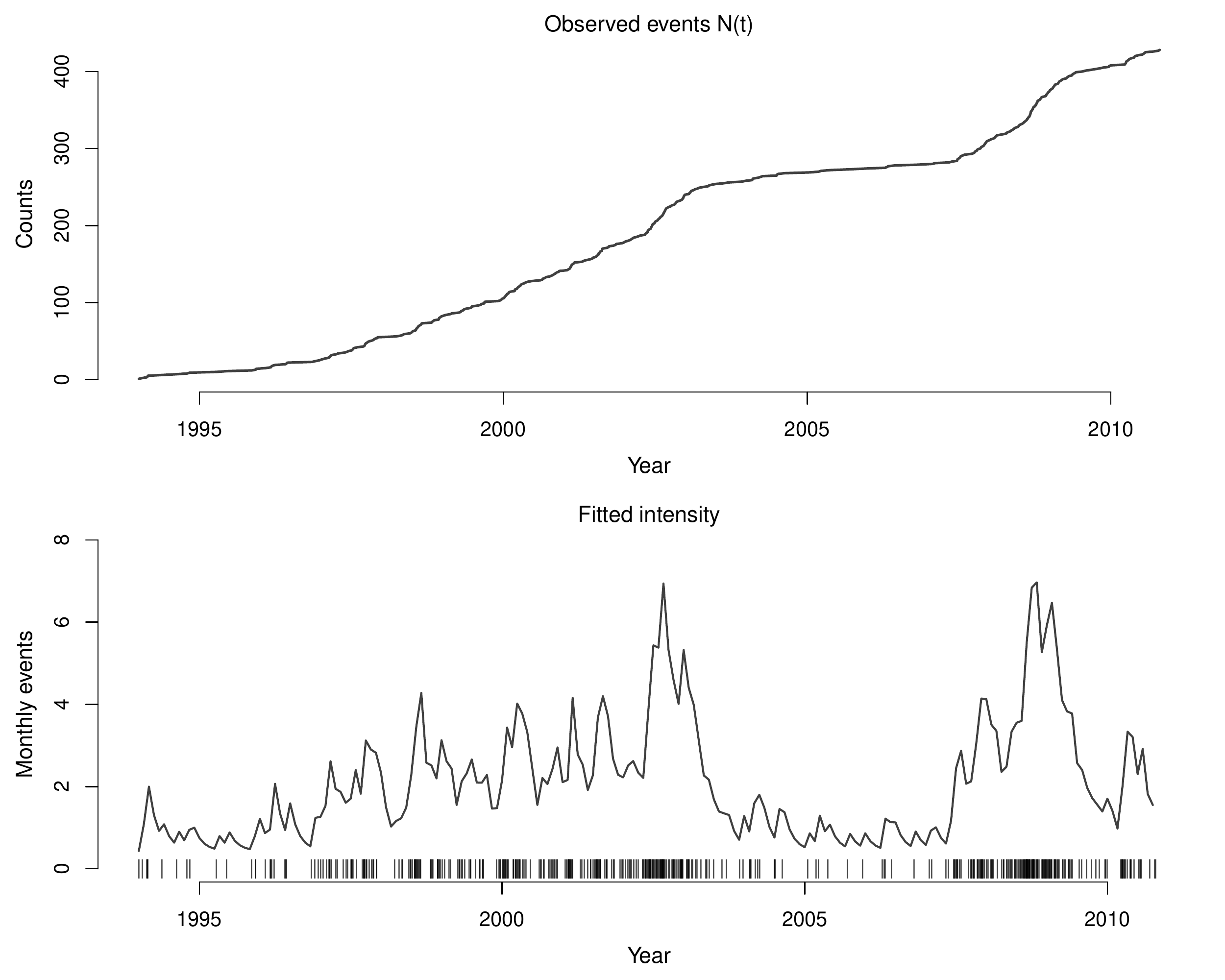}
\caption{Dow Jones Index data. The top panel shows the observed counts at times $\{t_i\}$ while the bottom panel shows the estimated intensity $\lambda(t;\hat{\theta})$ with event times marked as barcodes. The time resolution for the bottom panel is in months.}
\label{ref2figDJ1}
\end{figure}

As in \citet{ELL2011}, we consider Dow Jones Index (DJI) daily (log) returns
observed over the period January 1, 1994 to December 31, 2010. The event times
corresponding to extreme returns are given by the trading days where the
corresponding daily return is below the $10\%$ empirical quantile (negative
occurrences), resulting in $n_{T}=428$ events during the period of $T=6144$
days considered. Figure~\ref{ref2figDJ1} (top panel) shows the event times and
the associated counting process.

\begin{figure}[htp!]
\centering
\includegraphics[width=\textwidth]{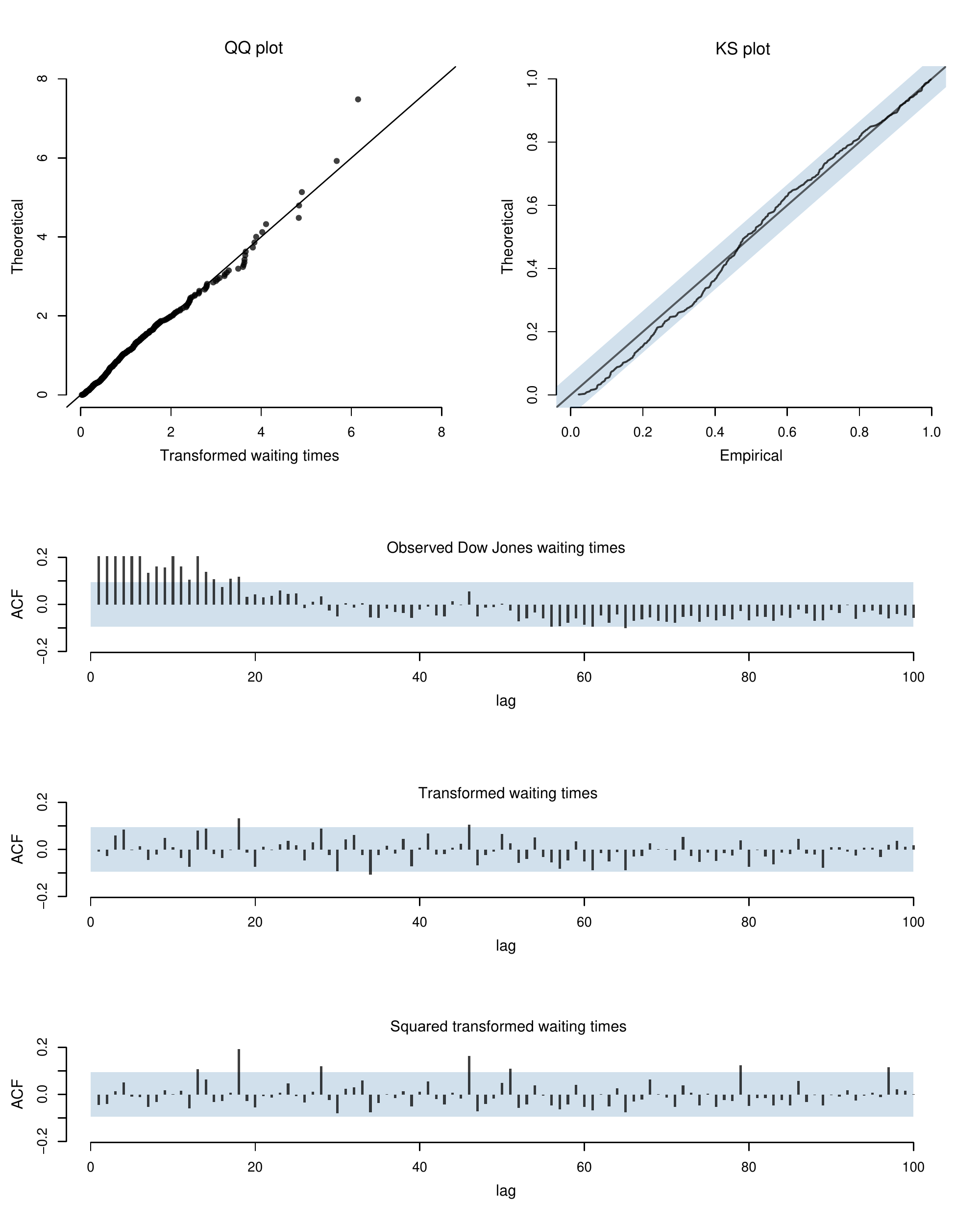}
\caption{Misspecification analysis for the DJI data. The top left panel shows a QQ-plot of the time transformed waiting times $\hat v_i=\Lambda(t_i,t_{i-1};\hat{\theta})$ against a unit exponential distribution. The top right panel shows the corresponding KS plot with 95\% confidence band in shaded blue. The three lower panels show the autocorrelations for the observed waiting times, the time transformed waiting times $\hat v_i$ and $\hat v_i^2$ with 95\% confidence band in shaded blue.}
\label{ref2secApp}
\end{figure}

To analyze the data, we consider a Hawkes model with intensity reparameterized
as
\begin{equation}
\lambda(t;\theta)=\mu+a\sum_{t_{i}<t}\gamma(t-t_{i};\theta)
\label{emp: intens}
\end{equation}
where $a$ is the branching ratio and $\gamma$ is the (exponential) kernel;
that is, $\gamma(t;\theta)=\beta\exp(-\beta t)$, see also
(\ref{eq: exp kernel}) and Section~\ref{Sec MC sim}. With parameter vector
$\theta=(\mu,a,\beta)^{\prime}$, the MLE $\hat{\theta}$ is obtained by
maximizing the log-likelihood in (\ref{eq log lik compact}) subject to
$\mu,\beta>0$, $0<a<1$ and with initial values from \citet{ELL2011}.
Estimation results are reported in Table~\ref{ref2tabDJ1}; the estimated
intensity is portrayed in the bottom panel of Figure~\ref{ref2figDJ1}. The MLE
$\hat{\theta}$ is very similar to \citet{ELL2011}, and we observe in
particular that the branching ratio $a$ appears to be well inside the
stationary region.

As previously emphasized, if the model is correctly specified, the transformed
waiting times should be i.i.d. $\mathcal{E}\left(  1\right)  $. Therefore, the
model fit can be evaluated by considering the estimated transformed waiting
times
\[
\hat{v}_{i}=\Lambda(t_{i},t_{i-1};\hat{\theta}),\quad i=1,2,...,n_{T},
\]
with $\Lambda$ defined in (\ref{eq: piece Lam}). Figure~\ref{ref2secApp}
contains QQ-plots and Kolmogorov-Smirnov (KS) plots, as well as sample
autocorrelograms and related tests. Based on these, we see no clear signs of
model misspecification. Precisely, the QQ plot of $\hat{v}_{i}$ against a unit
exponential distribution has no significant deviations from the identity line,
except a few quantiles in the extreme upper tail, as also confirmed by the KS
statistic p-value ($0.147$). Moreover, while the observed waiting times
$w_{i}$ are autocorrelated, this is not the case for the transformed waiting
times $\hat{v}_{i}$ (and its squares, $\hat{v}_{i}^{2}$).

We next compare the different bootstrap algorithms in terms of confidence
intervals for the parameters, and compare these with the asymptotic CIs.
With $\{\hat{\theta}_{T,i:b}^{\ast}\}_{b=1}^{B}$ the i.i.d. bootstrap
realizations of the $i$-th element of $\hat{\theta}_{T}^{\ast}$, the bootstrap
CIs reported are based on the empirical $\alpha/2$ and $1-\alpha/2$ quantiles
of the empirical distribution function of the $\hat{\theta}_{T,i:b}^{\ast}$'s.
In Table \ref{ref2tabDJ1}, while we find no noticeable difference between the
parametric and non-parametric bootstraps, the bootstrapped CIs based on the
FIB are less wide when compared to the asymptotic and RIB CIs (recall also
from the Monte Carlo results that in general the bootstrap coverage
probabilities are better than those associated to the asymptotic CIs). The
observed difference between the FIB and RIB CIs is likely to be caused by the
added randomness in the sequential computation of the RIB. Interestingly, the
FIB and RIB bootstrap CIs are further away from the non-stationary region
($a\geq1$) than the asymptotic CIs.

\begin{table}[t!]
\caption{\textsc{Estimated parameters and bootstrap $95\%$ confidence intervals for DJI data.}}
\smallskip 

\label{ref2tabDJ1}
\setlength{\tabcolsep}{0.6em}
\begin{tabular}{rc c c c c c}
\toprule
& $\hat{\theta}$ & Asymptotic & PRFB & NPFB & PRRB & NPRB\\
\midrule
$\mu$& 0.205 & [0.10; 0.31] & [0.19; 0.34] & [0.19; 0.34] & [0.11; 0.38] & [0.13; 0.40] \\ 
$a$ & 0.800 & [0.67; 0.93] & [0.67; 0.82] & [0.67; 0.82] & [0.58; 0.91] & [0.54; 0.89] \\ 
$\beta$	& 0.275 & [0.15; 0.40] & [0.14; 0.35] & [0.14; 0.30] & [0.18; 0.43] & [0.18; 0.42] \\ \bottomrule
\end{tabular}
\medskip

{\small Note: PRFB, NPFB, PRRB, and NPRB refer to parametric fixed
intensity, non-parametric fixed intensity, parametric recursive intensity and non-parametric recursive intensity bootstraps.}
\end{table}

\subsection{COVID-19 Tweets}
\begin{figure}[t!]
\centering
\includegraphics[width=.9\textwidth]{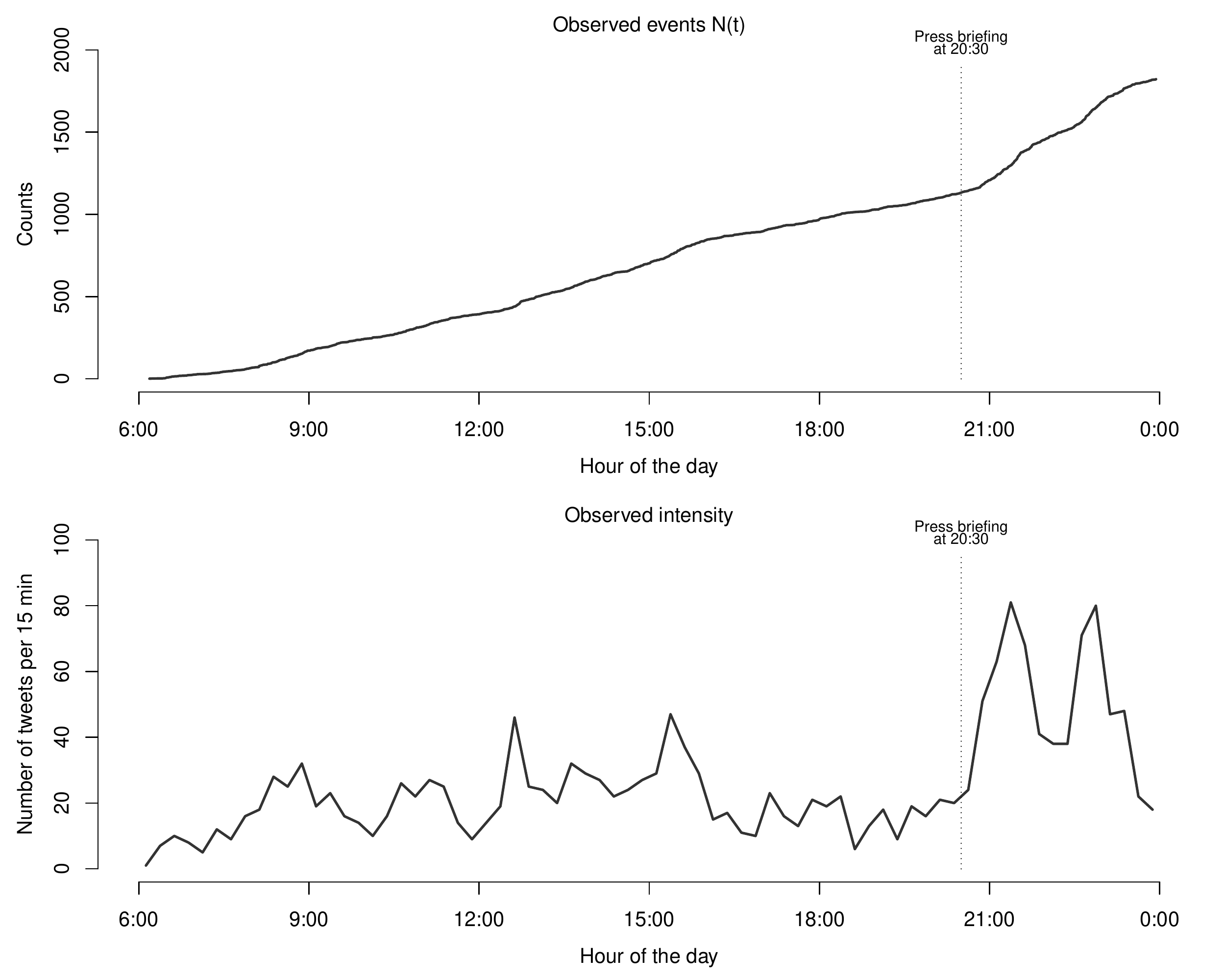}
\caption{Danish COVID-19 tweets data. The top panel shows the observed counts at times $\{t_i\}$, while the bottom panel shows the the number of events every $15$ minutes. The time resolution for the bottom panel is 15 minute intervals. The vertical dashed line shows the time of the press briefing.}
\label{ref2figCV1}
\end{figure}

We consider the arrival times of tweets related to the COVID-19 pandemic,
recorded on March 11 (06:00-00:00) 2020, when during a press briefing the
Danish Prime Minister at 20:30 announced the first lockdown of Denmark. In
total, there are $n_{T}=1822$ events from $1166$ unique individuals, with each
event time $\{t_{i}\}_{i=0}^{n_{T}}$ ($t_{0}=0$) measured with a time
resolution of $1$ second within the $T=18$ hours considered. In order to
analyze the effects of the announcement, we analyze the full sample, as well
as the pre-press briefing sample (06:00--20:30), and the post-press briefing
sample (20:30--00:00). In Figure~\ref{ref2figCV1}, we show the observed
counting process $N(t)$ for $t\in\lbrack0,T]$ as well as an initial proxy for
the intensity given by the number of events per $15$-minute intervals. It is
worth noticing that there is a surge in activity after 20:30, visible both in
the counting process and the increased intensity.

\begin{figure}[htp!]
\centering
\includegraphics[width=\textwidth]{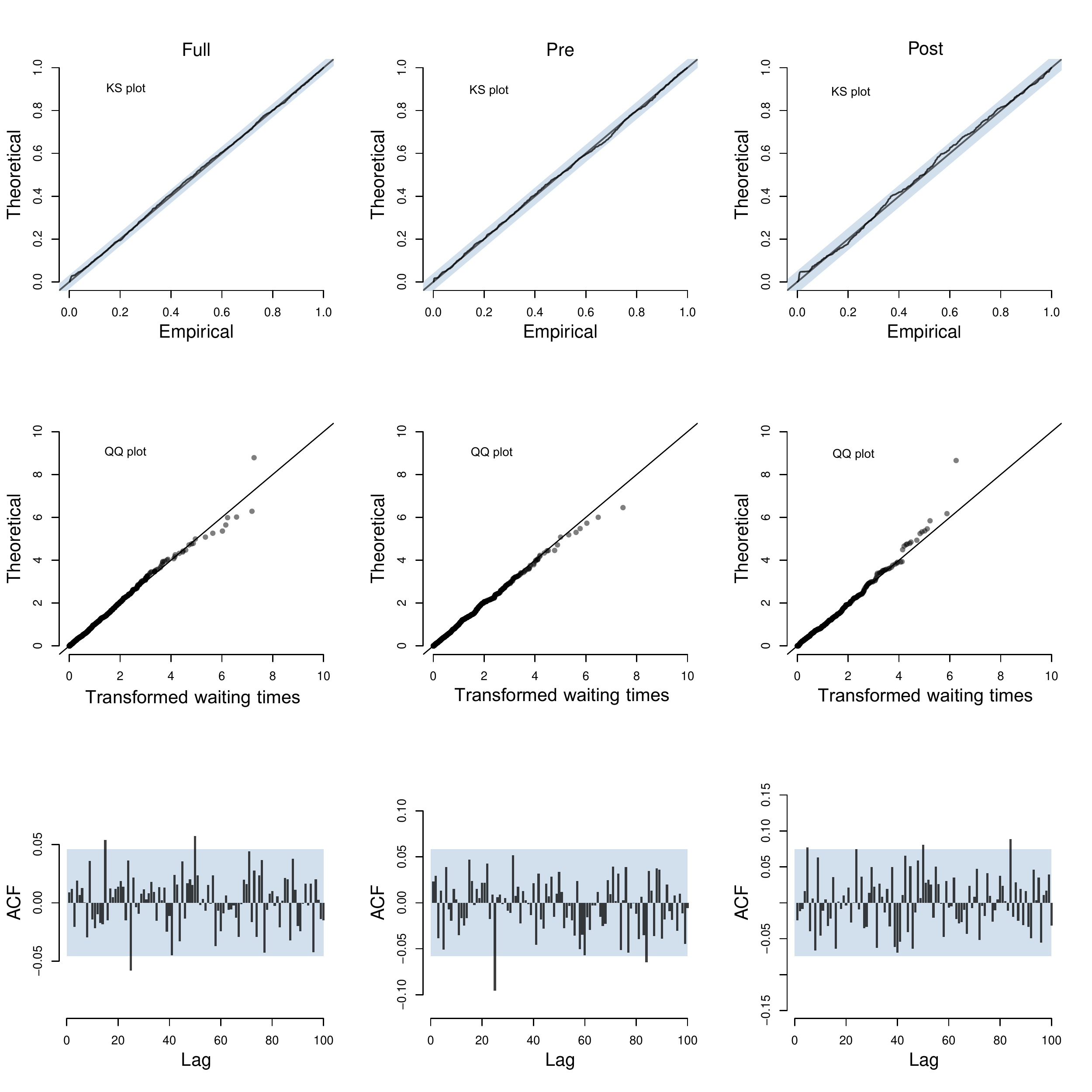}
\caption{Danish COVID-19 tweets data. The ``Full'', ``Pre'' and ``Post'' refer to the full sample and the samples pre- and post-announcement on March 11, 2021. The top row presents KS plots for these three time periods. The middle row presents QQ plots for these three time periods. The bottom row presents autocorrelations of the time transformed waiting times for these three periods.}
\label{ref2figCV2}
\end{figure}

As for the DJI data, we consider the Hawkes model with exponential kernel.
Based on the diagnostics (see Figure~\ref{ref2figCV2}), the model seems to be
well specified in all the three (sub)samples. However, we observe a large
difference between the estimates reported for the first subsample and for the
second subsample, see Table~\ref{ref2tabCV1}. In particular, the effect of the
response to the announcement is a substantial increase in the intensity. One
may also note that the estimated memory parameter $\beta$ for the full period
is between the estimates for the pre-announcement and post-announcement
periods. Table~\ref{ref2tabCV1} also reports asymptotic CIs and FIB CIs. As
can clearly be seen, the bootstrap CIs indicates the presence of
non-overlapping parameter estimates for the samples before and after the
announcement. This possibly reflects different types of dynamics in the two
samples, and indicates a structural break around the press briefing. We note
that this is not detectable by the standard misspecification tests for the
full sample, and is much less pronounced from the reported asymptotic CIs for
the three samples (in particular so for the baseline $\mu$).

\begin{table}[t]
\caption{\textsc{Estimated parameters, asymptotic and bootstrap $95\%$ confidence intervals for Twitter data.}}
\smallskip 

\label{ref2tabCV1}
\setlength{\tabcolsep}{0.7em}
\begin{tabular}{crc|rc|rc}
\multicolumn{7}{c}{Asymptotic confidence intervals}\\
\toprule 
& \multicolumn{2}{c|}{Full} & \multicolumn{2}{c|}{Pre} & \multicolumn{2}{c}{Post}\\
& $\hat{\theta}$ &CI& $\hat{\theta}$ &CI& $\hat{\theta}$&CI\\
\hline
$\mu$&12.19&[6.85; 21.70]&14.51&[7.64; 27.58]&47.09&[24.76; 89.56]\\ 
$a$ &0.88&[0.78; 0.94]& 0.82&[0.66; 0.92]&0.76&[0.57; 0.89]\\ 
$\beta$	&8.11&[6.11; 10.78]&5.53&[3.62; 8.43]&21.06&[10.89; 40.74]\\ 
\bottomrule
\end{tabular}
\medskip

\begin{tabular}{crc|rc|rc}
\multicolumn{7}{c}{Bootstrap confidence intervals}\\
\toprule 
& \multicolumn{2}{c|}{Full} & \multicolumn{2}{c|}{Pre} & \multicolumn{2}{c}{Post}\\
& $\hat{\theta}$& CI &$\hat{\theta}$ & CI & $\hat{\theta}$ & CI\\
\hline
$\mu$&12.21&[11.76; 17.02]&14.54&[13.58; 23.65]&47.07&[44.11; 89.24]\\ 
$a$&0.88&[0.84; 0.89]&0.82&[0.71; 0.84]&0.76&[0.55; 0.78]\\ 
$\beta$&8.12&[5.71; 9.31]&5.53&[2.64; 6.37]&21.07&[10.42; 41.11]\\ 
\bottomrule
\end{tabular}
\medskip

{\small Note: Non-parametric fixed intensity bootstrap is implemented; `Full', `Pre' and `Post' refer to the full sample and the samples pre- and post-announcement on March 11, 2021.}
\end{table}

In addition, we have also considered the power law kernel, where
$\gamma(t;\theta)$ in \eqref{emp: intens} is replaced by a power law, see
\eqref{eq: power law}. Interestingly, unreported results show that, in terms
of model misspecification, one is unable to discriminate between the two
models, and moreover that the estimates of the baseline $\mu$ and branching
ratio $a$ are virtually indistinguishable from those obtained using the
exponential kernel. Finally, estimation based on the power law kernel (unlike
the exponential kernel) is highly sensitive to initial values, which may
reflect the large correlation of the parameter estimators for power law kernels.

\section{Conclusions\label{sec: concl}}

In this paper we have discussed the theoretical foundations and practical
implementations of bootstrap inference for self-exciting point process models.
Applications of the bootstrap in order to improve upon the poor quality of
asymptotic approximations are scarce in the literature. Classic `recursive
intensity bootstrap' (RIB) schemes have been proposed in the recent
literature, although without proof of their first-order validity. RIB schemes
can also be quite involved to implement in practice, as they generally require
numerical integration for the recursive computation of the intensity for each
bootstrap repetition. To improve, we have introduced a new bootstrap scheme,
the `fixed intensity bootstrap' (FIB), where the conditional intensity is kept
fixed across bootstrap repetitions. By doing so, conditionally on the original
data the bootstrap data generating process follows a simple inhomogeneous
point process with known intensity; therefore, it is very simple to implement
and to use in practice. For both bootstrap schemes, we have provided a new
bootstrap (asymptotic) theory, which allows to assess bootstrap validity for
both bootstraps. Monte Carlo evidence supports the idea that the bootstrap is
a valid inference method when applied to point process models.

The results in the paper could be extended in several directions. On top of
the obvious extension to multivariate point process models, an interesting one
is how to deal with marked point process models. Marked (self-exciting)
processes are particularly useful in applications, as the intensity function
can be made dependent on a set of `marks' associated to past events (for
financial returns, the trading volumes; for energy prices, the magnitude of
price spikes; for tweets, the number of followers; for earthquakes modelling,
the magnitude of the earthquakes). In this context the proposed FIB seems very
powerful as re-sampling with a fixed intensity, even as a function of marks,
is feasible and easy to implement. As an example, consider briefly an
extension of the Hawkes model with exponential kernel in \eqref{eq: exp kernel}. One may include real-valued marks, or covariates, $y_{t}\in\mathbb{R}^{d}$
in the conditional intensity $\lambda(t;\theta)$ as for example,
\[
\lambda\left(  t;\theta\right)  =\mu\left(  y_{t}\right)  +\alpha\sum
_{t_{i}<t}\beta\left(  y_{t_{i}}\right)  \gamma\left(  t-t_{i}\right)  ,
\]
where $\mu,\beta:\mathbb{R}^{d}\mathbb{\rightarrow R}_{+}$, see e.g. \citet{CHH2015} 
for an
application to price spikes in electricity markets. Under the assumption of
`strongly exogenous' (or, ancillary) marks, similar to exogenous covariates in
discrete time Poisson autoregressions \citep[see][]{ACKR2016}
and with $\theta$ the parameters parameterizing the extended
Hawkes intensity, estimation and inference based on the FIB utilize the
original event times and marks, $\{t_{i},y_{t_{i}}\}_{i=1}^{N(T)}$.
Thus, in contrast to the RIB and other existing
recursive bootstraps, bootstrap inference based on FIB would not require
further assumptions (apart from stationarity) of the covariates.

A further extension is to develop model misspecification-robust bootstrap
methods. In particular, throughout the paper we have assumed that the model is
correctly specified. This assumption implies that the bootstrap can be
implemented parametrically by constructing bootstrap waiting times from an
i.i.d. sequence of mean one exponential random variables (the waiting times in
transformed time scale), as discussed in Sections~\ref{sec: FIB} and \ref{sec: RIB}. However,
misspecification of the model (in the simplest case, data are modelled as a
Poisson process, but the waiting times form a renewal process) may result in
i.i.d., but non-exponential (transformed) waiting times. Although in this case
the parametric bootstraps could fail, we believe that the non-parametric
bootstrap algorithms discussed in Section~\ref{sec NPAR} could serve as the
basis of novel misspecification-robust bootstrap methods. All these extensions
are left for future research.

\section*{Acknowledgements}

We are grateful to Torben Andersen (Co-Editor) and two anonymous referees for
many constructive comments and suggestions on an earlier version of the paper.
We have also benefited from discussions and feedback from seminar participants
at Saint-Petersburg State University (CEBA talks), Singapore Management
University, Macquarie University, as well as participants of the 9th Italian
Congress of Econometrics and Empirical Economics (University of Cagliari) and
the 2021 Virtual Workshop on Financial Econometrics (Durham University).

This research was supported by the Danish Council for Independent Research
(DSF Grant 015-00028B), the Center for Information and Bubble Studies,
University of Copenhagen, the Italian Ministry of University and Research
(PRIN 2017 Grant 2017TA7TYC) and the University of Sydney (Faculty Research
Future Fix 2020 Grant). Part of this paper was written while Giuseppe
Cavaliere was visiting the School of Economics of the University of Sydney;
financial support and hospitality are gratefully acknowledged. Finally, the
authors acknowledge the technical assistance provided by the Sydney
Informatics Hub of the University of Sydney for the high-performance computing
and cloud services.

\medskip

\appendix

\section*{Appendix\label{Sec Appendix}}

The appendix contains proofs and results for the bootstrap theory for the FIB
and RIB schemes. It is structured as follows.
Section~\ref{sec: Auxiallary results} contains a general bootstrap theory to
establish asymptotic properties of the bootstrap estimators, as well as
central limit theorem (CLT) for inhomogeneous Poisson processes.
Section~\ref{sec: FIB proofs} contains the proofs of Lemma
\ref{Lemma BS score and Hessian} and Theorem \ref{Th asy validity of the FIB}
for the FIB validity. Similarly, Section~\ref{sec RIB proofs} contains the
proofs of Lemma \ref{Lemma RIB BS score and Hessian} and
Theorem~\ref{Th asy validity of the RIB} for the RIB. Some auxiliary lemmas
for derivatives of the (bootstrap) likelihood are given in
Section~\ref{sec suppl aux lemmata}. Finally,
Section~\ref{sec suppl aux proofs} contains the proof of the two lemmas in
Section~\ref{sec: Auxiallary results}.

\setcounter{theorem}{0}\setcounter{equation}{0}\setcounter{lemma}{0}\renewcommand{\thelemma}{\thesection.\arabic{lemma}}\renewcommand{\thetheorem}{\thesection.\arabic{theorem}}\numberwithin{theorem}{section}
\numberwithin{lemma}{section} \numberwithin{remark}{section}

\section{Auxiliary results}

\label{sec: Auxiallary results}

\subsection{General asymptotic theory for bootstrap estimators}

\label{sec BS CAN} Before formulating the assumptions for Lemma~\ref{lem BS2},
we need to properly define a neighborhood $N(\theta)$ of $\theta$ where
$\theta\in\operatorname*{int}\Theta$. Without loss of generality, let
$\theta=(\theta_{1},...,\theta_{d})^{\prime}\in\Theta\subseteq\mathbb{R}^{d}$,
and assume that $\Theta$ is a product of intervals $I_{i}$, $i=1,...,d$, which
can be (sub-intervals of) $\mathbb{R},\mathbb{R}_{+}$ or $\mathbb{R}^{+}$.
That is, $\Theta=I_{1}\times...\times I_{d}$, and define $N(\theta)$ as
\begin{equation}
N(\theta)=[\theta_{1L},\ \theta_{1U}]\times...\times\lbrack\theta
_{mL},\ \theta_{mU}], \label{def N}%
\end{equation}
where $\theta_{iL}<\theta_{i}<\theta_{iU}$ for $i=1,2,...,d$. We make the
following assumption. \bigskip

\noindent Assumption A.1 Consider a bootstrap log-likelihood, or criterion
function $\ell_{T}^{\ast}(\theta)$, which is a function of the bootstrap
sample and the parameter $\theta\in\Theta\subseteq\mathbb{R}^{d}$. Assume that
$\ell_{T}^{\ast}(\theta)$ is thrice continuously differentiable in $\theta$,
and moreover that for the bootstrap true value $\theta_{T}^{\ast}$ it holds that:

\noindent(i) $\theta_{T}^{\ast}\rightarrow_{p}\theta_{\dagger}$, where
$\theta_{\dagger}\in\operatorname*{int}\Theta$;

\noindent(ii) $T^{-1/2}\partial\ell_{T}^{\ast}(\theta_{T}^{\ast}%
)/\partial\theta\overset{d^{\ast}}{\rightarrow}_{p}\mathcal{N(}0,\Omega
_{S}),\ \Omega_{S}>0$;

\noindent(iii) $-T^{-1}\partial^{2}\ell_{T}^{\ast}(\theta_{T}^{\ast}%
)/\partial\theta\partial\theta^{\prime}\overset{p^{\ast}}{\rightarrow}%
_{p}\Omega_{I}>0$;

\noindent(iv) with $\mathcal{N(}\theta_{\dagger})$ a neighborhood of
$\theta_{\dagger}$, see \eqref{def N},
\[
\max_{h,i,j=1,...,d}\sup_{\theta\in\mathcal{N(}\theta_{\dagger})}\left\vert
\frac{1}{T}\frac{\partial^{3}\ell_{T}^{\ast}(\theta)}{\partial\theta
_{h}\partial\theta_{i}\partial\theta_{j}}\right\vert \leq c_{T}^{\ast},
\]
where $c_{T}^{\ast}\overset{p^{\ast}}{\rightarrow}_{p}c$, $0<c<\infty$.

\begin{lemma}
\label{lem BS2} Assume that Assumption~A.1 holds. Then in a fixed open
neighborhood $U(\theta_{\dagger})$ of $\theta_{\dagger}$ the following holds
as $T\rightarrow\infty$:

\noindent(i) The probability, conditionally on the original data, that there
exists a unique maximum point $\hat{\theta}_{T}^{\ast}$ of $\partial\ell
_{T}^{\ast}(\theta)$ which solves the estimating equation $\partial\ell
_{T}(\hat{\theta}_{T}^{\ast})/\partial\theta=0$, converges in probability to one;

\noindent(ii) $\hat{\theta}_{T}^{\ast}-\theta_{T}^{\ast}\overset{p^{\ast}%
}{\rightarrow}_{p}0$;

\noindent(iii) $T^{1/2}(\hat{\theta}_{T}^{\ast}-\theta_{T}^{\ast}%
)\overset{d^{\ast}}{\rightarrow}_{p}\mathcal{N(}0,\Omega_{I}^{-1}\Omega
_{S}\Omega_{I}^{-1})$.
\end{lemma}

The proof of Lemma~\ref{lem BS2} is given in
Section~\ref{sec suppl aux proofs}.

\begin{remark}
$\overset{}{\text{ }}$

\medskip\noindent(i) Note that for the restricted bootstrap for testing the
simple null hypothesis $\theta=\bar{\theta}$, Assumption~A.1(i) is trivially
satisfied with $\theta_{\dagger}=\bar{\theta}$. In the general case,
$\theta_{T}^{\ast}=\tilde{\theta}_{T}$, where $\tilde{\theta}_{T}$ is an
estimator, restricted by the null hypothesis, obtained on the original data;
then, Assumption~A.1(i) is implied by establishing $\tilde{\theta}%
_{T}\rightarrow_{p}\theta_{\dagger}$, where under the null $\theta_{\dagger
}=\theta_{0}$ while under the alternative, $\theta_{\dagger}$ is a pseudo-true
value. For the unrestricted bootstrap, $\theta_{T}^{\ast}=\hat{\theta}_{T}$,
and Assumption~A.1(i) is implied by establishing the classic consistency
result, $\hat{\theta}_{T}\rightarrow_{p}\theta_{0}$.

\medskip\noindent(ii) As for the condition (iv) in Assumption~A.1 on the third
derivative of $\ell_{T}^{*}(\theta)$, this may be replaced by a uniform
requirement for the second order derivative of $\ell_{T}^{\ast}(\theta)$,
$\partial^{2}\ell_{T}^{\ast}(\theta)/\partial\theta\partial\theta^{\prime}$.
Specifically, Lemma~\ref{lem BS2} holds with condition (iv) in Assumption~A.1
replaced by the following condition:

\medskip\noindent($\text{iv}^{\ast}$) Assume that there exists a continuous
function, $f:\mathbb{R}^{d}\rightarrow\mathbb{R}^{d\times d}$ such that,
\[
\Vert T^{-1}\partial^{2}\ell_{T}^{\ast}(\theta)/\partial\theta\partial
\theta^{\prime}-f(\theta)\Vert\overset{p^{\ast}}{\rightarrow}_{p}0
\]
uniformly over $\theta\in\mathcal{N(}\theta_{\dagger})$;

\medskip\noindent see \citet[proof of Lemma A.1]{LRJ2011} for its
non-bootstrap equivalent. \hfill$\square$
\end{remark}

\subsection{Central limit theory for bootstrap point processes}

\label{sec BS LLN and CLT} The following lemma is a bootstrap extension of
Lemma 2 in \citet{O1978}. In Section~\ref{sec BS validity}, we consider a
bootstrap point process $N^{\ast}(t)$ whose conditional intensity, say
$\lambda(t)$, depends only on the original data.\footnote{Notice that the
distribution of $N^{\ast}$ (as well as the conditional intensity process
$\lambda(t)$) depends on $T$, the sample span of the data. Hence, we formally
have a triangular array of the form $\{N_{T}(t)$, $0\leq t\leq T$, $T\geq0\}$;
this is not essential and we hence suppress the triangular array notation.}
Consider an integral of the form
\begin{equation}
Y_{T}^{\ast}:=\int_{0}^{T}\xi_{T}(u)dM^{\ast}(u) \label{eq: def Ystar}%
\end{equation}
where $\xi(u)$ is a function of the original data. We have the following CLT.

\begin{lemma}
\label{Lemma inhomogeneous PP CLT} For all $T\geq0$, let $N^{\ast}(t)$ be a
bootstrap inhomogeneous Poisson process with conditional intensity
$\lambda_{T}(t)$ and let $\xi_{T}(t)$ be an $d$-dimensional stochastic
process, where both $\lambda_{T}(t)$ and $\xi_{T}(t)$ depend only on the
original data. Consider $Y_{T}^{\ast}$ defined in \eqref{eq: def Ystar} with
$M^{\ast}(t):=N^{\ast}(t)-\Lambda_{T}(t)$, where $\Lambda_{T}(t)=\int_{0}%
^{t}\lambda_{T}(u)du$. Then, with $h_{T}(t):=\xi_{T}(t)\xi_{T}(t)^{\prime
}\lambda_{T}(t)$, assume that,
\begin{align}
\frac{1}{T}\int_{0}^{T}h_{T}(t)dt  &  \rightarrow_{p}V<\infty
,\label{eq lemma inhomo PP cond var}\\
\frac{1}{T}\int_{0}^{T}\Vert h_{T}(t)\Vert^{1+\eta}dt  &  =O_{p}(1),
\label{eq lemma inhomo PP delta plus moments}%
\end{align}
for some $\eta>0$. Then if either (i) $\lambda_{T}\left(  t\right)
\geq\lambda_{L}>0$, or (ii) $\left\Vert \xi_{T}\left(  t\right)  \right\Vert
\leq c_{\xi}<\infty$, it holds that
\[
T^{-1/2}Y_{T}^{\ast}\overset{d^{\ast}}{\rightarrow}_{p}\mathcal{N}(0,V).
\]

\end{lemma}

The proof of Lemma~\ref{Lemma inhomogeneous PP CLT} is given in
Section~\ref{sec suppl aux proofs}.

\section{Proofs for the fixed-intensity bootstrap}

\label{sec: FIB proofs}

\subsection{Proof of Lemma~\ref{Lemma BS score and Hessian}}

\noindent We first consider the score at the true value, $S_{T}^{\ast}%
(\theta_{T}^{\ast})=\int_{0}^{T}\hat{\xi}(t)dM^{\ast}(t)$, see
\eqref{eq score at the true value FIB}. Conditionally on the data, $M^{\ast
}(t)$ is a $\mathcal{F}_{t}^{\ast}$-martingale and $\hat{\xi}(t)$ is a
predictable process. Therefore, $\hat{Y}(t):=\int_{0}^{t}\hat{\xi}(s)dM^{\ast
}(s)$, as a martingale transformation, is also a $\mathcal{F}_{t}^{\ast}%
$-martingale (under bootstrap probability) starting from $\hat{Y}(0)=0$. We
apply Lemma~\ref{Lemma inhomogeneous PP CLT} to show that $S_{T}^{\ast}%
(\theta_{T}^{\ast})=\hat{Y}(T)$ satisfies the CLT.

We first verify condition~\ref{eq lemma inhomo PP cond var}, with $h(t)$
replaced by $\hat{h}(t)=\hat{\xi}(t)\hat{\xi}(t)^{\prime}\hat{\lambda}(t)$ and
$V=I(\theta_{0})$. To do so, write $\int_{0}^{T}\hat{h}(t)dt$ as
\[
\int_{0}^{T}\hat{h}(t)dt=V_{1,T}+V_{2,T},
\]
where $V_{1,T}=T^{-1}\int_{0}^{T}h(t)dt$ and $V_{2,T}=T^{-1}\int_{0}^{T}%
(\hat{h}(t)-h(t))dt$. Under stationarity (Assumption 1(b)), predictability
(Assumption 1(c)) and finite variance of $h(t)$ (Assumption 2(b)), by Lemma 2
(eq. (3.3)) in \citet{O1978},
\begin{equation}
V_{1,T}\rightarrow_{p}E(h(t))=I(\theta_{0}). \label{eq limit for V1,T 2}%
\end{equation}
To show that $V_{2,T}\rightarrow_{p}0$, since $h(t,\theta)$ is continuously
differentiable in a neighborhood of $\theta_{0}$ (as implied by Assumption
2(a)) and $\theta_{T}^{\ast}-\theta_{0}=o_{p}(1)$, by the delta method, with
$A_{ij}$ denoting the $(i,j)$-th entry of a generic matrix $A$,
\[
T^{-1}\int_{0}^{T}(\hat{h}_{ij}(t)-h_{ij}(t))dt=\left(  T^{-1}\int_{0}%
^{T}\partial_{\theta_{0}}h_{ij}(t)dt\right)  ^{\prime}(\theta_{T}^{\ast
}-\theta_{0})(1+o_{p}(1))
\]
The proof of Lemma 2 in \citet{O1978} implies that the first term on
the right-hand side of the previous equation is of $O_{p}(1)$ provided
$E(\partial_{\theta_{0}}h_{ij}(t))$ is finite, which is implied by Assumption 2(a),(b).

We next verify \eqref{eq lemma inhomo PP delta plus moments}. As $h(t;\theta)$
is continuously differentiable around $\theta_{0}$ and $\eta>0$, by a Taylor
expansion around $\theta_{0}$ and using the fact that $\theta_{T}^{\ast
}\rightarrow_{p}\theta_{0}$, it trivially holds that $T^{-1}\int_{0}^{T}%
\Vert\hat{h}(t)\Vert^{1+\eta}dt=O_{p}(1)$, provided $E(|\partial_{\theta_{0}%
}h_{ij}(t)|^{1+\eta})<\infty$, which holds if, additionally, $E((\partial
_{\theta_{i}}\lambda(t;\theta))^{3})<\infty$; see
Lemma~\ref{Lemma result on derivative of h given third derivative} in
Section~\ref{sec: Lem derivative h}. Hence, $T^{-1/2}S_{T}^{\ast}(\theta
_{T}^{\ast})\overset{d^{\ast}}{\rightarrow}_{p}\mathcal{N(}0,I(\theta_{0}))$
by Lemma~ \ref{Lemma inhomogeneous PP CLT}.

Next consider the Hessian at the true value, $H_{T}^{\ast}(\theta_{T}^{\ast
})=\int_{0}^{T}\hat{\zeta}(t)dM^{\ast}(t)-\int_{0}^{T}\hat{h}(t)dt$, see
\eqref{eq Hessian at the true value FIB}. Taking expectation conditionally on
the data, it follows that
\[
E^{\ast}(-T^{-1}H_{T}^{\ast}(\theta_{T}^{\ast}))=T^{-1}\int_{0}^{T}\hat
{h}(t)dt=V^{\ast}(T^{-1/2}S_{T}^{\ast}(\theta_{T}^{\ast}))\rightarrow
_{p}I(\theta_{0}),
\]
as shown above. The convergence in \eqref{eq asymptotics for FIB Hessian} then
follows provided $T^{-1}H_{T}^{\ast}(\theta_{T}^{\ast})-E^{\ast}(T^{-1}%
H_{T}^{\ast}(\theta_{T}^{\ast}))=o_{p}^{\ast}(1)$, in probability. To see
this, notice that
\begin{align*}
T^{-1}H_{T}^{\ast}(\theta_{T}^{\ast})-E^{\ast}(T^{-1}H_{T}^{\ast}(\theta
_{T}^{\ast}))  &  =T^{-1}\int_{0}^{T}\hat{\zeta}(t)dM^{\ast}(t)\\
&  =T^{-1}\int_{0}^{T}\zeta(t)dM^{\ast}(t)+o_{p}(1).
\end{align*}
The last equality is again by delta method, noticing that $\zeta(t,\theta)$ is
continuously differentiable in a neighborhood of $\theta_{0}$ (as implied by
Assumption 2(a)) and $\theta_{T}^{\ast}-\theta_{0}=o_{p}(1)$. Specifically,
\begin{align*}
&  T^{-1}\int_{0}^{T}\hat{\zeta}_{ij}(t)dM^{\ast}(t)-T^{-1}\int_{0}^{T}%
\zeta_{ij}(t)dM^{\ast}(t)\\
&  \hspace{1cm}=\left(  T^{-1}\int_{0}^{T}\partial_{\theta_{0}}\zeta
_{ij}(t)dM^{\ast}(t)\right)  ^{\prime}((\theta_{T}^{\ast}-\theta_{0}%
)(1+o_{p}(1))=o_{p}(1)
\end{align*}
by Lemma 2 in \citet{O1978} under the condition that $E(|\partial_{\theta_{0}%
}\zeta_{ij}(t)|\lambda(t))<\infty$ as implied by Assumption 2(c). The term
$T^{-1}\int\zeta_{ij}(t)dM^{\ast}(t)$ has variance
\begin{align}
&  V^{\ast}\left(  T^{-1}\int_{0}^{T}\zeta_{ij}(t)dM^{\ast}(t)\right)
=T^{-2}\int_{0}^{T}\zeta_{ij}^{2}(t)\hat{\lambda}(t)dt \label{eq the variance}%
\\
&  \hspace{1cm}=T^{-2}\int_{0}^{T}\zeta_{ij}^{2}(t)\lambda(t)dt+o_{p}%
(T^{-1})\nonumber
\end{align}
using continuous differentiability of $\lambda(t;\theta)$ around $\theta_{0}$.
By the ergodic theorem, using $E(\zeta_{ij}^{2}(t)\lambda(t))<\infty$, we have
that $T^{-1}\int_{0}^{T}\zeta_{ij}^{2}(t)\lambda(t)dt$ converges in
probability, and hence the variance in \eqref{eq the variance} is of
$o_{p}(1)$. Taken together, these results imply $-T^{-1}H_{T}^{\ast}%
(\theta_{T}^{\ast})\rightarrow_{p}I(\theta_{0})$.

\subsection{Proof of Theorem \ref{Th asy validity of the FIB}}

The theorem follows by a straightforward application of Lemma~\ref{lem BS2}.
Specifically, Assumption A.0 is satisfied with $\theta^{\dagger}=\theta_{0}$
as by Assumption, $\theta_{T}^{\ast}\rightarrow_{p}\theta_{0}$. Assumptions
A.1 and A.2 follow from Lemma~\ref{Lemma BS score and Hessian} with
$\Omega_{I}=\Omega_{S}=I(\theta_{0})$. Finally, Assumption A.3 follows from
Assumption 2(c), which holds conditionally on the original data; this is shown
in the Supplement, Lemma~\ref{lemma 3rd derivativ par FIB}.

\section{Proofs for the recursive intensity bootstrap}

\label{sec RIB proofs}

\subsection{Proof of Lemma~\ref{Lemma RIB BS score and Hessian}}

Recall that, conditionally on the original sample, only $N^{\ast}(t)$ and the
associated event times $t_{1}^{\ast},...,t_{n_{T}^{\ast}}^{\ast}$ are random.
Moreover, conditionally on the original sample, $N^{\ast}(t)$ has conditional
intensity given by $\lambda^{\ast}(t;\theta_{T}^{\ast})$, which in contrast to
the FIB, is now a stochastic process even upon conditioning on the original
data. As a consequence, at the bootstrap true value $\theta_{T}^{\ast}$, and
with $\mathcal{F}_{t}^{\ast}$ denoting the filtration associated to
$\{N^{\ast}(s),s\leq t\}$, it holds that $E^{\ast}(dN^{\ast}(t)|\mathcal{F}%
_{t-}^{\ast})=\lambda^{\ast}(t;\theta_{T}^{\ast})dt$. Notice that $N^{\ast}$
as well as the intensity $\lambda^{\ast}$ depend on the original data through
$\theta_{T}^{\ast}$ only. Notice, finally, that $P(\theta_{T}^{\ast}\in
\Theta_{0})$ can be made arbitrarily close to one by picking $T$ large enough
(such that the bootstrap process can be made stationary upon proper choice of
the distribution of the initial values).

Consider first the score evaluated at $\theta_{T}^{\ast}$, see
\eqref{eq RIB score at true value}, which we write as
\[
S_{T}^{\ast}(\theta_{T}^{\ast})=\int_{0}^{T}\hat{\xi}^{\ast}(t)dM^{\ast
}(t),\quad\hat{\xi}^{\ast}(t):=\partial_{\theta}\log\lambda^{\ast}%
(t;\theta_{T}^{\ast}).
\]
By construction $S_{T}^{\ast}(\theta_{T}^{\ast})$ is a martingale difference
array. Assuming without loss of generality that $T$ is an integer we can
write
\[
T^{-1/2}S_{T}^{\ast}(\theta_{T}^{\ast})=T^{-1/2}\sum_{k=1}^{T}\varepsilon
_{T,k}^{\ast},\quad\varepsilon_{T,k}^{\ast}:=\int_{k-1}^{k}\hat{\xi}^{\ast
}(t)dM^{\ast}(t),
\]
where, conditionally on $\mathcal{F}_{k-}^{\ast}$, $E^{\ast}(\varepsilon
_{T,k}^{\ast}|\mathcal{F}_{k-}^{\ast})=0$. We show that $T^{-1/2}S_{T}^{\ast
}(\theta_{T}^{\ast})$ satisfies the conditions for the CLT for martingale
triangular arrays, see e.g. Theorem~3.4.10 in \citet{D2019}. The unconditional
variance is given by
\begin{align*}
\frac{1}{T}\sum_{k=1}^{T}E^{\ast}((\varepsilon_{T,k}^{\ast})^{2})  &
=\frac{1}{T}\sum_{k=1}^{T}E^{\ast}\left(  \left(  \int_{k-1}^{k}\hat{\xi
}^{\ast}(t)dM^{\ast}(t)\right)  ^{2}\right) \\
&  =\frac{1}{T}\sum_{k=1}^{T}E^{\ast}\left(  \int_{k-1}^{k}\hat{\xi}^{\ast
}(t)\hat{\xi}^{\ast}(t)^{\prime}\lambda^{\ast}(t;\theta_{T}^{\ast})dt\right)
\\
&  =E^{\ast}\left(  \int_{0}^{1}\hat{\xi}^{\ast}(t)\hat{\xi}^{\ast}%
(t)^{\prime}\lambda^{\ast}(t;\theta_{T}^{\ast})dt\right) \\
&  =E^{\ast}\left(  \int_{0}^{1}\hat{h}^{\ast}(t)dt\right)  =I(\theta
_{T}^{\ast}),
\end{align*}
where the last two equalities hold as, with probability tending to one,
$\theta_{T}^{\ast}\in\Theta_{0}$. Moreover, by condition
\eqref{eq extra condition RIB validity} and for $T$ large enough,
$I(\theta_{T}^{\ast})=E(h(t,\theta_{T}^{\ast}))\overset{p}{\rightarrow
}E(h(t,\theta_{0}))=:I(\theta_{0})$. To prove that the Lindeberg condition
holds in probability, notice that
\[
\frac{1}{T}\sum_{k=1}^{T}E^{\ast}\left(  \Vert\varepsilon_{T,k}^{\ast}%
\Vert^{2}\mathbb{I}(\Vert\varepsilon_{T,k}^{\ast}\Vert>\delta T^{1/2})\right)
=E^{\ast}\left(  \Vert\varepsilon_{T,k}^{\ast}\Vert^{2}\mathbb{I}%
(\Vert\varepsilon_{T,k}^{\ast}\Vert>\delta T^{1/2})\right)  \rightarrow_{p}0
\]
as, for all $\lambda\in\mathbb{R}^{d}$,
\begin{align*}
E^{\ast}\left(  (\lambda^{\prime}\varepsilon_{T,k}^{\ast})^{2}\mathbb{I}%
(\Vert\varepsilon_{T,k}^{\ast}\Vert>\delta T^{1/2})\right)   &  \leq E^{\ast
}((\lambda^{\prime}\varepsilon_{T,k}^{\ast})^{2})=\lambda^{\prime}E^{\ast
}(\varepsilon_{T,k}^{\ast}\varepsilon_{T,k}^{\ast\prime})\lambda\\
&  \leq cE\sup_{\theta\in\Theta_{0}}\Vert h(t;\theta)\Vert<\infty,
\end{align*}
again by \eqref{eq extra condition RIB validity}. Hence, $T^{-1/2}S_{T}^{\ast
}(\theta_{T}^{\ast})\overset{d^{\ast}}{\rightarrow}\mathcal{N(}0,I(\theta
_{0}))$, in probability.

Consider now the Hessian at the true value, given in
eq.~\eqref{eq RIB hessian at true value}. It holds that for $T$ large enough
\[
|T^{-1}H_{T}^{\ast}(\theta_{T}^{\ast})-T^{-1}H_{T}^{\ast}(\theta_{0})|\leq
\sup_{\theta\in\mathcal{N(}\theta_{0})}|T^{-1}H_{T}^{\ast}(\theta)-T^{-1}%
H_{T}^{\ast}(\theta_{0})|=o_{p}^{\ast}(1),
\]
in probability, where the last equality is found as in the proof of Theorem 3
in \citet{O1978}. As $T^{-1}H_{T}^{\ast}(\theta_{0})\overset{p^{\ast}%
}{\rightarrow}I(\theta_{0})$, the proof is completed.

\subsection{Proof of Theorem~\ref{Th asy validity of the RIB}}

As for Theorem~\ref{Th asy validity of the FIB}, we apply Lemma~\ref{lem BS2}
where Assumption A.0 is satisfied with $\theta^{\dag}=\theta_{0}$ and
Assumptions A.1 and A.2 follow from Lemma \ref{Lemma RIB BS score and Hessian}
with $\Omega_{I}=\Omega_{S}=I(\theta_{0})$. Finally, Assumption A.3 is
verified in Lemma~\ref{lemma 3rd derivativ par RIB} in
Section~\ref{sec 3rd der RIB}.

\section{Auxiliary lemmas}

\label{sec suppl aux lemmata}

\subsection{Derivative of $h(\cdot)$}

\label{sec: Lem derivative h}

\begin{lemma}
\label{Lemma result on derivative of h given third derivative} Consider
$h(\cdot)$ defined in Assumption 2(b). Under Assumptions 1 and 2, a sufficient
condition for $E(|\partial_{\theta_{0}}h_{ij}(t)|^{1+\eta})<\infty$ is that
$E((\partial_{\theta_{i}}\lambda(t;\theta))^{3})<\infty$.
\end{lemma}

\noindent Proof. Without loss of generality we consider the scalar case,
$\theta\in\mathbb{R}$. To simplify notation, $h(\theta)$ and $\lambda(\theta)$
are denoted by $h$ and $\lambda$, respectively. Consider
\[
\partial_{\theta}h^{1+\eta}=(1+\eta)h^{\eta}\partial_{\theta}h.
\]
As
\[
|\partial_{\theta}h|=\left\vert \frac{\left(  \partial_{\theta}\lambda\right)
\left(  \partial_{\theta}^{2}\lambda\right)  }{\lambda}-\frac{\left(
\partial_{\theta}\lambda\right)  ^{3}}{\lambda^{2}}\right\vert \leq\left\vert
\frac{\left(  \partial_{\theta}\lambda\right)  }{\lambda^{1/2}}\frac
{1}{\lambda^{1/2}}\left(  \partial_{\theta}^{2}\lambda\right)  \right\vert
+\left\vert \frac{\left(  \partial_{\theta}\lambda\right)  ^{3}}{\lambda^{2}%
}\right\vert ,
\]
we have the inequality
\[
|h^{\eta}\partial_{\theta}h|\leq\left\vert h^{\eta}\frac{\left(
\partial_{\theta}\lambda\right)  }{\lambda^{1/2}}\frac{1}{\lambda_{L}^{1/2}%
}\left(  \partial_{\theta}^{2}\lambda\right)  \right\vert +\left\vert h^{\eta
}\frac{\left(  \partial_{\theta}\lambda\right)  ^{3}}{\lambda^{2}}\right\vert
:=a_{1}+a_{2},
\]
with $a_{1},a_{2}$ implicitly defined. By the Cauchy-Schwartz inequality
\begin{align*}
(E(a_{1}))^{2}  &  \leq E\left(  \left(  h^{\eta}\left(  \frac{1}%
{\lambda^{1/2}}\partial_{\theta}\lambda\right)  \right)  ^{2}\right)  E\left(
\left(  \partial_{\theta}^{2}\lambda\right)  ^{2}\right) \\
&  =E\left(  h^{2\eta}\frac{1}{\lambda}\left(  \partial_{\theta}%
\lambda\right)  ^{2}\right)  E(\left(  \partial_{\theta}^{2}\lambda\right)
^{2})=E(h^{1+2\eta})E(\left(  \partial_{\theta}^{2}\lambda\right)
^{2})<\infty
\end{align*}
as, by choosing $\eta\leq\frac{1}{2}$, $E(h^{1+2\eta})<\infty$ (by
Assumption~2(b), $h$ has finite variance), and $E\left(  (\partial_{\theta
}^{2}\lambda)^{2}\right)  <\infty$ (by Assumption~2(a)). Consider now $a_{2}$.
Since
\[
h^{\eta}\frac{\left(  \partial_{\theta}\lambda\right)  ^{3}}{\lambda^{2}%
}=h^{\eta}\frac{\left(  \partial_{\theta}\lambda\right)  ^{3/2}}{\lambda
^{3/4}}\frac{\left(  \partial_{\theta}\lambda\right)  ^{3/2}}{\lambda^{5/4}%
}=h^{\frac{3}{4}+\eta}\frac{\left(  \partial_{\theta}\lambda\right)  ^{3/2}%
}{\lambda^{5/4}},
\]
we can consider the inequality
\[
E(a_{2})\leq\frac{1}{\lambda_{L}^{5/4}}E\left\vert h^{\frac{3}{4}+\eta}\left(
\partial_{\theta}\lambda\right)  ^{3/2}\right\vert
\]
which, by the Cauchy-Schwartz inequality, is bounded provided $E\left(
(h^{\frac{3}{4}+\eta})^{2}\right)  =E(h^{\frac{3}{2}+2\eta})<\infty$ and
$E\left(  (\partial_{\theta}\lambda)^{3}\right)  <\infty$. The former
inequality holds by the finite variance assumption on $h$ by choosing
$\eta<1/4$, while the latter holds by assumption.

\subsection{Third derivative of the FIB likelihood}

\label{sec 3rd der FIB}

\begin{lemma}
\label{lemma 3rd derivativ par FIB} Under the conditions of
Theorem~\ref{Th asy validity of the FIB} for the FIB,
\[
\sup_{\theta\in N_{\epsilon}\left(  \theta_{\dag}\right)  }|\partial
_{\theta_{i},\theta_{j},\theta_{k}}^{3}\tfrac{1}{T}l_{T}^{\ast}\left(
t;\theta\right)  |\leq c_{T}^{\ast}\overset{p^{\ast}}{\rightarrow}_{p}c.
\]

\end{lemma}

\noindent Proof. Without loss of generality, to simplify notation we assume
that $\theta$ is scalar. It follows that
\begin{align*}
\sup_{\theta\in\mathcal{N(}\theta_{0})}\left\vert \frac{1}{T}\partial_{\theta
}^{3}l_{T}^{\ast}\left(  \theta\right)  \right\vert  &  =\sup_{\theta
\in\mathcal{N(}\theta_{0})}\left\vert \frac{1}{T}\int_{0}^{T}\partial_{\theta
}^{3}\log\lambda(t,\theta)dN^{\ast}(t)-\frac{1}{T}\int_{0}^{T}\partial
_{\theta}^{3}\lambda(t,\theta)dt\right\vert \\
&  \leq\sup_{\theta\in\mathcal{N(}\theta_{0})}\left\vert \frac{1}{T}\int
_{0}^{T}\partial_{\theta}^{3}\log\lambda(t,\theta)dN^{\ast}(t)\right\vert \\
&  +\sup_{\theta\in\mathcal{N(}\theta_{0})}\left\vert \frac{1}{T}\int_{0}%
^{T}\partial_{\theta}^{3}\lambda(t,\theta)dt\right\vert =:A_{T}^{\ast}+B_{T}%
\end{align*}
where
\begin{align*}
B_{T}:=\sup_{\theta\in\mathcal{N(}\theta_{0})}\left\vert \frac{1}{T}\int
_{0}^{T}\partial_{\theta}^{3}\lambda(t,\theta)dt\right\vert  &  \leq\frac
{1}{T}\int_{0}^{T}c(t)dt\rightarrow_{p}E(c(t))\\
A_{T}^{\ast}:=\sup_{\theta\in\mathcal{N(}\theta_{0})}\left\vert \frac{1}%
{T}\int_{0}^{T}\partial_{\theta}^{3}\log\lambda(t,\theta)dN^{\ast
}(t)\right\vert  &  \leq\frac{1}{T}\int_{0}^{T}d(t)dN^{\ast}(t)\overset
{p^{\ast}}{\rightarrow}_{p}E(d(t)\lambda(t)).
\end{align*}
The first convergence follows from the LLN for stationary and ergodic
processes. To see the second convergence, consider the following
decomposition:
\begin{equation}
\frac{1}{T}\int_{0}^{T}d(t)dN^{\ast}(t)=\frac{1}{T}\int_{0}^{T}d(t)dM^{\ast
}(t)+\frac{1}{T}\int_{0}^{T}d(t)\hat{\lambda}(t)dt \label{eq app deco 1}%
\end{equation}
where $T^{-1}\int_{0}^{T}d(t)dM^{\ast}(t)=O_{p}^{\ast}(T^{-1/2})$, in
probability. By a Taylor expansion,
\[
\frac{1}{T}\int_{0}^{T}d(t)\hat{\lambda}(t)dt=\frac{1}{T}\int_{0}%
^{T}d(t)\lambda(t)dt+\frac{1}{T}\int_{0}^{T}d(t)\partial_{\theta}%
\lambda(t)dt(\theta_{T}^{\ast}-\theta_{0})(1+o_{p}(1))
\]
where
\[
\frac{1}{T}\int_{0}^{T}d(t)\lambda(t)dt\rightarrow_{p}E(d(t)\lambda(t))<\infty
\]
by Assumption 2(c). To see that the last term on the right hand side is
negligible, it suffices no show that
\[
E|d(t)\partial_{\theta}\lambda(t)|<\infty,
\]
which holds as
\[
E|d(t)\partial_{\theta}\lambda(t)|=E\left\vert \left(  \lambda(t)d(t)\right)
\left(  \frac{\partial_{\theta}\lambda(t)}{\lambda(t)}\right)  \right\vert
<\infty
\]
by the Cauchy-Schwarz inequality, as $E\left(  (\lambda(t)d(t))^{2}\right)
<\infty$ and $E((\frac{\partial_{\theta}\lambda(t)}{\lambda(t)})^{2}%
)<(\lambda_{L})^{-1}E(\partial_{\theta}\lambda(t))^{2}<\infty$. For the first
term in \eqref{eq app deco 1}, we have (see also the proof of
Lemma~\ref{Lemma BS score and Hessian})
\begin{align*}
V^{\ast}\left(  \frac{1}{T}\int_{0}^{T}d(t)dM^{\ast}(t)\right)   &  =\frac
{1}{T^{2}}\int_{0}^{T}d(t)^{2}\hat{\lambda}(t)dt=\frac{1}{T^{2}}\int_{0}%
^{T}d(t)^{2}\lambda(t)dt\\
&  +\frac{1}{T^{2}}\int_{0}^{T}d(t)^{2}\partial_{\theta}\lambda(t)dt\left(
(\hat{\theta}-\theta_{0})(1+o_{p}(1))\right) \\
&  =O_{p}(T^{-1})
\end{align*}
as $\int_{0}^{T}d(t)^{2}\lambda(t)dt\leq\lambda_{L}^{-1}\int_{0}^{T}%
d(t)^{2}\lambda^{2}(t)dt=O_{p}(T)$, and
\[
\frac{1}{T^{2}}\int_{0}^{T}d(t)^{2}\partial_{\theta}\lambda(t)dt\leq\frac
{1}{T}\sup_{t\in\lbrack0,T]}|d(t)\lambda(t)|\frac{1}{T\lambda_{L}}\int_{0}%
^{T}d(t)|\partial_{\theta}\lambda(t)|dt
\]
where for any $\varepsilon>0$,
\begin{align*}
P\left(  \frac{1}{T}\sup_{t\in\lbrack0,T]}|d(t)\lambda(t)|>\varepsilon\right)
&  =\int_{0}^{T}\left(  P(|d(t)\lambda(t)|>\varepsilon T\right)  dt\\
&  \leq T\frac{E|d(t)\lambda(t)|^{2}}{\varepsilon^{2}T^{2}}=O_{p}(T^{-1})
\end{align*}
under the stated conditions and $T^{-1}\int_{0}^{T}d(t)|\partial_{\theta
}\lambda(t)|dt=O_{p}(1)$ as shown above.

\subsection{Third derivative of the RIB likelihood}

\label{sec 3rd der RIB}

\begin{lemma}
\label{lemma 3rd derivativ par RIB} Under the conditions of
Theorem~\ref{Th asy validity of the RIB} for the RIB,
\[
\sup_{\theta\in N_{\epsilon}(\theta_{0})}|\partial_{\theta_{i},\theta
_{j},\theta_{k}}^{3}\frac{1}{T}l_{T}^{\ast}(t;\theta)|\leq c_{T}^{\ast
}\overset{p^{\ast}}{\rightarrow}_{p}c.
\]

\end{lemma}

\noindent Proof. Similarly to the proof of
Lemma~\ref{lemma 3rd derivativ par FIB}, we have
\begin{align*}
\sup_{\theta\in\mathcal{N(}\theta_{0})}\left\vert \frac{1}{T}\partial_{\theta
}^{3}l_{T}^{\ast}(\theta)\right\vert  &  =\sup_{\theta\in\mathcal{N(}%
\theta_{0})}\left\vert \frac{1}{T}\int_{0}^{T}\partial_{\theta}^{3}\log
\lambda^{\ast}(t,\theta)dN^{\ast}(t)-\frac{1}{T}\int_{0}^{T}\partial_{\theta
}^{3}\lambda^{\ast}(t,\theta)dt\right\vert \\
&  \leq\sup_{\theta\in\mathcal{N(}\theta_{0})}\left\vert \frac{1}{T}\int
_{0}^{T}\partial_{\theta}^{3}\log\lambda^{\ast}(t,\theta)dN^{\ast
}(t)\right\vert \\
&  +\sup_{\theta\in\mathcal{N(}\theta_{0})}\left\vert \frac{1}{T}\int_{0}%
^{T}\partial_{\theta}^{3}\lambda^{\ast}(t,\theta)dt\right\vert =:A_{T}^{\ast
}+B_{T}^{\ast}.
\end{align*}
Notice that, in contrast to the FIB, the term $B_{T}^{\ast}$ now depends on
the bootstrap data. The reminder of the proof follows the same lines of the
proof of Lemma~\ref{lemma 3rd derivativ par FIB}. The difference is that,
since conditionally on the data, the intensity of the bootstrap process is a
stochastic process, in order to bound the third order derivatives of
$\lambda^{\ast}(t,\theta)$ and $\log\lambda^{\ast}(t,\theta)$, the additional
condition in Assumption 2(c$^{\ast}$) is required.

\section{Miscellaneous proofs for bootstrap theory\label{sec suppl aux proofs}%
}

\subsection{Proof of Lemma~\ref{lem BS2}}

The proof is based on the proof of Lemma 1 in
\citet{JR2004} [henceforth, JR04], which is modified here for the bootstrap.
With $\tilde{\ell}_{T}(\theta)=-\frac{2}{T}\ell_{T}^{\ast}(\theta)$,
$\partial\tilde{\ell}_{T}(\theta_{T}^{\ast})=\left.  \partial\tilde{\ell}%
_{T}(\theta)\partial\theta^{\prime}\right\vert _{\theta=\theta_{T}^{\ast}}$
and $\partial^{2}\tilde{\ell}_{T}(\theta^{\ast})=\left.  \partial^{2}%
\tilde{\ell}_{T}(\theta)/\partial\theta\partial\theta^{\prime}\right\vert
_{\theta=\theta_{T}^{\ast}}$, using Assumption~A.1(iv) it follows as in JR04
that for any $v_{1},v_{2}\in\mathbb{R}^{d}$, $\theta\in\mathcal{N(}%
\theta_{\dagger})$ and $T$ large enough, such that $\theta_{T}^{\ast}%
\in\mathcal{N(}\theta_{\dagger})$ (by Assumption~A.1(i)),
\begin{equation}
\left\vert v_{1}^{\prime}\left(  \partial^{2}\tilde{\ell}_{T}(\theta
)-\partial^{2}\tilde{\ell}_{T}(\theta_{T}^{\ast})\right)  v_{2}\right\vert
\leq\left\Vert v_{1}\right\Vert \left\Vert v_{2}\right\Vert \left\Vert
\theta-\theta_{T}^{\ast}\right\Vert \tilde{c}_{T}^{\ast},
\label{upper bound vs2}%
\end{equation}
where $\tilde{c}_{T}^{\ast}=2d^{3/2}c_{T}^{\ast}$. Next, by continuity,
$\tilde{\ell}_{T}(\theta)$ attains its minimum in any compact neighborhood
$K(\theta_{\dagger},r)=\{\theta|\ \Vert\theta-\theta_{\dagger}\Vert\leq
r\}\subseteq\mathcal{N(}\theta_{\dagger})$ of $\theta_{\dagger}$. Similar to
JR04, we next show that with probability tending to one, in probability, as
$T\rightarrow\infty,$ $\tilde{\ell}_{T}(\theta)$ cannot obtain its minimum on
the boundary of $K(\theta_{\dagger},r)$ and that $\tilde{\ell}_{T}(\theta)$ is
convex in the interior of $K(\theta_{\dagger},r)$, $\operatorname*{int}%
K(\theta_{\dagger},r)$. Specifically, with $v_{\theta}^{\ast}:=(\theta
-\theta_{T}^{\ast}),$ and $\bar{\theta}$ on the line from $\theta$ to
$\theta_{T}^{\ast}$, Taylor's formula gives
\begin{align}
&  \tilde{\ell}_{T}(\theta)-\tilde{\ell}_{T}(\theta_{T}^{\ast})=\partial
\tilde{\ell}_{T}(\theta_{T}^{\ast})v_{\theta}^{\ast}+\tfrac{1}{2}v_{\theta
}^{\ast\prime}\partial^{2}\tilde{\ell}_{T}(\bar{\theta})v_{\theta}^{\ast
}=\label{Taylors formula vs2}\\
&  \partial\tilde{\ell}_{T}(\theta_{T}^{\ast})v_{\theta}^{\ast}+\tfrac{1}%
{2}v_{\theta}^{\ast\prime}\left[  \tilde{\Omega}_{I}+(\partial^{2}\tilde{\ell
}_{T}(\theta_{T}^{\ast})-\tilde{\Omega}_{I})+(\partial^{2}\tilde{\ell}%
_{T}(\bar{\theta})-\partial^{2}\tilde{\ell}_{T}(\theta_{T}^{\ast}))\right]
v_{\theta}^{\ast},\nonumber
\end{align}
where $\tilde{\Omega}_{I}=2\Omega_{I}$. Denote by $\delta_{T}^{\ast}$ and
$\rho>0$, the smallest eigenvalues of the matrix $[\partial^{2}\tilde{\ell
}_{T}(\theta_{T}^{\ast})-\tilde{\Omega}_{I}]$ and $\tilde{\Omega}_{I}$,
respectively, where $\delta_{T}^{\ast}\overset{p^{\ast}}{\rightarrow}_{p}0$ by
Assumption~A.1(iii) and the fact that the smallest eigenvalue of a $d\times d$
symmetric matrix $M$ is continuous in $M$. Taken together, Assumption~A.1(ii)
and (iv), with $\tilde{c}=2k^{3/2}c$, and the uniform upper bound in
\eqref{upper bound vs2} imply that
\begin{align*}
\inf_{\theta:\Vert\theta-\theta_{\dagger}\Vert=r}[\tilde{\ell}_{T}%
(\theta)-\tilde{\ell}_{T}(\theta_{T}^{\ast})]  &  \geq-\Vert\partial
\tilde{\ell}_{T}(\theta_{T}^{\ast})\Vert\left\Vert v_{\theta}^{\ast
}\right\Vert +\tfrac{1}{2}\left\Vert v_{\theta}^{\ast}\right\Vert ^{2}\left[
\rho+\delta_{T}^{\ast}-\tilde{c}_{T}^{\ast}r\right] \\
&  \overset{p^{\ast}}{\rightarrow}_{p}\tfrac{1}{2}[\rho-\tilde{c}r]r^{2}%
=:\eta,
\end{align*}
where we have used $\left\Vert v_{\theta}^{\ast}\right\Vert \rightarrow
_{p}r=\left\Vert \theta-\theta_{\dagger}\right\Vert $ by Assumption~A.1(i).
Hence, if $r<\rho/\tilde{c}$ then $\inf_{\theta:\Vert\theta-\theta_{\dagger
}\Vert=r}[\tilde{\ell}_{T}(\theta)-\tilde{\ell}_{T}(\theta_{T}^{\ast}%
)]\geq\eta>0$ with probability tending to one (in probability). As
$\tilde{\ell}_{T}(\theta)|_{\theta=\theta_{T}^{\ast}}-\tilde{\ell}_{T}%
(\theta_{T}^{\ast})=0$, this implies that the probability that $\tilde{\ell
}_{T}(\theta)$ attains its minimum on the boundary of $K(\theta_{\dagger},r)$
tends to zero (in probability). Next, for $\theta\in K(\theta_{\dagger},r)$
and $v\in\mathbb{R}^{d}$, rewriting $v^{\prime}\partial^{2}\tilde{\ell}%
_{T}(\theta)v$ as in \eqref{Taylors formula vs2} one finds
\[
v^{\prime}\partial^{2}\tilde{\ell}_{T}(\theta)v\geq\Vert v\Vert^{2}%
(\rho+\delta_{T}^{\ast}-r\tilde{c}_{T}^{\ast})\overset{p^{\ast}}{\rightarrow
}_{p}\Vert v\Vert^{2}(\rho-r\tilde{c}).
\]
Hence, if $r<\rho/\tilde{c}$ the probability that, conditionally on the data,
$\tilde{\ell}_{T}(\theta)$ is strongly convex in the interior of
$K(\theta_{\dagger},r)$ tends to $1$ in probability, and therefore it has at
most one stationary point (with probability tending to one). As in JR04, this
establishes Part~(i) of the lemma with $U(\theta_{\dagger}%
)=\operatorname*{int}K(\theta_{\dagger},r)$: $\hat{\theta}_{T}^{\ast}$ is the
unique minimum point of $\tilde{\ell}_{T}(\theta)$ in $U(\theta_{\dagger})$,
and $\partial\tilde{\ell}_{T}(\hat{\theta}_{T}^{\ast})=0$ implies
$\partial\ell_{T}^{\ast}(\hat{\theta}_{T}^{\ast})=0$. Likewise, it follows
that $\hat{\theta}_{T}^{\ast}-\theta_{\dagger}\overset{p^{\ast}}{\rightarrow
}_{p}0$, which establishes Part~(ii) of the lemma, as by Assumption~A.1(i),
\[
\left\Vert \hat{\theta}_{T}^{\ast}-\theta_{T}^{\ast}\right\Vert \leq\left\Vert
\hat{\theta}_{T}^{\ast}-\theta_{\dagger}\right\Vert +\left\Vert \theta
_{\dagger}-\theta_{T}^{\ast}\right\Vert \overset{p^{\ast}}{\rightarrow}_{p}0.
\]
As to Part~(iii) of the lemma, note that by Assumption~A.1(ii) and Taylor's
formula for $\partial\tilde{\ell}_{T}(\theta)/\partial\theta_{j},j=1,\dots
,d$,
\begin{equation}
T^{1/2}\partial\tilde{\ell}_{T}(\theta_{T}^{\ast})=(\tilde{\Omega}_{I}%
+Q_{T}(\bar{\theta}))T^{1/2}(\hat{\theta}_{T}^{\ast}-\theta_{T}^{\ast}),
\label{Score vs2}%
\end{equation}
where $Q_{T}(\bar{\theta}):=\partial^{2}\tilde{\ell}_{T}(\bar{\theta}%
)-\tilde{\Omega}_{I}$, $\bar{\theta}$ being a point on the line from
$\theta_{T}^{\ast}$ to $\hat{\theta}_{T}^{\ast}$. Next, $Q_{T}(\bar{\theta
})\overset{p^{\ast}}{\rightarrow}_{p}0$ by \eqref{upper bound vs2},
\begin{align*}
|v_{1}^{\prime}Q_{T}(\bar{\theta})v_{2}|  &  \leq|v_{1}^{\prime}\left(
\partial^{2}\tilde{\ell}_{T}(\bar{\theta})-\partial^{2}\tilde{\ell}_{T}%
(\theta_{T}^{\ast})\right)  v_{2}|+|v_{1}^{\prime}(\partial^{2}\tilde{\ell
}_{T}(\theta_{T}^{\ast})-\tilde{\Omega}_{I})v_{2}|\\
&  \leq\Vert v_{1}\Vert\Vert v_{2}\Vert\Vert\bar{\theta}-\theta_{T}^{\ast
}\Vert\tilde{c}_{T}^{\ast}+|v_{1}^{\prime}(\partial^{2}\tilde{\ell}_{T}%
(\theta_{T}^{\ast})-\tilde{\Omega}_{I})v_{2}|\overset{p^{\ast}}{\rightarrow
}_{p}0
\end{align*}
since $\bar{\theta}-\theta_{T}^{\ast}\overset{p^{\ast}}{\rightarrow}_{p}0$,
$\tilde{c}_{T}^{\ast}\overset{p^{\ast}}{\rightarrow}_{p}\tilde{c}<\infty$ and
applying Assumption~A.1(iii). Hence, using Assumption~A.1(ii),
\eqref{Score vs2} gives the desired,
\[
T^{1/2}(\hat{\theta}_{T}^{\ast}-\theta_{T}^{\ast})\overset{d^{\ast}%
}{\rightarrow}_{p}\mathcal{N(}0,\tilde{\Omega}_{I}^{-1}2^{2}\Omega_{S}%
\tilde{\Omega}_{I}^{-1})=\mathcal{N(}0,\Omega_{I}^{-1}\Omega_{S}\Omega
_{I}^{-1}).
\]
This completes the proof.

\subsection{Proof of Lemma~\ref{Lemma inhomogeneous PP CLT}}

Assume w.l.g. that $T$ is an integer and let $\varepsilon_{T,k}^{\ast}%
:=\int_{k-1}^{k}\xi_{T}(t)dM^{\ast}(t)$, such that $Y_{T}^{\ast}=\sum
_{k=1}^{T}\varepsilon_{T,k}^{\ast}$. Clearly, $\varepsilon_{T,k}^{\ast}$
defined a martingale difference array [mda] with average conditional variance
given by
\begin{align*}
&  \frac{1}{T}\sum_{k=1}^{T}E^{\ast}((\varepsilon_{T,k}^{\ast})^{2}%
|\mathcal{F}_{k-}^{\ast})\\
&  =\frac{1}{T}\sum_{k=1}^{T}E^{\ast}\left(  \left.  \left(  \int_{k-1}^{k}%
\xi_{T}(t)dM^{\ast}(t)\right)  \left(  \int_{k-1}^{k}\xi_{T}(t)dM^{\ast
}(t)\right)  ^{\prime}\right\vert \mathcal{F}_{k-}^{\ast}\right) \\
&  =\frac{1}{T}\sum_{k=1}^{T}E^{\ast}\left(  \left.  \int_{k-1}^{k}\xi
_{T}(t)\xi_{T}(t)^{\prime}[dM^{\ast}(t)]^{2}\right\vert \mathcal{F}_{k-}%
^{\ast}\right) \\
&  =\frac{1}{T}\sum_{k=1}^{T}E^{\ast}\left(  \left.  \int_{k-1}^{k}\xi
_{T}(t)\xi_{T}(t)^{\prime}E^{\ast}([dM^{\ast}(t)]^{2}|\mathcal{F}_{t-}^{\ast
})\right\vert \mathcal{F}_{k-}^{\ast}\right) \\
&  =\frac{1}{T}\sum_{k=1}^{T}E^{\ast}\left(  \left.  \int_{k-1}^{k}\xi
_{T}(t)\xi_{T}(t)^{\prime}\lambda_{T}(t)dt\right\vert \mathcal{F}_{k-}^{\ast
}\right) \\
&  =\frac{1}{T}\sum_{k=1}^{T}\int_{k-1}^{k}\xi_{T}(t)\xi_{T}(t)^{\prime
}\lambda_{T}(t)dt=\frac{1}{T}\int_{0}^{T}h_{T}(t)dt\rightarrow_{p}V
\end{align*}
by Assumption (\ref{eq lemma inhomo PP cond var}). By the CLT for mda's, see
e.g. Corollary 3.1 in \cite{HH1980}, $T^{-1/2}Y_{T}^{\ast}\overset{d^{\ast}%
}{\rightarrow}_{p}\mathcal{N(}0,V)$, provided that the Lindeberg condition
holds on $\varepsilon_{T,k}^{\ast}$, which we prove next.

Consider, for some $\eta>0$,
\[
L_{T}^{\ast}:=\frac{1}{T}\sum_{k=1}^{T}E^{\ast}(\Vert\varepsilon_{T,k}^{\ast
}\Vert^{2}\mathbb{I}(\Vert\varepsilon_{T,k}^{\ast}\Vert>T^{1/2}\delta
)|\mathcal{F}_{k-1}^{\ast}),
\]
for which it follows that,
\begin{align*}
L_{T}^{\ast}  &  \leq\frac{1}{T^{1+\eta}\delta^{2\eta}}\sum_{k=1}^{T}E^{\ast
}\left(  \Vert\varepsilon_{T,k}^{\ast}\Vert^{2(1+\eta)}|\right) \\
&  =\frac{1}{T^{\eta}\delta^{2\eta}}\frac{1}{T}\sum_{k=1}^{T}E^{\ast}\left(
\left\Vert \int_{k-1}^{k}\xi_{T}(t)dM^{\ast}(t)\right\Vert ^{2(1+\eta
)}\right)  .
\end{align*}
Next, by Lemma A.2 in \citet{CY2017} with $p=\log_{2}(2(1+\eta))$, it holds
(with $c$ a generic constant),
\begin{align*}
&  E^{\ast}\left(  \left\Vert \int_{k-1}^{k}\xi_{T}(t)dM^{\ast}(t)\right\Vert
^{2(1+\eta)}\right) \\
\leq\  &  cE^{\ast}\left(  \int_{k-1}^{k}\left\Vert \xi_{T}(t)\right\Vert
^{2(1+\eta)}\lambda_{T}(t)dt\right)  +cE^{\ast}\left\vert \int_{k-1}^{k}%
\Vert\xi_{T}(t)\Vert^{2}\lambda_{T}(t)dt\right\vert ^{1+\eta}\\
=\  &  c\int_{k-1}^{k}\Vert\xi_{T}(t)\Vert^{2(1+\eta)}\lambda_{T}%
(t)dt+c\left(  \int_{k-1}^{k}\Vert\xi_{T}(t)\Vert^{2}\lambda_{T}(t)dt\right)
^{1+\eta}\\
\leq\  &  c\int_{k-1}^{k}\Vert\xi_{T}(t)\Vert^{2\left(  1+\eta\right)
}\lambda_{T}(t)dt+c\int_{k-1}^{k}\Vert\xi_{T}(t)\Vert^{2(1+\eta)}\lambda
_{T}(t)^{1+\eta}dt
\end{align*}
Consider the second term:%
\[
\frac{1}{T^{1+\eta}}\sum_{k=1}^{T}\int_{k-1}^{k}\Vert\xi_{T}(t)\Vert
^{2(1+\eta)}\lambda_{T}(t)^{1+\eta}dt=\frac{1}{T^{1+\eta}}\int_{0}^{T}\Vert
h_{T}(t)\Vert^{1+\eta}dt=O_{p}\left(  T^{-\eta}\right)
\]
by finite variance of $h_{T}(t)$. Next, for the first term,
assume that $\lambda_{T}(t)\geq\lambda_{L}>0$. Then
\begin{align*}
\frac{1}{T^{1+\eta}}\int_{0}^{T}\Vert\xi_{T}(t)\Vert^{2(1+\eta)}\lambda
_{T}(t)dt  &  =\frac{1}{T^{1+\eta}}\int_{0}^{T}\Vert h_{T}(t)\Vert^{1+\eta
}\frac{1}{\lambda_{T}^{\eta}(t)}dt\\
&  \leq\frac{1}{\lambda_{L}^{\eta}T^{1+\eta}}\int_{0}^{T}\Vert h_{T}%
(t)\Vert^{1+\eta}dt=O_{p}\left(  T^{-\eta}\right)  .
\end{align*}
Finally, assuming instead for the first term $\left\Vert \xi_{T}\left(
t\right)  \right\Vert \leq c_{\xi}<\infty$, it follows that,%
\[
E\left[  \Vert\xi_{T}(t)\Vert^{2(1+\eta)}\lambda_{T}(t)\right]  \leq c_{\xi
}^{2(1+\eta)}E\left(  \lambda_{T}^{2}\left(  t\right)  \right)  ^{1/2}%
<\infty,
\]
and hence $\int_{0}^{T}\Vert\xi_{T}(t)\Vert^{2(1+\eta)}\lambda_{T}%
(t)dt=O_{p}\left(  T\right)  $ as desired.

\end{document}